\documentclass{article}
\usepackage[utf8]{inputenc}
\usepackage{amsmath,amssymb,amsthm}
\usepackage{graphicx}
\usepackage{epstopdf}
\usepackage{mathtools}
\usepackage{fancyhdr}
\usepackage{hyperref}
\usepackage{cleveref}
\usepackage{csquotes}
\usepackage{float}
\usepackage{verbatim}
\usepackage{caption}
\usepackage{subcaption}
\usepackage{listings}
\usepackage{color}
\usepackage{cases}
\usepackage[T1]{fontenc}
\usepackage{setspace}
\usepackage{times}
\usepackage{bm}
\usepackage{algorithm}
\usepackage{algorithmicx}
\usepackage{algpseudocode}
\usepackage{adjustbox}
\usepackage{multirow}
\DeclareCaptionLabelFormat{andtable}{#1~#2  \&  \tablename~\thetable}
\usepackage[a4paper,left=1.6cm,right=1.6cm,top=1.9cm,bottom=3.7cm]{geometry}

\DeclarePairedDelimiter\abs{\lvert}{\rvert}%
\newcommand*\diff{\mathop{}\!\mathrm{d}}

\newcommand\myeq{\mathrel{\overset{\makebox[0pt]{\mbox{\normalfont\tiny\sffamily def}}}{=}}}
\DeclareMathOperator*{\argmin}{arg\,min}

\DeclarePairedDelimiter\floor{\lfloor}{\rfloor}

\algnewcommand\algorithmicinput{\textbf{Input:}}
\algnewcommand\INPUT{\item[\algorithmicinput]}
\algnewcommand\algorithmicoutput{\textbf{Output:}}
\algnewcommand\OUTPUT{\item[\algorithmicoutput]}

\usepackage[english]{babel}
 
\usepackage[backend=biber,
 style=authoryear,
 uniquename=false,
 maxbibnames=99,
 firstinits=true,
 dashed=false,
 maxcitenames=2,
 uniquelist=false
]{biblatex}
\DeclareNameAlias{author}{last-first}
\renewbibmacro{in:}{}

\newtheorem{theorem}{Theorem}[section]

\newtheorem{lemma}[theorem]{Lemma}
\newtheorem{assumption}{Assumption}
\newtheorem{proposition}{Proposition}

\newtheorem{theorem1}{Theorem}
\newtheorem{proofpart}{Part}[theorem1]

\theoremstyle{remark}
\newtheorem*{remark}{Remark}

\theoremstyle{definition}

\title{On Convergence Rate of the Generalized Diversity Subsampling Method} 
\author{
  Shang, Boyang\\
  \texttt{Northwestern University (PhD graduate on Sep 2022\footnote{This research was conducted when the author was a Ph.D. student at Northwestern University.})}\\
  \texttt{City University of Hong Kong (Postdoc, Nov 2022 - Nov 2023)}\\
}
\date{}

\begin{document}
\maketitle
\begin{abstract}
\cite{shang2022diversity} proposes a useful algorithm, named generalized Diversity Subsampling (g-DS) algorithm, to select a subsample following some target probability distribution from a finite data set and demonstrates its effectiveness numerically.  While the asymptotic performances of g-DS when the true data distribution is known was discussed in \cite{shang2022diversity}, it remains an interesting question how the estimation errors in the density estimation step, which is an unavoidable step to use g-DS in real-world data sets, influences its asymptotic performance. In this paper, we study the pointwise convergence rate of probability density function (p.d.f) the g-DS subsample to the target p.d.f value, as the data set size approaches infinity, under consideration of the pointwise bias and variance of the estimated data p.d.f. 
\end{abstract}


\section{Introduction}
Selecting a subsample following some target distribution from a finite data set is a useful procedure, for instance, to select a diverse subsample (\cite{shang2022diversity} or a sequential subsample that gradually incorporates information gained from previous experiments (\cite{joseph2021supervised}). The asymptotic performances of such an approach guide how close a selected subsample will be to the target, under the influence of each step of the algorithm, when the data set size approaches infinity. In this paper, we focus on (a simplified version of) the generalized Diversity Subsampling (g-DS) algorithm proposed by \cite{shang2022diversity} and study its asymptotic behavior. 

We introduce a few necessary notations here and then summarize a simplified g-DS algorithm (\cite{shang2022diversity}). Let $\boldsymbol{X}_1, \cdots, \boldsymbol{X}_N$ be identically and independently distributed (i.i.d) with random vector $\boldsymbol{X} \in \mathbb{R}^q$. Suppose that $\boldsymbol{X}$ follows some unknown distribution with p.d.f $f(\boldsymbol{X})$. The g-DS algorithm aims to select a subsample of size $n$ from $\{\boldsymbol{X}_1, \cdots, \boldsymbol{X}_N\}$ such that the selected subsample follows (as closely as possible) a desired distribution with a known target p.d.f $g(\boldsymbol{X})$. It mainly consists of two steps:
\begin{enumerate}
    \item Estimate the unknown data p.d.f $f(\boldsymbol{x})$ as $\hat{f}(\boldsymbol{x})$ using $\{\boldsymbol{X}_1, \cdots, \boldsymbol{X}_N\}$;
    \item Sample without replacement from $\{\boldsymbol{X}_1, \cdots, \boldsymbol{X}_N\}$ a subset of size $n$, with the probability of each $\boldsymbol{X}_i$ being selected proportional to $\frac{g(\boldsymbol{X}_i)}{\hat{f}(\boldsymbol{X}_i)}$, $i = 1, \cdots, N$.
\end{enumerate}
The performances of the g-DS algorithm are closely related to the accuracy of the estimated $\hat{f}(\boldsymbol{x})$. Although \cite{shang2022diversity} discusses (by citing Theorem 2 in \cite{skare2003improved}) the asymptotic performances of g-DS when $\hat{f}(\boldsymbol{x})$ perfectly equals $f(\boldsymbol{x})$, in practice $\hat{f}(\boldsymbol{x})$ will always contain stochastic errors and it is of practical interest to learn the impact of this error on the asymptotic performances of g-DS. And this paper focuses on addressing this issue. Note that the g-DS algorithm proposed in \cite{shang2022diversity} contains a few other steps to further improve the algorithm's numerical performance; we only focus on the two major steps listed above for simplicity. In the next section, we specify our assumptions on $\hat{f}(\boldsymbol{x})$ and study the pointwise convergence of the p.d.f of the subsample selected by g-DS as $N$ approaches infinity.

\section{Assumptions and Main Results}\label{sec-results}
 In this section, we study the asymptotic properties of the g-DS algorithm under an assumed specific form of $\hat{f}(\boldsymbol{x})$ (\Cref{assump-fhat}) and a few regularity conditions (\Cref{assump-reg}). The main result is stated in \Cref{ds-convergence}. Inspired by the asymptotic properties of KDE (\cite{chen2017tutorial}), \Cref{assump-fhat} assumes that, for any fixed $N$, $\hat{f}(\boldsymbol{x})$ has a (potentially non-zero) bias and a Gaussian stochastic error. 

\begin{assumption}[Convergence of the Estimated Density]\label{assump-fhat}
\textcolor{black}{
Let $\boldsymbol{X}^N = (\boldsymbol{X}_i : i = 1,2,\cdots,N)$, where $\boldsymbol{X}_1,\boldsymbol{X}_2,\cdots,\boldsymbol{X}_N \in \mathbb{R}^q$ are independently and identically distributed (i.i.d.) copies of random vector $\boldsymbol{X}$ having density function $f(\boldsymbol{x})$ with support $S \subset \mathbb{R}^q$. Let $g(\boldsymbol{x}) = c\tilde{g}(\boldsymbol{x})$ be the desired density function for the selected subsample that is known up to a constant $c \in \mathbb{R}^+$ and let the support of $g(\boldsymbol{x})$ be $S_g \subset S$. Let $\hat{f}_N(\boldsymbol{X}_i) \myeq{} f(\boldsymbol{X}_i) \left(1 + \tau_N(\boldsymbol{X}_i) + \sigma_N(\boldsymbol{X}_i)W_i \right)$ denote the estimator of $f(\boldsymbol{X}_i)$ using $\boldsymbol{X}^N$, for $i = 1, \cdots, N$. Here, $\tau_N(\boldsymbol{x})$ is a real-valued deterministic function on $S$ of form $\tau_N(\boldsymbol{x}) \myeq \tau(\boldsymbol{x}) + N^{-r_1} a_1(\boldsymbol{x})$ and $\abs{\tau(\boldsymbol{x})} < \varepsilon < 1, \forall \boldsymbol{x} \in S$. $\sigma_N(\boldsymbol{x})$ is a real-valued deterministic function on $S$ of form  $\sigma_N(\boldsymbol{x}) \myeq N^{-r_2} b_1(\boldsymbol{x})$. \textcolor{black}{For simplicity, we let $r_1 > \frac{1}{3}$ and $r_2 > \frac{1}{3}$. Also,} $W_i \stackrel{i.i.d}{\sim} \mathcal{N}(0,1)$, the standard normal distribution, for $i = 1, \cdots, N$ and ${\{W_i\}}_{i=1}^{N}$ are independent of $\boldsymbol{X}^N$. In addition, assume $\abs{a_1(\boldsymbol{x})} < \alpha < 1 - \varepsilon$,  and $\abs{b_1(\boldsymbol{x})} < \beta \in \mathbb{R}^+$, $\forall \boldsymbol{x} \in S$.
}
\end{assumption}

\textcolor{black}{
 Under \Cref{assump-fhat}, the bias is $E[\hat{f}_N(\boldsymbol{x})] - f(\boldsymbol{x}) = f(\boldsymbol{x})\tau_N(\boldsymbol{x}) \rightarrow f(\boldsymbol{x})\tau(\boldsymbol{x})$, as $N \rightarrow +\infty$. So when $\tau_N(\boldsymbol{x}) = 0$, $\forall \boldsymbol{x} \in S$, $\hat{f}_N(\boldsymbol{x})$ is an unbiased estimator of $f(\boldsymbol{x})$. When $\tau(\boldsymbol{x}) = 0$, $\forall \boldsymbol{x} \in S$, $\hat{f}_N(\boldsymbol{x})$ is asymptotically unbiased. Note that the independence conditions on ${\{W_i\}}_{i=1}^N$ and $\boldsymbol{X}^N$ is violated in practice for the KDE. One justification is that, as $N \rightarrow \infty$ and the KDE bandwidth shrinks accordingly, the dependence of $W_i$ on $\boldsymbol{X}^N \setminus \{\boldsymbol{X}_i\}$ decreases. }

\begin{assumption}[Regularity Conditions]\label{assump-reg}
\textcolor{black}{
Let $\boldsymbol{X}^N$, $f(\boldsymbol{x})$, $g(\boldsymbol{x})$, $S$, $S_g$, $a_1(\boldsymbol{x})$, $b_1(\boldsymbol{x})$ and $\tau(\boldsymbol{x})$ be as in \Cref{assump-fhat}. Let \Cref{assump-fhat} hold. 
We assume the following:
\begin{itemize}
    \item $E[\frac{g^m(\boldsymbol{X})}{f^m(\boldsymbol{X})}] = \int_{S} \frac{g^m(\boldsymbol{x})}{f^{m-1}(\boldsymbol{x})}\diff x < +\infty$, for any $m = 1, 2, \cdots$ (note that  when $m = 2$, this implies $\sigma^2_{i, k-i} < \infty$ under \Cref{assump-fhat})
    \item The expectation of $\frac{g(\boldsymbol{X_1)}}{f(\boldsymbol{X}_1)}
    \frac{1}{1 - \big[
    \abs{\tau(\boldsymbol{X}_1)} + N^{-r_1}\abs{a_1(\boldsymbol{X}_1)} + N^{-r_2 }\abs{b_1(\boldsymbol{X}_1) W_1}
    \big]}$  is finite \textcolor{black}{for any finite but large $N$}.
\end{itemize}
}
\end{assumption}

\begin{theorem}[Convergence of the generalized DS algorithm (\cite{shang2022diversity})]\label{ds-convergence}
\textcolor{black}{Let $\boldsymbol{X}^N$, $\boldsymbol{X}$, $f(\boldsymbol{x})$, $S$, $g(\boldsymbol{x})$, $S_g$, $\tilde{g}(\boldsymbol{x})$, $c$ and $n$ be as in the statement of \Cref{assump-fhat}, and suppose \Cref{assump-fhat} and \Cref{assump-reg} hold. \textcolor{black}{For simplicity, let $n \geq 2$.} For each $N = 1, 2, \cdots$, let $(\boldsymbol{Z}_N^k: k = 1,2,\cdots,n)$ be drawn sequentially from $\boldsymbol{X}^N$, without replacement, according to the conditional p.m.f \Cref{pmf-ds-1}(for $k = 1$)}
\begin{equation}\label{pmf-ds-1}\color{black}
    P(\boldsymbol{Z}_N^1 = \boldsymbol{X}_i | \boldsymbol{X}^N) = \frac{\frac{\tilde{g}(\boldsymbol{x})}{\hat{f}_N(\boldsymbol{X}_i)}}{\sum_{j=1}^{N} \frac{\tilde{g}(\boldsymbol{x})}{\hat{f}_N(\boldsymbol{X_j})}},
\end{equation}
\textcolor{black}{and conditional p.m.f \Cref{pmf-ds-2} (for $k \geq 2$)}
\begin{equation}\label{pmf-ds-2}\color{black}
    P(\boldsymbol{Z}_N^k = \boldsymbol{X}_i | \boldsymbol{X}^N, \boldsymbol{Z}_N^1, \cdots, \boldsymbol{Z}_N^{k-1}) = \frac{\frac{\tilde{g}(\boldsymbol{X}_i)}{\hat{f}_N(\boldsymbol{X}_i)}}{\sum_{j=1}^N \frac{\tilde{g}(\boldsymbol{X}_j)}{\hat{f}_N(\boldsymbol{X}_j)} - \sum_{j=1}^{k-1} \frac{\tilde{g}(\boldsymbol{Z}_N^j)}{\hat{f}_N(\boldsymbol{Z}_N^j)}},
\end{equation}
\textcolor{black}{respectively. For \Cref{pmf-ds-1}, $i=1,2,\cdots,N$, and for \Cref{pmf-ds-2}, $i \in \{1,2,\cdots,N\} \setminus \{ j_1^N,j_2^N,\cdots,j_{k-1}^N\}$. Here $j_1^N, j_2^N, \cdots, j_n^N$ are the indices of the sampled observations in $\boldsymbol{X}^N$, i.e., $\boldsymbol{Z}_N^k = \boldsymbol{X}_{j_k^N}$, for $k= 1,2,\cdots,n$. Denote the joint p.d.f of $\boldsymbol{Z}_N^1, \cdots, \boldsymbol{Z}_N^{n}$ as $p_N^n(\boldsymbol{z}_1, \cdots, \boldsymbol{z}_{n})$. Let }
\textcolor{black}{$\boldsymbol{z}_{1:n} \myeq (\boldsymbol{z}_1, \cdots, \boldsymbol{z}_n) \in S^n$, and} 
\color{black}{
\begin{align}
\mathcal{I}(\boldsymbol{z}_{1:n}) &= \prod_{k=1}^{n}\left( 1 + \tau(\boldsymbol{z}_k) \right)
\label{I-def}
\\
   A_1(\boldsymbol{z}_{1:n}) &= 
   \sum_{j=1}^{n} a_1(\boldsymbol{z}_j) \prod_{\stackrel{k=1}{k \neq j}}^{n}  \left( 1 + \tau(\boldsymbol{z}_k) \right)
   \label{A1-def}
   \\
  A_2(\boldsymbol{z}_{1:n}) & =
  \sum_{\stackrel{\{i,j\}\subset \{1,\cdots, n\}}{i<j}}  a_1(\boldsymbol{z}_i)a_1(\boldsymbol{z}_j)\prod_{\stackrel{k=1}{k\notin \{i,j\}}}^{n} \left(1 + \tau(\boldsymbol{z}_k) \right) ,
  \label{A2-def}
\end{align}
}
\textcolor{black}{where $\prod_{k \in \emptyset} (1 + \tau(\boldsymbol{z}_k)) \myeq 0$. Also denote $\mu_{i, k-i} \myeq E_{(\boldsymbol{X}_1, W_1)}\left[
{(-1)}^k\binom{k}{i}\frac{g(\boldsymbol{X}_1)}{f(\boldsymbol{X}_1)}\frac{a_1^{i}(\boldsymbol{X}_1) b_1^{k-i}(\boldsymbol{X}_1)}{({1+\tau(\boldsymbol{X}_1))}^{k+1}}W_1^{k-i}
\right]$ for $k =0, 1, 2$, $i = 0, \cdots, k$ and denote  $\sigma^2_{0, 0} \myeq Var\left[
\frac{g(\boldsymbol{X}_1)}{f(\boldsymbol{X}_1)}\frac{ 1}{{1+\tau(\boldsymbol{X}_1)}}
\right]$. Then}
\begin{equation}\label{ptwise-err-ds}
\begin{aligned}
    &\frac{p_N^n(\boldsymbol{z}_1, \cdots, \boldsymbol{z}_{n})}{\prod_{k=1}^n g(\boldsymbol{z}_k)} -1 \\ 
  =&
    \mu_{0,0}^{-n}\mathcal{I}^{-1}(\boldsymbol{z}_{1:n})-1
    \\
    -&
    N^{-r_1}\left\{
    n\mu_{0,0}^{-n-1}\mu_{1,0}\mathcal{I}^{-1}(\boldsymbol{z}_{1:n})
    +
    \mu^{-n}_{0,0}\mathcal{I}^{-2}(\boldsymbol{z}_{1:n})A_1(\boldsymbol{z}_{1:n})
    \right\}\\
    +&
    N^{-2r_1}\Big\{
    -n\mu_{0,0}^{-n-1}\mu_{2,0}\mathcal{I}^{-1}(\boldsymbol{z}_{1:n})
    + \frac{1}{2}n(n+1)\mu_{0,0}^{-n-2}\mu_{1,0}^2\mathcal{I}^{-1}(\boldsymbol{z}_{1:n}) \\
    &\qquad
    +n\mu_{0,0}^{-n-1}\mu_{1,0}\mathcal{I}^{-2}(\boldsymbol{z}_{1:n})A_1(\boldsymbol{z}_{1:n}) - \mu_{0,0}^{-n}\mathcal{I}^{-2}(\boldsymbol{z}_{1:n})A_2(\boldsymbol{z}_{1:n})\\
    &
    \qquad +\mu_{0,0}^{-n} \mathcal{I}^{-3}(\boldsymbol{z}_{1:n})A_1^2(\boldsymbol{z}_{1:n})
    \Big\}\\
    +&
    N^{-2r_2}\Big\{
    -n\mu_{0,0}^{-n-1}\mu_{0,2}\mathcal{I}^{-1}(\boldsymbol{z}_{1:n})  +
    \mu_{0,0}^{-n}\mathcal{I}^{-3}(\boldsymbol{z}_{1:n})\sum_{j=1}^n b_1^2(\boldsymbol{z}_j)\prod_{\stackrel{k=1}{k \neq j}}^{n}{\left(1 + \tau(\boldsymbol{z}_k)\right)}^2 \Big \}\\
    +&
    N^{-1}
    \Big[ 
    \frac{n(n+1)}{2}\mu_{0,0}^{-n} +
    \frac{1}{2}n(n+1)\mu_{0,0}^{-n-2}\sigma_{0,0}^2\\
    &\qquad 
    -\mu_{0,0}^{-n-1}\left(
    \sum_{i=1}^{n}i\frac{g(\boldsymbol{z}_i)}{f(\boldsymbol{z}_i)}{\left( 
    1 + \tau(\boldsymbol{z}_i)
    \right)}^{-1}
    \right)
    \Big]\mathcal{I}^{-1}(\boldsymbol{z}_{1:n})
    \\
    +&
    o(N^{-1}).
\end{aligned}
\end{equation}
\end{theorem}

See \Cref{append-proof-ds} for the proof of \Cref{ds-convergence}.

\begin{remark}
    \Cref{ds-convergence} agrees with the Theorem 2 in \cite{skare2003improved} (and in \cite{shang2022diversity}) when $\hat{f}(\boldsymbol{x})$ equals $f(\boldsymbol{x})$.
\end{remark}

\begin{remark}
    \Cref{ds-convergence} indicates that the convergence rates of the bias and various of $\hat{f}_N(\boldsymbol{x})$ to $0$ both influence the convergence rate of the g-DS algorithm. In short, the g-DS algorithm converges at rate of $O(N^{-\min\{r_1,2r_2,1\}})$, when $\hat{f}_N(\boldsymbol{x})$ follows the form in \Cref{assump-fhat} and its bias and variance converge to $0$ at rates $O(N^{-r_1})$ and $O(N^{-2r_2})$ respectively. The g-DS algorithm fails to converge if $\hat{f}_N(\boldsymbol{x})$ is not asymptotically unbiased.
\end{remark}

\section{Discussion}
This work studies the influence of estimation errors in the density estimation step on the convergence of the g-DS algorithm. As mentioned before, the assumptions on $\hat{f}$ currently cannot be satisfied by existing density estimation methods, such as KDE or Gaussian Mixture Models. We would be interested in exploring the convergence of g-DS using different density estimators in the future.  

\section*{Acknowledgement}
The author is grateful to Prof. Daniel W. Apley and Prof. Sanjay Mehrotra from Northwestern University for reading the first draft of this theorem and to Prof. Sanjay Mehrotra for providing helpful suggestions on organizing its proof better. 

\appendix
\section{Proof of \Cref{ds-convergence}}\label{append-proof-ds}
\textcolor{black}{
Before proving \Cref{ds-convergence}, we first introduce a convenient lemma. 
\begin{lemma}\label{pdf-ds}
For $k = 1, \cdots, n$, let $\boldsymbol{X}_{k:N} \myeq (\boldsymbol{X}_k, \cdots, \boldsymbol{X}_N)$, $W_{k:N} \myeq (W_k, \cdots, W_N)$, $\boldsymbol{z}_{k:n} \myeq (\boldsymbol{z}_k, \cdots, \boldsymbol{z}_n)$  and $E_{(\boldsymbol{X}_{(n+1):N},W_{1:N})}$ indicate that the expectation is taken w.r.t the joint distribution of $\boldsymbol{X}_{(n+1):N}$ and $W_{1:N}$. For \Cref{ds-convergence}, one has the p.d.f of $\boldsymbol{Z}_N^1, \cdots, \boldsymbol{Z}_N^n$ as
\begin{equation}\label{pdf-ds-eq}
\begin{split}
    &p_N^n(\boldsymbol{z}_1, \cdots, \boldsymbol{z}_n)\\ = 
    &\frac{N!}{(N-n)!} \left(\prod_{k=1}^{n} f(\boldsymbol{z}_k)\right) E_{(\boldsymbol{X}_{(n+1):N}, W_{1:N})} \left[ 
    \prod_{k=1}^{n} P\left( \boldsymbol{Z}_N^k = \boldsymbol{X}_k \middle| \boldsymbol{X}_{k:n} = \boldsymbol{z}_{k:n}, \boldsymbol{X}_{(n+1):N}, W_{k:N} \right)
    \right].
\end{split}
\end{equation}
\end{lemma}
}

\begin{proof}[Proof of \Cref{pdf-ds}]\label{proof-of-pdf-ds}
\color{black}
We first derive the c.d.f of $\boldsymbol{Z}_N^{1:n} \myeq (\boldsymbol{Z}_N^1, \cdots, \boldsymbol{Z}_N^n)$. For convenience, we denote the $j$th component of \textcolor{black}{$\boldsymbol{Z}_N^k$} as ${(\boldsymbol{Z}_N^k)}_j$ for $k = 1, \cdots, n$ and $j = 1, \cdots, q$. Then, 
\begin{equation}\label{cdf-sample}
\begin{split}
    &P\left( {\left(\boldsymbol{Z}_N^1\right)}_1 \leq z_{1,1}, \cdots, {\left(\boldsymbol{Z}_N^1\right)}_q \leq z_{1,q}, \cdots,
    {\left(\boldsymbol{Z}_N^n\right)}_1 \leq z_{n,1}, \cdots, {\left(\boldsymbol{Z}_N^n\right)}_q \leq z_{n,q}\right)\\
    =&
    \color{black}{
    \int_{-\infty}^{z_{1,1}}\cdots \int_{-\infty}^{z_{1,q}} \cdots
    \int_{-\infty}^{z_{n,1}}\cdots \int_{-\infty}^{z_{n,q}}
    \prod_{k=1}^n \prod_{j=1}^{q} \boldsymbol{I}_{\left\{ 
    y_{k,j} \leq z_{k,j}
    \right\}}
    }\\
    &\qquad
    \color{black}{
    \diff F_{\boldsymbol{Z}_N^{1:n}}\left({\left\{{\left(\boldsymbol{Z}_N^1\right)}_j = y_{1,j}\right\}}_{j=1}^q, \cdots, {\left\{{\left(\boldsymbol{Z}_N^n\right)}_j = y_{n,j}\right\}}_{j=1}^q \right)
    }
    \\
    =&
    E_{\boldsymbol{Z}_N^{1:n}} \left[
    \prod_{k=1}^n \prod_{j=1}^{q} \boldsymbol{I}_{\left\{ 
    {\left(\boldsymbol{Z}_N^k\right)}_j \leq z_{k,j}
    \right\}}
    \right]\\
    = &
    E_{(\boldsymbol{X}_{1:N},W_{1:N})}\left[
    E_{\boldsymbol{Z}_N^{1:n}|(\boldsymbol{X}_{1:N},W_{1:N})}\left[ 
    \prod_{k=1}^n \prod_{j=1}^{q} \boldsymbol{I}_{\left\{ 
    {\left(\boldsymbol{Z}_N^k\right)}_j \leq z_{k,j}
    \right\}}
    \middle|
    \boldsymbol{X}_{1:N},W_{1:N}
    \right]
    \right],
\end{split}
\end{equation}
where $E_{\boldsymbol{Z}_N^{1:n}|(\boldsymbol{X}_{1:N},W_{1:N})}$ indicates that the expectation is w.r.t the conditional distribution of $\boldsymbol{Z}_N^{1:n}$ given $(\boldsymbol{X}_{1:N},W_{1:N})$.

Noticing that, \textcolor{black}{by design, $\boldsymbol{Z}_N^{1:n}$ are selected from $\boldsymbol{X}_{1:N}$ without replacement. Thus}
\begin{equation}\label{innter-exp-perm}
\begin{split}
   &E_{\boldsymbol{Z}_N^{1:n}|(\boldsymbol{X}_{1:N},W_{1:N})}\left[ 
    \prod_{k=1}^n \prod_{j=1}^{q} \boldsymbol{I}_{\left\{ 
    {\left(\boldsymbol{Z}_N^k\right)}_j \leq z_{k,j}
    \right\}}
    \middle|
    \boldsymbol{X}_{1:N},W_{1:N}
    \right] \\= 
    &\sum_{(i_1, \cdots, i_n) \subset \mathcal{P}(1:N)} \left( \prod_{k=1}^{n}\prod_{j=1}^{q} \boldsymbol{I}_{\{
    X_{i_k, j} \leq z_{k,j}
    \}}
    \right)
    P\left( {\{\boldsymbol{Z}_N^k = \boldsymbol{X}_{i_k}\}}_{k=1}^n
    \middle|
    \boldsymbol{X}_{1:N}, W_{1:N}
    \right),
    \end{split}
\end{equation}
where $\mathcal{P}(1:N)$ denotes all permutations of set $\{1, \cdots, N \}$. \textcolor{black}{Here $X_{i_k, j}$ denotes the $j$th element of $\boldsymbol{X}_{i_k}$.} Plugging \Cref{innter-exp-perm} into \Cref{cdf-sample} and noticing that $\boldsymbol{X}_{1:N}$ and $W_{1:N}$ are both i.i.d and that they are mutually independent, one has
\begin{equation}\label{cdf-sample-2}
    \begin{split}
       & P\left( {\left(\boldsymbol{Z}_N^1\right)}_1 \leq z_{1,1}, \cdots, {\left(\boldsymbol{Z}_N^1\right)}_q \leq z_{1,q}, \cdots,
    {\left(\boldsymbol{Z}_N^n\right)}_1 \leq z_{n,1}, \cdots, {\left(\boldsymbol{Z}_N^n\right)}_q \leq z_{n,q}\right)\\
    =&
    \frac{N!}{(N-n)!} E_{(\boldsymbol{X}_{1:N}, W_{1:N})} \left[ 
    \left( \prod_{k=1}^{n}\prod_{j=1}^{q} \boldsymbol{I}_{\{
    X_{k, j} \leq z_{k,j}
    \}}
    \right)
    P\left( {\{\boldsymbol{Z}_N^k = \boldsymbol{X}_{k}\}}_{k=1}^n
    \middle|
    \boldsymbol{X}_{1:N}, W_{1:N}
    \right)
    \right]\\
    =&
    \frac{N!}{(N-n)!} \int_{-\infty}^{z_{n,q}}\cdots \int_{-\infty}^{z_{n,1}}\cdots \int_{-\infty}^{z_{1,q}}\cdots \int_{-\infty}^{z_{1,1}} \\
    & \quad \quad 
    E_{(\boldsymbol{X}_{(n+1):N},W_{1:N})} \left[
    P\left( {\{\boldsymbol{Z}_N^k = \boldsymbol{X}_{k}\}}_{k=1}^n
    \middle|
    \boldsymbol{X}_{1:n} = \boldsymbol{x}_{1:n},
    \boldsymbol{X}_{(n+1):N},
    W_{1:N}
    \right)
    \right]
    \left(\prod_{k=1}^{n}f(\boldsymbol{x}_k)\right)\\
    &\quad \quad
    \diff x_{1,1} \cdots \diff x_{1,q}
    \cdots
    \diff x_{n,1} \cdots \diff x_{n,q}.
    \end{split}
\end{equation}
In the last line of \Cref{cdf-sample-2}, we use Fubini's Theorem (Theorem 9.1 in \cite{gut2005probability}) to interchange the orders of the integration and expectation operators. 

Applying Leibniz integral rule to \Cref{cdf-sample-2} yields
\begin{equation}\label{pdf-sample}
    \begin{split}
        &p_N^n(\boldsymbol{z}_1, \cdots, \boldsymbol{z}_n) \\
        =&
        \frac{\partial^{nq}}{\partial z_{n,q} \cdots \partial z_{n,1}, \cdots, \partial z_{1,q}, \cdots, \partial z_{1,1}}
        P\left( {\left(\boldsymbol{Z}_N^1\right)}_1 \leq z_{1,1}, \cdots, {\left(\boldsymbol{Z}_N^1\right)}_q \leq z_{1,q}, \cdots,
    {\left(\boldsymbol{Z}_N^n\right)}_1 \leq z_{n,1}, \cdots, {\left(\boldsymbol{Z}_N^n\right)}_q \leq z_{n,q}\right)\\
    =& \frac{N!}{(N-n)!} \left(\prod_{k=1}^{n}f(\boldsymbol{z}_k)\right)
    E_{(\boldsymbol{X}_{(n+1):N},W_{1:N})} \left[
    P\left( {\{\boldsymbol{Z}_N^k = \boldsymbol{X}_{k}\}}_{k=1}^n
    \middle|
    \boldsymbol{X}_{1:n} = \boldsymbol{z}_{1:n},
    \boldsymbol{X}_{(n+1):N},
    W_{1:N}
    \right)
    \right].
   \end{split}
\end{equation}
By the definition of conditional probability, \textcolor{black}{\Cref{assump-fhat}, and the facts that for $k = 2, \cdots, n$, $\boldsymbol{Z}_N^k$ is chosen among $\boldsymbol{X}_{1:N} \setminus \{\boldsymbol{Z}_N^{1:(k-1)}\}$, we have,}
\begin{equation}\label{a-fact-1}
    \begin{split}
        &P\left( 
        \boldsymbol{Z}_N^k = \boldsymbol{X}_k \middle|
        {\{\boldsymbol{Z}_N^j = \boldsymbol{X}_j\}}_{j=1}^{k-1}, 
        \boldsymbol{X}_{1:n} = \boldsymbol{z}_{1:n},
    \boldsymbol{X}_{(n+1):N},
    W_{1:N}
        \right)\\
        =&
       P\left(
       \boldsymbol{Z}_N^k = \boldsymbol{X}_k \middle|
       \boldsymbol{X}_{k:n} = \boldsymbol{z}_{k:n},
    \boldsymbol{X}_{(n+1):N},
    W_{k:N}
       \right).
    \end{split}
\end{equation}
Thus
\begin{equation}\label{chain-prob}
    \begin{split}
        &
        P\left( {\{\boldsymbol{Z}_N^k = \boldsymbol{X}_{k}\}}_{k=1}^n
    \middle|
    \boldsymbol{X}_{1:n} = \boldsymbol{z}_{1:n},
    \boldsymbol{X}_{(n+1):N},
    W_{1:N}
    \right)
        \\=
        &
        \prod_{k=1}^{n} P\left( \boldsymbol{Z}_N^k = \boldsymbol{X}_k \middle| \boldsymbol{X}_{k:n} = \boldsymbol{z}_{k:n}, \boldsymbol{X}_{(n+1):N}, W_{k:N} \right).
    \end{split}
\end{equation}
The results then follows.
\end{proof}

\textcolor{black}{The following lemma about the moments of shifted reciprocal of normal random variables is needed to prove \Cref{ds-convergence}.
\begin{lemma}\label{SRnormal}
Let $W \sim \mathcal{N}(0,1)$ and $Z = \frac{1}{p + \sigma W}$, where $p \in \mathbb{R}$, $p \neq 0$ and $\sigma \neq 0$. Then, for $j = 2, 3, \cdots$, we have $E\left[ Z^j \right] = {(j-1)}^{-1}\sigma^{-2}\left( pE\left[
Z^{j-1}\right] -E\left[Z^{j-2} \right]\right)$.
\end{lemma}
}

\begin{proof}[Proof of \Cref{SRnormal}]\label{proof-sr-normal}
\color{black}
For any $j = 2, 3, \cdots$, one has $(p + \sigma x)^{-j} = -{(j-1)}^{-1}\sigma^{-1}\frac{\diff {(p+\sigma x)}^{-{(j-1)}}}{\diff x}$. Let $\phi(x) = \frac{1}{\sqrt{2\pi}}\exp{\{-x^2\}}$ denote the p.d.f of a standard normal random variable.  Then it is easy to verify that $\diff \phi(x) = -x\phi(x)\diff{x}$. Thus, 
\begin{equation}
    \begin{split}
        E\left[ Z^j \right] & = \int_{\mathbb{R}} {(p+\sigma x)}^{-j} \phi(x) \diff x \\
        &= -{(j-1)}^{-1}\sigma^{-1} \int_{\mathbb{R}} \phi(x) \diff {(p + \sigma x)}^{-(j-1)}\\
        &= {(j-1)}^{-1} \sigma^{-1} \int_{\mathbb{R}}  {(p+\sigma x)}^{-(j-1)} (-x) \phi(x) \diff x \\
        &= -{(j-1)}^{-1} \sigma^{-1} \int_{\mathbb{R}}  {(p+\sigma x)}^{-(j-2)} \sigma^{-1} \left( 1 - \frac{p}{p+\sigma x} \right) \phi(x) \diff x \\
        &= {(j-1)}^{-1} \sigma^{-2}  \left( pE\left[
Z^{j-1}\right] -E\left[Z^{j-2} \right]\right).
    \end{split}
\end{equation}
\end{proof}

\textcolor{black}{
\begin{lemma}\label{xSRNormal}
    Let $W \sim \mathcal{N}(0,1)$, $p \in \mathbb{R}$, $p \neq 0$ and $\sigma \neq 0$. Then for $j = 1, 2, \cdots$, 
    \begin{equation}\label{xsrnormal-eq}
    E\left[ W{\left( \frac{1}{p + \sigma W}\right)}^j 
    \right] = -\sigma j E\left[ {\left( \frac{1}{p + \sigma W}\right)}^{j+1}\right].
    \end{equation}
\end{lemma}
}

\begin{proof}[Proof of \Cref{xSRNormal}]\label{proof-x-sr-normal}
\color{black}
\begin{equation}
    \begin{split}
        E\left[ W{\left( \frac{1}{p + \sigma W}\right)}^j 
    \right] &= \sigma^{-1}E\left[ {\left( \frac{1}{p + \sigma W}\right)}^{j-1} \left( 1 - \frac{p}{p + \sigma W}\right)
    \right]\\
    &=-\sigma^{-1}\left(
    pE\left[ {\left( \frac{1}{p + \sigma W}\right)}^{j}
    \right]-
    E\left[ {\left( \frac{1}{p + \sigma W}\right)}^{j-1}
    \right]
    \right)\\
    &=-\sigma j E\left[ {\left( \frac{1}{p + \sigma W}\right)}^{j+1}\right],
    \end{split}
\end{equation}
where in the last equality we applied \Cref{SRnormal}.
\end{proof}

\textcolor{black}{
\begin{lemma}\label{momentsSRNormal}
For $h(W) = \frac{1}{1+\tau(\boldsymbol{z}) + N^{-r_1}a_1(\boldsymbol{z}) + N^{-r_2}b_1(\boldsymbol{z})W}$, where $\boldsymbol{z} \in \mathbb{R}^q$ and $\tau(\cdot)$, $a_1(\cdot)$, $b_1(\cdot)$, $r_1$, $r_2$, and $N$ are as stated in \Cref{ds-convergence}. $W$ is a standard normal random variable. We have for a finite $j \in \mathbb{Z}^+$, with $N$ \textcolor{black}{approaching} $+\infty$,
\begin{equation}\label{expSRNk}
    E [h^j(W)] \sim \sum_{k=j-1}^\infty N^{-(k-(j-1))2r_2}\alpha_j(k)\frac{b_1^{2(k-(j-1))}(\boldsymbol{z})}{{\{ 
    1 + \tau(\boldsymbol{z}) +N^{-r_1}a_1(\boldsymbol{z})
    \}}^{2(k+1)-j}},
\end{equation}
where $\sim$ indicates asymptotic equivalence as $N \rightarrow \infty$ and
\begin{equation}\label{akj}
    \alpha_j(k) = \left( 2k - \left( 2\left\lfloor\frac{j-1}{2}\right\rfloor+1\right) \right)!! \frac{1}{(j-1)!}2^{\left\lfloor\frac{j-1}{2}\right\rfloor} {\left\{ 
    \left(k - \left\lfloor\frac{j}{2}\right\rfloor\right)
    \left( k - \left(\left\lfloor\frac{j}{2}\right\rfloor + 1\right)\right)
    \cdots
    \left( k-\left(j-2\right) \vphantom{\frac{j}{2}} \right)
    \right\}}^{\boldsymbol{I}_{\left\{j > 2\right\}}}.
\end{equation}
Here $(k)!! \myeq  k\times (k-2) \times \cdots$ denotes the double factorial and $\floor{x}$ denotes the largest integer value no larger than $x$. $\boldsymbol{I}_{\{j>2\}}$ is the indicator function equalling $1$ if $j > 2$ and $0$ otherwise.
\end{lemma}
}

\begin{proof}[Proof of \Cref{momentsSRNormal}]\label{proof-m-sr-normal}
\color{black}
We prove the results by induction. When $j = 1$, 
$h(W)$ is a shifted reciprocal function of a normal random variable whose expectation is, by \cite{lecomte2013exact}, $\frac{\sqrt{2}}{N^{-r_2}\abs{b_1(\boldsymbol{z}_i)}}D(\frac{1+\tau(\boldsymbol{z}_i)+N^{-r_1}a_1(\boldsymbol{z}_i)}{\sqrt{2}N^{-r_2}\abs{b_1(\boldsymbol{z}_i)}})$, where $D(\cdot)$ is the Dawson function (\cite{temme2010error}).  \cite{hummer1964expansion} provides an asymptotic expansion of $E[h(W)]$ at $N = \infty$,
\begin{equation}\label{e-hw1}
\begin{split}
    E[h(W)] &\sim \frac{\sqrt{2}}{N^{-r_2}\abs{b_1(\boldsymbol{z})}}\sum_{k=0}^{\infty}(2k-1)!!2^{-k-1}{\left(\frac{\sqrt{2}N^{-r_2}\abs{b_1(\boldsymbol{z})}}{1+\tau(\boldsymbol{z})+N^{-r_1}a_1(\boldsymbol{z})} \right)}^{2k+1}\\
    &= \sum_{k=0}^{\infty} N^{-k(2r_2)} (2k-1)!!
    \frac{b_1^{2k}(\boldsymbol{z})}{{\left(
    1 + \tau(\boldsymbol{z}) + N^{-r_1}a_1(\boldsymbol{z})
    \right)}^{2k+1}},
\end{split}
\end{equation}
which agrees with the claimed result, i.e. \Cref{expSRNk}.

When $j = 2$, applying \Cref{SRnormal} and \Cref{e-hw1}, we have
\begin{equation}\label{e-hw2}
    \begin{split}
        E[h^2(W)] &= N^{2r_2}b_1^{-2}(\boldsymbol{z})\left( 
        \left(1 + \tau(\boldsymbol{z}) + N^{-r_1}a_1(\boldsymbol{z}) \right)E[h(W)] - 1
        \right)\\
        &\sim
        \sum_{k=1}^{\infty} N^{-(k-1)2r_2} (2k-1)!! 
        \frac{b_1^{2(k-1)}(\boldsymbol{z})}{{\left(
    1 + \tau(\boldsymbol{z}) + N^{-r_1}a_1(\boldsymbol{z})
    \right)}^{2k}},
    \end{split}
\end{equation}
which also agrees with the claimed result in \Cref{expSRNk}.

Now suppose that the claimed result holds at $j = n-1$ and $j = n-2$ for some $n \geq 3$, we prove that \Cref{expSRNk} holds for $j = n$. 

First observe that, by \Cref{akj}, for each $j = 1, 2, \cdots$, $\alpha_j(j-1) = 1$: When $j = 1$, $\alpha_1(0) = (-1)!! = 1$; When $j = 2$, $\alpha_2(1) = 1!! = 1$; When $j > 3$ and $j$ is even, $a_j(j-1)=(j-1)!!\frac{1}{(j-1)!}2^{\frac{j-2}{2}} \left(\frac{j}{2}-1\right)! = 1$; When $j > 3$ and $j$ is odd, $a_j(j-1) = (j-2)!!\frac{1}{(j-1)!}2^{\frac{j-1}{2}}\left( \frac{j-1}{2}\right)! = 1$.

Now by \Cref{SRnormal}, 
\begin{equation}\label{e-hwn}
    \begin{split}
        &E[h^n(W)] \\
        =& \frac{1}{n-1}N^{2r_2}b_1^{-2}(\boldsymbol{z})\left( 
        \left(1 + \tau(\boldsymbol{z}) + N^{-r_1}a_1(\boldsymbol{z}) \right)E[h^{(n-1)}(W)] - E[h^{(n-2)}(W)]
        \right)\\
        \sim& \frac{1}{n-1}N^{2r_2}b_1^{-2}(\boldsymbol{z})\Bigg(
        \left(1 + \tau(\boldsymbol{z}) + N^{-r_1}a_1(\boldsymbol{z}) \right)
        \sum_{k=n-2}^\infty N^{-(k-(n-2))2r_2}\alpha_{n-1}(k)\frac{b_1^{2(k-(n-2))}(\boldsymbol{z})}{{\{ 
    1 + \tau(\boldsymbol{z}) +N^{-r_1}a_1(\boldsymbol{z})
    \}}^{2(k+1)-(n-1)}} \\
    &\qquad -
    \sum_{k=n-3}^\infty N^{-(k-(n-3))2r_2}\alpha_{n-2}(k)\frac{b_1^{2(k-(n-3))}(\boldsymbol{z})}{{\{ 
    1 + \tau(\boldsymbol{z}) +N^{-r_1}a_1(\boldsymbol{z})
    \}}^{2(k+1)-(n-2)}}
        \Bigg)\\
        =&\frac{1}{n-1}N^{2r_2}b_1^{-2}(\boldsymbol{z})\Bigg(
        \frac{\alpha_{n-1}(n-2)}{{\{ 
    1 + \tau(\boldsymbol{z}) +N^{-r_1}a_1(\boldsymbol{z})
    \}}^{n-2}} -\frac{\alpha_{n-2}(n-3)}{{\{ 
    1 + \tau(\boldsymbol{z}) +N^{-r_1}a_1(\boldsymbol{z})
    \}}^{n-2}} +\\
    &\qquad
    \sum_{k=n-1}^\infty N^{-(k-(n-2))2r_2}\alpha_{n-1}(k)\frac{b_1^{2(k-(n-2))}(\boldsymbol{z})}{{\{ 
    1 + \tau(\boldsymbol{z}) +N^{-r_1}a_1(\boldsymbol{z})
    \}}^{2(k+1)-n}}- \\
    &\qquad
    \sum_{k=n-2}^\infty N^{-(k-(n-3))2r_2}\alpha_{n-2}(k)\frac{b_1^{2(k-(n-3))}(\boldsymbol{z})}{{\{ 
    1 + \tau(\boldsymbol{z}) +N^{-r_1}a_1(\boldsymbol{z})
    \}}^{2(k+1)-(n-2)}}
        \Bigg)\\
        =&
        \frac{1}{n-1} N^{2r_2}b_1^{-2}(\boldsymbol{z})\Bigg(
        \sum_{k=n-1}^\infty N^{-(k-(n-2))2r_2}\alpha_{n-1}(k)\frac{b_1^{2(k-(n-2))}(\boldsymbol{z})}{{\{ 
    1 + \tau(\boldsymbol{z}) +N^{-r_1}a_1(\boldsymbol{z})
    \}}^{2(k+1)-n}}- \\
    &\qquad
    \sum_{k=n-1}^\infty N^{-(k-(n-2))2r_2}\alpha_{n-2}(k-1)\frac{b_1^{2(k-(n-2))}(\boldsymbol{z})}{{\{ 
    1 + \tau(\boldsymbol{z}) +N^{-r_1}a_1(\boldsymbol{z})
    \}}^{2(k+1)-n}}\\
    =&
    \sum_{k=n-1}^{\infty} N^{-(k-(n-1))2r_2}\left( 
    \frac{1}{n-1}\left(\alpha_{n-1}(k)- \alpha_{n-2}(k-1) \right)
    \right)
    \frac{b_1^{2(k-(n-1))}(\boldsymbol{z})}{{\{ 
    1 + \tau(\boldsymbol{z}) +N^{-r_1}a_1(\boldsymbol{z})
    \}}^{2(k+1)-n}}
        \Bigg),
    \end{split}
\end{equation}
where in the fourth equality we used the observation that $a_{n-1}(n-2) = a_{n-2}(n-3) = 1$. To conclude the results in \Cref{momentsSRNormal}, we just need to show that $\alpha_n(k) = \frac{1}{n-1}\left(\alpha_{n-1}(k)- \alpha_{n-2}(k-1) \right)$. Using \Cref{akj}, we have
\begin{equation}\label{diffalpha}
\begin{split}
    &\alpha_{n-1}(k)- \alpha_{n-2}(k-1)\\
    =&
    \left( 2k - \left( 2\left\lfloor\frac{n-2}{2}\right\rfloor+1\right) \right)!! \frac{1}{(n-2)!}2^{\left\lfloor\frac{n-2}{2}\right\rfloor}\times
    \\&\qquad {\left\{ 
    \left(k - \left\lfloor\frac{n-1}{2}\right\rfloor\right)
    \left( k - \left(\left\lfloor\frac{n-1}{2}\right\rfloor + 1\right)\right)
    \cdots
    \left( k-\left(n-3\right) \vphantom{\frac{j}{2}} \right)
    \right\}}^{\boldsymbol{I}_{\left\{n > 3\right\}}}-\\
    &\qquad
    \left( 2(k-1) - \left( 2\left\lfloor\frac{n-3}{2}\right\rfloor+1\right) \right)!! \frac{1}{(n-3)!}2^{\left\lfloor\frac{n-3}{2}\right\rfloor}\times
    \\&\qquad
    {\left\{ 
    \left(k -1 - \left\lfloor\frac{n-2}{2}\right\rfloor\right)
    \left( k -1- \left(\left\lfloor\frac{n-2}{2}\right\rfloor + 1\right)\right)
    \cdots
    \left( k-1-\left(n-4\right) \vphantom{\frac{j}{2}} \right)
    \right\}}^{\boldsymbol{I}_{\left\{n > 4\right\}}}
    \end{split}
\end{equation}

When $n > 4$ and $n$ is even, \Cref{diffalpha} yields
\begin{equation}\label{diffalpha-neven}
    \begin{split}
        &\alpha_{n-1}(k)- \alpha_{n-2}(k-1)\\
    =&\left(2k-(n-1) \right)!!\frac{1}{(n-3)!}2^{\frac{n-4}{2}}\left( k-\frac{n}{2}\right)\left( k-\left(\frac{n}{2}+1\right)\right)\cdots\left(k-(n-3)\vphantom{\frac{n}{2}}\right)\left\{ 
    \frac{1}{n-2}2\left(k - \left( \frac{n}{2}-1\right)\right)-1
    \right\}\\
    =&
    \left(2k-(n-1) \right)!!\frac{1}{(n-2)!}2^{\frac{n-2}{2}}\left( k-\frac{n}{2}\right)\left( k-\left(\frac{n}{2}+1\right)\right)\cdots\left(k-(n-2)\vphantom{\frac{n}{2}}\right).
    \end{split}
\end{equation}
Using \Cref{akj}, we see that $\alpha_n(k) = \frac{1}{n-1}\left(\alpha_{n-1}(k)- \alpha_{n-2}(k-1) \right)$ in this case. 

When $n > 4$ and $n$ is odd, \Cref{diffalpha} yields
\begin{equation}\label{diffalpha-odd}
    \begin{split}
        &\alpha_{n-1}(k)- \alpha_{n-2}(k-1)\\
    =&\left(2k-n \right)!!\frac{1}{(n-3)!}2^{\frac{n-3}{2}}\left( k-\frac{n-1}{2}\right)\left( k-\left(\frac{n-1}{2}+1\right)\right)\cdots\left(k-(n-3)\vphantom{\frac{n}{2}}\right)\left\{ 
    \frac{2k-(n-2)}{n-2}-1
    \right\}\\
    =&
    \left(2k-n \right)!!\frac{1}{(n-2)!}2^{\frac{n-1}{2}}\left( k-\frac{n-1}{2}\right)\left( k-\left(\frac{n-1}{2}+1\right)\right)\cdots\left(k-(n-2)\vphantom{\frac{n}{2}}\right).
    \end{split}
\end{equation}
Using \Cref{akj}, we again see that $\alpha_n(k) = \frac{1}{n-1}\left(\alpha_{n-1}(k)- \alpha_{n-2}(k-1) \right)$. 

When $n = 4$, $\alpha_{n-1}(k)- \alpha_{n-2}(k-1) = (2k-3)!!(k-2)$ and when $n = 3$, $\alpha_{n-1}(k)- \alpha_{n-2}(k-1) = (2k-3)!!2(k-1)$. In both cases, one can easily verify the relation $\alpha_n(k) = \frac{1}{n-1}\left(\alpha_{n-1}(k)- \alpha_{n-2}(k-1) \right)$ using \Cref{akj}.

Now that we have showed $\alpha_n(k) = \frac{1}{n-1}\left(\alpha_{n-1}(k)- \alpha_{n-2}(k-1) \right)$ for all $n \geq 3$, provided that \Cref{expSRNk} and \Cref{akj} hold for $j= n-1$ and $j = n-2$. Together with \Cref{e-hwn}, the proof completes. 
\end{proof}

\begin{lemma}\label{lemma-prod-U}
\color{black}
    Assume \Cref{assump-fhat} and \Cref{assump-reg}. \textcolor{black}{Let $h_{i,k-i}(\boldsymbol{X}_1,W_1) \myeq {(-1)}^k\binom{k}{i}\frac{g(\boldsymbol{X}_1)}{f(\boldsymbol{X}_1)}\frac{a_1^{i}(\boldsymbol{X}_1) b_1^{k-i}(\boldsymbol{X}_1)}{({1+\tau(\boldsymbol{X}_1))}^{k+1}}W_1^{k-i}$, for $k = 0, 1, \cdots$ and $i = 0, \cdots, k$ and $\mu_{i, k-i} \myeq E[h_{i,k-i}(\boldsymbol{X}_1,W_1)]$}. For any finite $p \in \mathbb{Z}$ and $p \geq 2$, let $k_p = 0, 1, \cdots$, $j_p = 0, \cdots, k_p$. Then for a large $N$ ($N \rightarrow \infty$) and a fixed finite $n \in \mathbb{Z}_{\geq 1}$, we have
    \begin{equation}
        E\left[\prod_{t=1}^{p} \left\{  \sum_{i=n+1}^{N} \left(
        h_{j_t, k_t-j_t} (\boldsymbol{X}_i, W_i) - \mu_{j_t, k_t-j_t}
        \right)
        \right\}\right] = 
        O(N^{\floor{\frac{p}{2}}}).
    \end{equation}
\end{lemma}

\begin{proof}[Proof of \Cref{lemma-prod-U}]\label{proof-of-lemma-prod-U}
\color{black}
Let $\mathcal{P}(n)$ denote the set of all permutations of $\{1, \cdots, n\}$, and for $t = 1, \cdots, p$, denote \newline $i(t) \myeq \argmin_{m \in \{n+1,\cdots, N\}} \left\{ \sum_{j=n+1}^{m} p_j \geq t \right\}$, where $\sum_{j=n+1}^{N} p_j = p$ and $p_j \in \mathbb{Z}_{\geq 0}$. Then we have
\begin{equation}\label{prod-u-pf-eq1}
    \begin{split}
       &E\left[ \prod_{t=1}^{p} \left\{  \sum_{i=n+1}^{N} \left(
        h_{j_t, k_t-j_t} (\boldsymbol{X}_i, W_i) - \mu_{j_t, k_t-j_t}
        \right)
        \right\}\right] \\
        =&
        \sum_{\stackrel{\sum_{j=n+1}^{N} p_j = p}{p_j \in \mathbb{Z}_{\geq 0}}} \binom{p}{p_{n+1}, \cdots, p_N} \sum_{\{\xi_1, \cdots, \xi_p \} \in \mathcal{P}(p)}  \\
        &\qquad
        \left\{ 
        E\left[
        \prod_{t=1}^{p_{n+1}} 
         \left(
        h_{j_{\xi_t}, k_{\xi_t}-j_{\xi_t}} (\boldsymbol{X}_{i(t)}, W_{i(t)}) - \mu_{j_{\xi_t}, k_{\xi_t}-j_{\xi_t}}
        \right)
        \right]\boldsymbol{1}_{\{p_{n+1} \geq 1\}} + \boldsymbol{1}_{\{p_{n+1} = 0\}}
        \right\} \times \\
        &\qquad
        \prod_{\stackrel{l=n+2}{p_l \geq 1}}^{N} E\left[
        \prod_{t = \sum_{u=n+1}^{l-1} p_u + 1}^{\sum_{u=n+1}^{l} p_u}
         \left(
        h_{j_{\xi_t}, k_{\xi_t}-j_{\xi_t}} (\boldsymbol{X}_{i(t)}, W_{i(t)}) - \mu_{j_{\xi_t}, k_{\xi_t}-j_{\xi_t}}
        \right)
        \right]\\
        =&
         \sum_{\stackrel{\sum_{j=n+1}^{N} p_j = p}{p_j \in \mathbb{Z}_{\geq 2}\; \text{or}\; p_j = 0}} \binom{p}{p_{n+1}, \cdots, p_N} \sum_{\{\xi_1, \cdots, \xi_p \} \in \mathcal{P}(p)}  \\
        &\qquad
        \left\{ 
        E\left[
        \prod_{t=1}^{p_{n+1}} 
         \left(
        h_{j_{\xi_t}, k_{\xi_t}-j_{\xi_t}} (\boldsymbol{X}_{i(t)}, W_{i(t)}) - \mu_{j_{\xi_t}, k_{\xi_t}-j_{\xi_t}}
        \right)
        \right]\boldsymbol{1}_{\{p_{n+1} \geq 1\}} + \boldsymbol{1}_{\{p_{n+1} = 0\}}
        \right\} \times \\
        &\qquad
        \prod_{\stackrel{l=n+2}{p_l \geq 1}}^{N} E\left[
        \prod_{t = \sum_{u=n+1}^{l-1} p_u + 1}^{\sum_{u=n+1}^{l} p_u}
         \left(
        h_{j_{\xi_t}, k_{\xi_t}-j_{\xi_t}} (\boldsymbol{X}_{i(t)}, W_{i(t)}) - \mu_{j_{\xi_t}, k_{\xi_t}-j_{\xi_t}}
        \right)
        \right]\\
        =& \sum_{u=1}^{\floor{\frac{p}{2}}} \binom{N-n}{u} \sum_{\stackrel{\sum_{j=n+1}^{n+u} p_j = p}{p_j \in \mathbb{Z}_{\geq 2}}} \binom{p}{p_{n+1},\cdots, p_{n+u}}
        \sum_{\{\xi_1, \cdots, \xi_p \} \in \mathcal{P}(p)}  \\
        &\qquad
        \left\{ 
        E\left[
        \prod_{t=1}^{p_{n+1}} 
         \left(
        h_{j_{\xi_t}, k_{\xi_t}-j_{\xi_t}} (\boldsymbol{X}_{i(t)}, W_{i(t)}) - \mu_{j_{\xi_t}, k_{\xi_t}-j_{\xi_t}}
        \right)
        \right]\boldsymbol{1}_{\{p_{n+1} \geq 1\}} + \boldsymbol{1}_{\{p_{n+1} = 0\}}
        \right\} \times \\
        &\qquad
        \prod_{\stackrel{l=n+2}{p_l \geq 1}}^{n+u} E\left[
        \prod_{t = \sum_{u=n+1}^{l-1} p_u + 1}^{\sum_{u=n+1}^{l} p_u}
         \left(
        h_{j_{\xi_t}, k_{\xi_t}-j_{\xi_t}} (\boldsymbol{X}_{i(t)}, W_{i(t)}) - \mu_{j_{\xi_t}, k_{\xi_t}-j_{\xi_t}}
        \right)
        \right]
    \end{split}
\end{equation}
where $\mathbb{Z}_{\geq 2}$ denotes all positive integers equal to or larger than $2$. The last but one equality in \Cref{prod-u-pf-eq1} uses the facts that $\boldsymbol{X}_{1:N}$ are i.i.d and that the value of $i(t)$ remain unchanged when $t$ increases from $\sum_{u=n+1}^{l-1} p_u + 1$ to $\sum_{u=n+1}^{l} p_u$, $\forall l = n+2, \cdots, N$ with $p_l \geq 1$ (the same can be said when $t$ increases from $1$ to $p_{n+1}$, when $p_{n+1} \geq 1$.)

By \Cref{assump-reg}, $E\left[
        \prod_{t=1}^{p_{n+1}} 
         \left(
        h_{j_{\xi_t}, k_{\xi_t}-j_{\xi_t}} (\boldsymbol{X}_{i(t)},W_{i(t)}) - \mu_{j_{\xi_t}, k_{\xi_t}-j_{\xi_t}}
        \right)
        \right] = O(1)$ and \newline $E\left[
        \prod_{t = \sum_{u=n+1}^{l-1} p_u + 1}^{\sum_{u=n+1}^{l} p_u}
         \left(
        h_{j_{\xi_t}, k_{\xi_t}-j_{\xi_t}} (\boldsymbol{X}_{i(t)}, W_{i(t)}) - \mu_{j_{\xi_t}, k_{\xi_t}-j_{\xi_t}}
        \right)
        \right] = O(1)$. Hence, for large $N$ and fixed $n$, we have 
 \begin{equation}
       E\left[ \prod_{t=1}^{p} \left\{  \sum_{i=n+1}^{N} \left(
        h_{j_t, k_t-j_t} (\boldsymbol{X}_i, W_i) - \mu_{j_t, k_t-j_t}
        \right)
        \right\}\right]
         = O\left(\binom{N-n}{\floor{\frac{p}{2}}}\right) = O(N^{\floor{\frac{p}{2}}}),
\end{equation}
  which completes the proof of \Cref{lemma-prod-U}.
\end{proof}

\textcolor{black}{Now we are ready to prove \Cref{ds-convergence}}.

\begin{proof}[Proof of \Cref{ds-convergence}]\label{proof-of-ds}

\color{black}
\begin{proofpart}
\textbf{An expression for $\frac{p_N^n(\boldsymbol{z}_1, \cdots, \boldsymbol{z}_n)}{\prod_{k=1}^{n} g(\boldsymbol{z}_k)}-1$.}

For the generalized DS algorithm as described in \cite{shang2022diversity}, $\forall k \in \{1, \cdots, n\}$, \textcolor{black}{using the assumption that $g(\boldsymbol{z}_k) = c\tilde{g}(\boldsymbol{z}_k)$ for some constant $c$}, we have
\begin{equation}\label{prob-ds-gen}
\begin{split}
    &P\left( \boldsymbol{Z}_N^k = \boldsymbol{X}_k \middle| \boldsymbol{X}_{k:n} = \boldsymbol{z}_{k:n}, \boldsymbol{X}_{(n+1):N}, W_{k:N} \right)\\
    =&
    \frac{
    \frac{\tilde{g}(\boldsymbol{z}_k)}{\hat{f}_N(\boldsymbol{z}_k)}}{
    \sum_{i=k}^n
    \frac{\tilde{g}(\boldsymbol{z}_i)}{\hat{f}_N(\boldsymbol{z}_i)} +
    \sum_{i=n+1}^{N}
    \frac{\tilde{g}(\boldsymbol{X}_i)}{\hat{f}_N(\boldsymbol{X}_i)}
    }\\
    =&
    \frac{
    \frac{g(\boldsymbol{z}_k)}{\hat{f}_N(\boldsymbol{z}_k)}}{
    \sum_{i=k}^n
    \frac{g(\boldsymbol{z}_i)}{\hat{f}_N(\boldsymbol{z}_i)} +
    \sum_{i=n+1}^{N}
    \frac{g(\boldsymbol{X}_i)}{\hat{f}_N(\boldsymbol{X}_i)}
    }.
\end{split}
\end{equation}

For $k = 1, \cdots, n$, denote 
\begin{equation}\label{f-Nk}
\color{black}
\begin{split}
    s_{N,k}(\boldsymbol{X}_{(n+1):N}, W_{k:N}) = &\frac{1}{N-k+1} \frac{\hat{f}_N(\boldsymbol{z}_k)}{f(\boldsymbol{z}_k)} \left(
     \sum_{i=k}^{n}
    \frac{g(\boldsymbol{z}_i)}{\hat{f}_N(\boldsymbol{z}_i)}
    + \sum_{i=n+1}^{N}
    \frac{g(\boldsymbol{X}_i)}{\hat{f}_N(\boldsymbol{X}_i)}
    \right)\\
    =&
    \frac{1}{1-\frac{k-1}{N}} \frac{\hat{f}_N(\boldsymbol{z}_k)}{f(\boldsymbol{z}_k)} \left(\frac{1}{N}
     \sum_{i=k}^{n}
    \frac{g(\boldsymbol{z}_i)}{\hat{f}_N(\boldsymbol{z}_i)}
    + \frac{1}{N}\sum_{i=n+1}^{N}
    \frac{g(\boldsymbol{X}_i)}{\hat{f}_N(\boldsymbol{X}_i)}
    \right)
    .
\end{split}
\end{equation}

By \Cref{pdf-ds}, \Cref{f-Nk} and the fact that $\frac{N!}{(N-n)!} = \prod_{k=1}^{n} (N-k+1)$, one gets
\begin{equation}\label{ptws-err-ds}
\color{black}
\begin{split}
&\frac{p_N^n(\boldsymbol{z}_1, \cdots, \boldsymbol{z}_n)}{\prod_{k=1}^{n} g(\boldsymbol{z}_k)}-1 \\
=&
    \frac{1}{\prod_{k=1}^{n} g(\boldsymbol{z}_k)}\frac{N!}{(N-n)!} \left(\prod_{k=1}^{n} f(\boldsymbol{z}_k)\right) E_{(\boldsymbol{X}_{(n+1):N}, W_{1:N})} \left[ 
    \prod_{k=1}^{n} P\left( \boldsymbol{Z}_N^k = \boldsymbol{X}_k \middle| \boldsymbol{X}_{k:n} = \boldsymbol{z}_{k:n}, \boldsymbol{X}_{(n+1):N}, W_{k:N} \right)
    \right]-1\\
    =&
    \frac{1}{\prod_{k=1}^{n} g(\boldsymbol{z}_k)}\frac{N!}{(N-n)!} \left(\prod_{k=1}^{n} f(\boldsymbol{z}_k)\right) E_{(\boldsymbol{X}_{(n+1):N}, W_{1:N})} \left[ 
    \prod_{k=1}^{n}
    \frac{
    \frac{g(\boldsymbol{z}_k)}{\hat{f}_N(\boldsymbol{z}_k)}}{
    \sum_{i=k}^n
    \frac{g(\boldsymbol{z}_i)}{\hat{f}_N(\boldsymbol{z}_i)} +
    \sum_{i=n+1}^{N}
    \frac{g(\boldsymbol{X}_i)}{\hat{f}_N(\boldsymbol{X}_i)}
    }
    \right]-1\\
    =&
    E_{(\boldsymbol{X}_{(n+1):N}, W_{1:N})} \left[ 
    \prod_{k=1}^{n} (N-k+1) \prod_{k=1}^n \frac{f(\boldsymbol{z}_k)}{\hat{f}_N(\boldsymbol{z}_k)}
    \prod_{k=1}^{n}
    \frac{
    1}{
    \sum_{i=k}^n
    \frac{g(\boldsymbol{z}_i)}{\hat{f}_N(\boldsymbol{z}_i)} +
    \sum_{i=n+1}^{N}
    \frac{g(\boldsymbol{X}_i)}{\hat{f}_N(\boldsymbol{X}_i)}
    }
    \right]-1
    \\
    =&
    E_{(\boldsymbol{X}_{(n+1):N}, W_{1:N})} \left[ 
    \prod_{k=1}^{n}
    \frac{
    1}{
    \frac{1}{N-k+1}
    \frac{\hat{f}_N(\boldsymbol{z}_k)}{f(\boldsymbol{z}_k)}
    \left(
    \sum_{i=k}^n
    \frac{g(\boldsymbol{z}_i)}{\hat{f}_N(\boldsymbol{z}_i)} +
    \sum_{i=n+1}^{N}
    \frac{g(\boldsymbol{X}_i)}{\hat{f}_N(\boldsymbol{X}_i)}
    \right)
    }
    \right]-1
    \\
    =&
     E_{(\boldsymbol{X}_{(n+1):N}, W_{1:N})} \left[    
    \frac{
    1}{\prod_{k=1}^{n} \left(
    \frac{1}{1-\frac{k-1}{N}}
    \frac{\hat{f}_N(\boldsymbol{z}_k)}{f(\boldsymbol{z}_k)}
    \left(\frac{1}{N}
    \sum_{i=k}^n
    \frac{g(\boldsymbol{z}_i)}{\hat{f}_N(\boldsymbol{z}_i)} +
    \frac{1}{N}
    \sum_{i=n+1}^{N}
    \frac{g(\boldsymbol{X}_i)}{\hat{f}_N(\boldsymbol{X}_i)}
    \right)
    \right)
    }
    \right]-1
    \\
    =&
    E_{(\boldsymbol{X}_{(n+1):N},W_{1:N})} \left[ 
    {\left(
    \prod_{k=1}^{n}
    s_{N,k}(\boldsymbol{X}_{(n+1):N}, W_{k:N})
    \right)}^{-1}
    \right]
    -1.
\end{split}
\end{equation}
\end{proofpart}

\begin{proofpart}
\textbf{An asymptotic expansion of $\prod_{k=1}^{n}
    s_{N,k}(\boldsymbol{X}_{(n+1):N}, W_{k:N})$.}
    
By \Cref{f-Nk}, 
\begin{equation}\label{prod-f-Nk}
\prod_{k=1}^{n}
    s_{N,k}(\boldsymbol{X}_{(n+1):N}, W_{k:N}) = 
    \left( 
    \prod_{k=1}^{n} \frac{1}{1-\frac{k-1}{N}}
    \right)
    \left(
    \prod_{k=1}^{n} \frac{\hat{f}_N(\boldsymbol{z}_k)}{f(\boldsymbol{z}_k)}
    \right)
    \prod_{k=1}^{n}
    \left(
    \frac{1}{N}
    \sum_{i=k}^{n}
    \frac{g(\boldsymbol{z}_i)}{\hat{f}_N(\boldsymbol{z}_i)}
    + 
    \frac{1}{N}
    \sum_{i=n+1}^{N}
    \frac{g(\boldsymbol{X}_i)}{\hat{f}_N(\boldsymbol{X}_i)}
    \right).
\end{equation}

Denote
\begin{equation}
    T_N(\boldsymbol{X}_{(n+1):N},W_{(n+1):N}) \myeq \frac{1}{N}
    \sum_{i=n+1}^{N}
    \frac{g(\boldsymbol{X}_i)}{\hat{f}_N(\boldsymbol{X}_i)}, 
\end{equation} 
and let
\begin{align}
    &\prod_{k=1}^{n} \frac{1}{1-\frac{k-1}{N}} =
    1 + N^{-1}\frac{n(n-1)}{2} + Rmd_0 \label{term0}\\
    &\prod_{k=1}^{n} \frac{\hat{f}_N(\boldsymbol{z}_k)}{f(\boldsymbol{z}_k)} =
    \mathcal{I}(\boldsymbol{z}_{1:n}) +  N^{-r_1}A_1(\boldsymbol{z}_{1:n}) + N^{-r_2}B_1(\boldsymbol{z}_{1:n}, W_{1:n})+\nonumber\\
    &\qquad
    N^{-2r_1}A_2(\boldsymbol{z}_{1:n}) + N^{-2r_2}B_2(\boldsymbol{z}_{1:n}, W_{1:n}) +\nonumber \\
    &\qquad
    N^{-r_1-r_2}\Theta(\boldsymbol{z}_{1:n},W_{1:n}) + Rmd_1(W_{1:n})
    \label{term1}
    \\
    &\prod_{k=1}^{n}
    \left(
    \frac{1}{N}
    \sum_{i=k}^{n}
    \frac{g(\boldsymbol{z}_i)}{\hat{f}_N(\boldsymbol{z}_i)}
    + 
    T_N(\boldsymbol{X}_{(n+1):N},W_{(n+1):N})
    \right) =
    T_N^n(\boldsymbol{X}_{(n+1):N},W_{(n+1):N}) + \nonumber\\
    &\qquad
    N^{-1}T_N^{n-1}(\boldsymbol{X}_{(n+1):N},W_{(n+1):N})
    \sum_{i=1}^{n}i\frac{g(\boldsymbol{z}_i)}{f(\boldsymbol{z}_i)(1+\tau(\boldsymbol{z}_i))} + \nonumber\\
    &\qquad
    Rmd_2(\boldsymbol{X}_{(n+1):N},W_{1:N})
    ,\label{term2}
\end{align}
where $\mathcal{I}(\boldsymbol{z}_{1:n})$, $A_1(\boldsymbol{z}_{1:n})$, $A_2(\boldsymbol{z}_{1:n})$ are as defined in \Cref{I-def,A1-def,A2-def}, and 
\begin{align}
    B_1(\boldsymbol{z}_{1:n},W_{1:n}) &= \sum_{j=1}^{n} W_j b_1(\boldsymbol{z}_j) \prod_{\stackrel{k=1}{k \neq j}}^{n} \left( 1 + \tau(\boldsymbol{z}_k) \right) \label{B1-def}\\
    B_2(\boldsymbol{z}_{1:n},W_{1:n}) &=
    \sum_{\stackrel{\{i,j\}\subset \{1,\cdots, n\}}{i<j}} W_i W_j b_1(\boldsymbol{z}_i) b_1(\boldsymbol{z}_j) \prod_{\stackrel{k=1}{k \notin \{i,j \}}}^{n} \left( 1 + \tau(\boldsymbol{z}_k)
    \right) \label{B2-def}\\
    \Theta(\boldsymbol{z}_{1:n},W_{1:n}) &=  \sum_{\stackrel{\{i,j\}\subset \{1,\cdots, n\}}{i<j}}
    \left[ 
    W_j a_1(\boldsymbol{z}_i) b_1(\boldsymbol{z}_j) 
    + W_i a_1 (\boldsymbol{z}_j)
    b_1(\boldsymbol{z}_i)
    \right]
    \prod_{\stackrel{k=1}{k \notin \{i,j \}}}^{n} \left( 1 + \tau(\boldsymbol{z}_k) \right)
    \label{AB-def}.
\end{align}
Here  $\prod_{k \in \emptyset} (1 + \tau(\boldsymbol{z}_k)) = 0$, as mentioned in \Cref{ds-convergence}. The remainder terms are expressed in the following,
\begin{align}
Rmd_0 &= 
N^{-2}\sum_{k=1}^{n} {(k-1)}^2 \frac{1}{1-N^{-1}(k-1)} +\nonumber \\
&\qquad
\sum_{t=2}^{n} \sum_{\stackrel{\{i_1, \cdots, i_t\} \subset\{1, \cdots, n\}}{i_1 < i_2 < \cdots < i_t}} \prod_{j=1}^{t}\left\{ 
N^{-1}(i_j-1) + N^{-2} {(i_j-1)}^2 \frac{1}{1-N^{-1}(i_j-1)}
\right\} \nonumber\\
&=
O(N^{-2})
\label{rmd0}\\
Rmd_1(W_{1:n}) &= \sum_{t=3}^n
\sum_{\stackrel{\{i_1,\cdots,i_t\}\subset \{1,\cdots,n\}}{i_1 < i_2 <\cdots<i_t}}\left\{
\prod_{j=1}^{t}\left[ N^{-r_1}a_1(\boldsymbol{z}_{i_j}) + N^{-r_2} b_1(\boldsymbol{z}_{i_j})W_{i_j} \right]
\right\} \cdot \nonumber\\
&\qquad \qquad
\prod_{\stackrel{j =1}{j \notin \{i_1,\cdots,i_t\}}}^n (1 + \tau(\boldsymbol{z}_j)) \nonumber\\
& \stackrel{w.p. 1}{=}
O(N^{-3\min\{r_1,r_2\}})
\label{rmd1}\\
Rmd_2(\boldsymbol{X}_{(n+1):N},W_{1:N}) &= 
N^{-1} T_N^{n-1}(\boldsymbol{X}_{(n+1):N},W_{(n+1):N}) \sum_{i=1}^{n}i
\frac{g(\boldsymbol{z}_i)}{f(\boldsymbol{z}_i)(1+\tau(\boldsymbol{z}_i))}R(\boldsymbol{z}_i,W_i)+\nonumber\\
&\qquad 
\sum_{i=2}^{n} N^{-i}
T_N^{n-i}(\boldsymbol{X}_{(n+1):N},W_{(n+1):N})
\sum_{\stackrel{\{k_1, \cdots, k_i\}\subset \{1, \cdots, n\}}{k_1 < k_2 < \cdots < k_i}}
\prod_{j=1}^{i} \gamma_{k_j}(\boldsymbol{z}_{k_j:n},W_{k_j:n})
,
\label{rmd2}
\end{align}
where
\begin{align}
R(\boldsymbol{z}_i,W_i)  &= -\frac{N^{-r_1}a_1(\boldsymbol{z}_i)+N^{-r_2}b_1(\boldsymbol{z}_i)W_i}{1 + \tau(\boldsymbol{z}_i)+N^{-r_1}a_1(\boldsymbol{z}_i) + N^{-r_2}b_1(\boldsymbol{z}_i)W_i} 
\stackrel{w.p. 1}{=} O(N^{-\min\{r_1, r_2\}})
,\qquad \text{for}\; i = 1, \cdots, n \label{Rziwi}\\
\gamma_k(\boldsymbol{z}_{k:n},W_{k:n}) 
&= \sum_{i=k}^{n} \frac{g(\boldsymbol{z}_i)}{f(\boldsymbol{z}_i)(1+\tau(\boldsymbol{z}_i))}\left( 
1 + R(\boldsymbol{z}_i,W_i)
\right)
\stackrel{w.p. 1}{=} O(1)
,\qquad \text{for} \; k = 1, \cdots, n. \label{gammak}
\end{align}
\textcolor{black}{Note in \Cref{rmd1}, that last equality holds w.p 1 because for any realization of $W_{1:n} = w_{1:n}$, $Rmd_1(w_{1:n}) = O(N^{-3\min\{r_1,r_2\}})$. Similarly for \Cref{Rziwi,gammak}.}

One gets
\begin{equation}\label{asym-f-Nk}
    \begin{split}
        &\prod_{k=1}^{n}
    s_{N,k}(\boldsymbol{X}_{(n+1):N}, W_{k:N}) \\
    =&
    T_N^n(\boldsymbol{X}_{(n+1):N},W_{(n+1):N})\mathcal{I}(\boldsymbol{z}_{1:n}) +N^{-r_1}T_N^n(\boldsymbol{X}_{(n+1):N},W_{(n+1):N}) A_1(\boldsymbol{z}_{1:n})\\
    +&
    N^{-r_2}T_N^n(\boldsymbol{X}_{(n+1):N},W_{(n+1):N}) B_1(\boldsymbol{z}_{1:n},W_{1:n}) \\
    +& 
    N^{-2r_1} T_N^n(\boldsymbol{X}_{(n+1):N},W_{(n+1):N}) A_2(\boldsymbol{z}_{1:n})
    + N^{-2r_2} T_N^n(\boldsymbol{X}_{(n+1):N},W_{(n+1):N}) B_2(\boldsymbol{z}_{1:n},W_{1:n})\\
    +& 
    N^{-(r_1+r_2)}
    T_N^n(\boldsymbol{X}_{(n+1):N},W_{(n+1):N})
    \Theta(\boldsymbol{z}_{1:n},W_{1:n})
    \\
    +&
    N^{-1} \Bigg\{
    T_N^n(\boldsymbol{X}_{(n+1):N},W_{(n+1):N}) \frac{n(n-1)}{2} \mathcal{I}(\boldsymbol{z}_{1:n})
    + \\
    &\quad 
    T_N^{n-1}(\boldsymbol{X}_{(n+1):N},W_{(n+1):N})
    \left(
    \sum_{i=1}^n i\frac{g(\boldsymbol{z}_i)}{f(\boldsymbol{z}_i)} {\left(
    1 + \tau(\boldsymbol{z}_i)
    \right)}^{-1}
    \right)
    \mathcal{I}(\boldsymbol{z}_{1:n})
    \Bigg\}\\
    +&
    Rmd_3(\boldsymbol{X}_{(n+1):N},W_{1:N}),
    \end{split}
\end{equation}

where the remainder term $Rmd_3(\boldsymbol{X}_{(n+1):N},W_{1:N})$ is written as
\begin{equation}\label{rmd3}
\begin{split}
    &Rmd_3(\boldsymbol{X}_{(n+1):N},W_{1:N}) \\=&T_0 T_1(W_{1:n}) Rmd_2(\boldsymbol{X}_{(n+1):N},W_{1:N}) + T_0 Rmd_1(W_{1:n}) T_2(\boldsymbol{X}_{(n+1):N},W_{(n+1):N}) \\
    +&T_0 Rmd_1(W_{1:n}) Rmd_2(\boldsymbol{X}_{(n+1):N},W_{1:N})
    +Rmd_0 T_1(W_{1:n}) T_2(\boldsymbol{X}_{(n+1):N},W_{(n+1):N})
    + Rmd_0 T_1(W_{1:n}) Rmd_2(\boldsymbol{X}_{(n+1):N},W_{1:N})
    \\+
    &Rmd_0 Rmd_1(W_{1:n}) T_2(\boldsymbol{X}_{(n+1):N},W_{(n+1):N})
    + Rmd_0 Rmd_1(W_{1:n}) Rmd_2(\boldsymbol{X}_{(n+1):N},W_{1:N})
    ,
    \end{split}
\end{equation}
where (refer to \Cref{term0,term1,term2})
\begin{align}
    T_0 &= 1 + N^{-1}\frac{n(n-1)}{2} = O(1), \label{T0}\\
    T_1(W_{1:n}) &= \mathcal{I}(\boldsymbol{z}_{1:n}) +  N^{-r_1}A_1(\boldsymbol{z}_{1:n}) + N^{-r_2}B_1(\boldsymbol{z}_{1:n}, W_{1:n})+\nonumber\\
    &\qquad
    N^{-2r_1}A_2(\boldsymbol{z}_{1:n}) + N^{-2r_2}B_2(\boldsymbol{z}_{1:n}, W_{1:n}) +\nonumber \\
    &\qquad
    N^{-r_1-r_2}\Theta(\boldsymbol{z}_{1:n},W_{1:n}) \nonumber \\
    &\stackrel{w.p. 1}{=} O(1), \label{T1}\\
    T_2(\boldsymbol{X}_{(n+1):N},W_{(n+1):N}) &=  T_N^n(\boldsymbol{X}_{(n+1):N},W_{(n+1):N}) + \nonumber\\
    &\qquad
    N^{-1}T_N^{n-1}(\boldsymbol{X}_{(n+1):N},W_{(n+1):N})
    \sum_{i=1}^{n}i\frac{g(\boldsymbol{z}_i)}{f(\boldsymbol{z}_i)(1+\tau(\boldsymbol{z}_i))}. \label{T2}
\end{align}
\end{proofpart}
 
\begin{proofpart}
\textbf{An asymptotic expansion for $T_N^p(\boldsymbol{X}_{(n+1):N},W_{(n+1):N})$, $p = 1, 2, \cdots$}    

Let $T_N \myeq T_N(\boldsymbol{X}_{(n+1):N},W_{(n+1):N})$.

\textcolor{black}{First consider that as $N \rightarrow +\infty$,},
\begin{equation}\label{taylor-g-fhat}
\color{black}
    \begin{split}
        &\frac{g(\boldsymbol{X}_1)}{\hat{f}_N(\boldsymbol{X}_1)}\\
        =&\frac{g(\boldsymbol{X}_1)}{f(\boldsymbol{X}_1)(1+\tau(\boldsymbol{X}_1))}\frac{1}{1+N^{-r_1}\frac{a_1(\boldsymbol{X}_1)}{1+\tau(\boldsymbol{X}_1)} + N^{-r_2}\frac{b_1(\boldsymbol{X}_1)}{1+\tau(\boldsymbol{X}_1)}W_1}\\
        =&\frac{g(\boldsymbol{X}_1)}{f(\boldsymbol{X}_1)(1+\tau(\boldsymbol{X}_1))}\sum_{k=0}^{\infty}
        {(-1)}^k{\left(N^{-r_1}\frac{a_1(\boldsymbol{X}_1)}{1+\tau(\boldsymbol{X}_1)} + N^{-r_2}\frac{b_1(\boldsymbol{X}_1)}{1+\tau(\boldsymbol{X}_1)}W_1 \right)}^k\\
        =&\frac{g(\boldsymbol{X}_1)}{f(\boldsymbol{X}_1)(1+\tau(\boldsymbol{X}_1))}\sum_{k=0}^{\infty}{(-1)}^k
        \sum_{i=0}^{k} \binom{k}{i} {\left( 
        N^{-r_1}\frac{a_1(\boldsymbol{X}_1)}{1+\tau(\boldsymbol{X}_1)}
        \right)}^i
        {\left( 
        N^{-r_2}\frac{b_1(\boldsymbol{X}_1)}{1+\tau(\boldsymbol{X}_1)}W_1
        \right)}^{k-i}\\
        =&
        \sum_{k=0}^{\infty}\sum_{i=0}^{k} N^{-ir_1-(k-i)r_2} {(-1)}^k \binom{k}{i} \frac{g(\boldsymbol{X}_1)}{f(\boldsymbol{X}_1)}
        \frac{a_1^i(\boldsymbol{X}_1)b_1^{k-i}(\boldsymbol{X}_1)}{{\left(1+\tau(\boldsymbol{X}_1)\right)}^{k+1}}W_1^{k-i},
    \end{split}
\end{equation}
\textcolor{black}{
which holds for any realizations of $(\boldsymbol{X}_1, W_1)$. For convenience, $\forall k = 0, 1, \cdots; \; i = 0, \cdots, k$, let} 
\begin{equation}\label{h_i_k}
\color{black}
    h_{i,k-i}(\boldsymbol{X}_1,W_1) \myeq {(-1)}^k\binom{k}{i}\frac{g(\boldsymbol{X}_1)}{f(\boldsymbol{X}_1)}\frac{a_1^{i}(\boldsymbol{X}_1) b_1^{k-i}(\boldsymbol{X}_1)}{({1+\tau(\boldsymbol{X}_1))}^{k+1}}W_1^{k-i}.
\end{equation}
\textcolor{black}{
Then we have}
\begin{equation}\label{g2fhat-final}
\color{black}
    \frac{g(\boldsymbol{X}_1)}{\hat{f}_N(\boldsymbol{X}_1)} = \sum_{k=0}^{\infty} \sum_{i=0}^k N^{-ir_1-(k-i)r_2}h_{i,k-i}(\boldsymbol{X}_1,W_1).
\end{equation}
\textcolor{black}{
Note that \Cref{g2fhat-final} also holds when replacing $(\boldsymbol{X_1}, W_1)$ with any realizations of $(\boldsymbol{X_i}, W_i)$, for $i = 2, \cdots, N$. Also, $\forall k = 0, 1, \cdots; \; i = 0, \cdots, k$, we denote $\mu_{i, k-i} = E\left[ h_{i,k-i}(\boldsymbol{X}_1,W_1) \right]$ and $\sigma^2_{i, k-i} = Var\left[ h_{i,k-i}(\boldsymbol{X}_1,W_1) \right]$. Note that $\sigma^2_{i, k-i}$ is finite by \Cref{assump-reg}, and so is $\mu_{i, k-i}$.}

By \Cref{g2fhat-final}, the assumption that $\boldsymbol{X}_{n+1}, \cdots, \boldsymbol{X}_N$ are i.i.d and that $W_{n+1}, \cdots, W_N$ are i.i.d, we get,
\begin{equation}\label{tN-series}
\begin{split}
   T_N = 
   &
   \frac{1}{N}\sum_{i=n+1}^{N}\frac{g(\boldsymbol{X}_i)}{\hat{f}_N(\boldsymbol{X}_i)}\\
   =&
   \frac{1}{N}\sum_{i=n+1}^N \left\{\sum_{k=0}^{\infty} \sum_{j=0}^k N^{-jr_1-(k-j)r_2} h_{j,k-j}(\boldsymbol{X}_i,W_i)
   \right\}\\
   =&
   \sum_{k=0}^{\infty} \sum_{j=0}^k N^{-jr_1-(k-j)r_2}
   \frac{1}{N}\sum_{i=n+1}^N  h_{j,k-j}(\boldsymbol{X}_i,W_i)
   \\
   =&
   \sum_{k=0}^{\infty} \sum_{j=0}^k N^{-jr_1-(k-j)r_2}
   \frac{1}{N}\sum_{i=n+1}^N \Big( h_{j,k-j}(\boldsymbol{X}_i,W_i) - E_{(\boldsymbol{X}_i, W_i)} \left[ 
   h_{j,k-j}(\boldsymbol{X}_i,W_i)
   \right] +  \\
   & \quad \quad
   E_{(\boldsymbol{X}_i, W_i)} \left[ 
   h_{j,k-j}(\boldsymbol{X}_i,W_i)
   \right] \Big)
   \\
   =&
   \sum_{k=0}^{\infty} \sum_{j=0}^k N^{-jr_1-(k-j)r_2} \frac{1}{N}\sum_{i=n+1}^N \left( h_{j,k-j}(\boldsymbol{X}_i,W_i) - \mu_{j,k-j}
   \right) + \\
   &\quad \quad
   \sum_{k=0}^{\infty} \sum_{j=0}^k N^{-jr_1-(k-j)r_2} (1-N^{-1}n) \mu_{j,k-j},
\end{split}
\end{equation}
w.p. 1. Here the third equality holds because for each $i = n+1, \cdots, N$, $\sum_{k=0}^{\infty} \sum_{j=0}^k N^{-jr_1-(k-j)r_2} h_{j,k-j}(\boldsymbol{X}_i,W_i)= \frac{g(\boldsymbol{X}_i)}{\hat{f}_N(\boldsymbol{X}_i)} < \infty$ w.p. 1.

Now for fixed and finite $k = 0, 1, \cdots$ and $j = 0, \cdots, k$, we study the term $\frac{1}{N}\sum_{i=n+1}^N \left( h_{j,k-j}(\boldsymbol{X}_i,W_i) - \mu_{j,k-j} \right)$. For convenience, denote $U_{j, k-j} \myeq \frac{1}{\sqrt{N-n}} \sum_{i=n+1}^N \sigma_{j, k-j}^{-1} \left( h_{j,k-j}(\boldsymbol{X}_i,W_i) - \mu_{j,k-j} \right)$, where $\sigma_{j, k-j} $ is the standard deviation of $h_{j,k-j}(\boldsymbol{X}_i,W_i)$, as defined before. Then the p.d.f of $U_{j, k-j}$ can be obtained via Edgeworth series (\cite{wallace1958asymptotic}), precisely,
\begin{equation}\label{edgeworth}
    f_{U_{j, k-j}}(x) = \phi(x)\left\{1 + N^{-\frac{1}{2}}\frac{1}{6}\lambda_{3,j,k-j} H_3(x)
    + N^{-1} \left( 
    \frac{1}{24}\lambda_{4,j,k-j} H_4(x) + \frac{1}{72}\lambda_{3,j,k-j}^2H_6(x)
    \right)
    +O(N^{-\frac{3}{2}})
    \right\}. 
\end{equation}
Here $\phi(\cdot)$ is the p.d.f of the standard normal distribution. For $i = 1, 2, \cdots$, $\lambda_{i,j,k-j} 
\myeq \frac{\kappa_{i,j,k-j}}{\sigma_{j,k-j}^i}$ , where $\kappa_{i,j,k-j}$ is the $i$th cumulant of $h_{j, k-j}(\boldsymbol{X}_{n+1}, W_{n+1})$; and $H_i(\cdot)$ is the Hermite polynomial of order $i$. Clearly $E_{(\boldsymbol{X}_{(n+1):N},W_{(n+1):N})}[U_{j, k-j}] = 0$. It is easy to compute that $E_{(\boldsymbol{X}_{(n+1):N},W_{(n+1):N})}[U^2_{j, k-j}] = 1 $ and by \Cref{lemma-prod-U}, $E_{(\boldsymbol{X}_{(n+1):N},W_{(n+1):N})}[U^3_{j, k-j}] = O(N^{-{\frac{1}{2}}})$.

Thus, w.p. 1, 
\begin{equation}\label{expan-tN-term1}
\begin{split}
    \frac{1}{N}\sum_{i=n+1}^N \left( h_{j,k-j}(\boldsymbol{X}_i,W_i) - \mu_{j,k-j} \right) =& \sigma_{j, k-j} \frac{\sqrt{N-n}}{N} U_{j, k-j} \\=
    &
    N^{-\frac{1}{2}}\left(1 - N^{-1}\frac{n}{2}-N^{-2}\frac{n^2}{8}{(1-\xi_N)}^{-\frac{3}{2}} \right) \sigma_{j, k-j} U_{j, k-j},
\end{split}
\end{equation}
where $\xi_N \in [0,\frac{n}{N}]$ is a constant, and hence $\xi_N = O(N^{-1})$.

By \Cref{expan-tN-term1},  $T_N$ can be written as, w.p. 1,
\begin{equation}\label{tN-asym}
    \begin{split}
        T_N =&T_N(\boldsymbol{X}_{(n+1):N},W_{(n+1):N})
        \\=& 
        \mu_{0,0} + N^{-r_1}\mu_{1,0}\\
        &+
        N^{-{2r_1}}\mu_{2,0} + N^{-2r_2}\mu_{0,2}  \\
        &+
        N^{-\frac{1}{2}}\sigma_{0,0}U_{0,0} + 
        N^{-\left(r_1 + \frac{1}{2}\right)}\sigma_{1,0}U_{1,0}
        +
        N^{-\left(r_2 + \frac{1}{2}\right)}\sigma_{0,1}U_{0,1}\\
        &-
        N^{-1}n\mu_{0,0} + Rmd_4(\boldsymbol{X}_{(n+1):N}, W_{(n+1):N}),
        \end{split}
\end{equation}
where we used the fact that $\mu_{0,1} = \mu_{1,1} = 0$ since $W_1$ is independent of $\boldsymbol{X}_1$ with an expectation of $0$. The remainder term is as follows,
\begin{equation}\label{rmd4}
\begin{split}
    &Rmd_4(\boldsymbol{X}_{(n+1):N}, W_{(n+1):N}) \\
    =&
    \sum_{k=0}^{\infty} \sum_{j=0}^{k} N^{-jr_1-(k-j)r_2-\frac{1}{2}}
    \left[
    I_{\{k \geq 2\}} - N^{-1}\frac{n}{2} - N^{-2}\frac{n^2}{8} {(1-\xi_N)}^{-\frac{3}{2}}
    \right]\sigma_{j, k-j} U_{j, k-j} +\\
    &\qquad 
    \sum_{k=0}^{\infty}\sum_{j=0}^{k} N^{-jr_1-(k-j)r_2}
    \left(
    I_{\{ k \geq 3 \}} + I_{\{ k \geq 1 \}} N^{-1}n
    \right)\mu_{j, k-j}\\
    \stackrel{w.p. 1}{=}&
    \color{black}{O(N^{-3\min\{r_1,r_2,\frac{1}{2}\}})},
\end{split}
\end{equation}
\textcolor{black}{of which the last equality holds w.p. $1$ because it holds for any realizations of $U_{j, k-j}$, for $k = 0, 1, \cdots$, $j = 0, \cdots, k$.}

By \Cref{tN-asym}, one gets $\forall p \in \mathbb{Z}^+$,
\begin{equation}\label{tnp-asym}
    \begin{split}
        T_N^p 
        =&T_N^p(\boldsymbol{X}_{(n+1):N},W_{(n+1):N})
        \\=&
        \mu_{0,0}^p + N^{-r_1} p\mu_{0,0}^{p-1}\mu_{1,0} \\
        &+
        N^{-2r_1}\left\{p\mu_{0,0}^{p-1}\mu_{2,0} + \binom{p}{2}\mu_{0,0}^{p-2} \mu_{1,0}^2\right\}\\
        &+
        N^{-2r_2} 
        p\mu_{0,0}^{p-1}\mu_{0,2} 
        \\
        &+
        N^{-\frac{1}{2}}p\mu_{0,0}^{p-1}\sigma_{0,0}U_{0,0}\\
        &+
        N^{-(r_1+\frac{1}{2})}\left\{
        p\mu_{0,0}^{p-1}\sigma_{1,0}U_{1,0}
        +\binom{p}{2} \mu_{0,0}^{p-2}2\mu_{1,0}\sigma_{0,0}U_{0,0}
        \right\}\\
        &+
        N^{-(r_2+\frac{1}{2})}
        p\mu_{0,0}^{p-1}\sigma_{0,1}U_{0,1}
        \\
        &+
        N^{-1}\left\{
        -p\mu_{0,0}^pn + \binom{p}{2}\mu_{0,0}^{p-2}\sigma_{0,0}^2U_{0,0}^2
        \right\}\\
        &+ Rmd_5(\boldsymbol{X}_{(n+1):N},W_{(n+1):N}; p),\\
        \stackrel{w.p. 1}{=}&
        O(1)
    \end{split}
\end{equation}
where the remainder term is
\begin{equation}\label{rmd5}
\begin{split}
    Rmd_5(\boldsymbol{X}_{(n+1):N},W_{(n+1):N}; p) &= 
    \mu_{0,0}^{p-1}pRmd_4(\boldsymbol{X}_{(n+1):N},W_{(n+1):N}) +\\
    &\qquad 
    \sum_{k=2}^{p}\mu_{0,0}^{p-k}\binom{p}{k} \Bigg[
        N^{-r_1}\mu_{1,0}
        +
        N^{-{2r_1}}\mu_{2,0} + N^{-2r_2}\mu_{0,2} +\\& \qquad
        N^{-\frac{1}{2}}\sigma_{0,0} U_{0,0} + 
        N^{-\left(r_1 + \frac{1}{2}\right)}\sigma_{1,0}U_{1,0}
        +
        N^{-\left(r_2 + \frac{1}{2}\right)}\sigma_{0,1}U_{0,1}-
        \\&\qquad
        N^{-1}n\mu_{0,0} + Rmd_4(\boldsymbol{X}_{(n+1):N}, W_{(n+1):N})
    \Bigg]^k\\
    &\qquad-
    \mu_{0,0}^{p-2}\binom{p}{2}\left\{
    N^{-2r_1}\mu_{1,0}^2 + N^{-1}\sigma_{0,0}^2U_{0,0}^2 + N^{-r_1-\frac{1}{2}}2\mu_{1,0}\sigma_{0,0}U_{0,0}.
    \right\}\\
    &\stackrel{w.p. 1}{=}
    O(N^{-3\min\{r_1,r_2,\frac{1}{2}\}}).
\end{split}
\end{equation}
\end{proofpart}

\begin{proofpart}
\textbf{An Asymptotic expansion for $\frac{p_N^n(\boldsymbol{z}_1, \cdots, \boldsymbol{z}_n)}{\prod_{k=1}^{n} g(\boldsymbol{z}_k)}-1$.}

Plugging \Cref{tnp-asym} back into \Cref{asym-f-Nk}, one gets,

\begin{equation}\label{f_Nk-asym-new}
    \begin{split}
        &\prod_{k=1}^{n}
    s_{N,k}(\boldsymbol{X}_{(n+1):N}, W_{k:N}) \\
    =&\mu_{0,0}^n\mathcal{I}(\boldsymbol{z}_{1:n})\\
    +&
    N^{-r_1}\left\{
    n\mu_{0,0}^{n-1}\mu_{1,0}\mathcal{I}(\boldsymbol{z}_{1:n})
    + \mu_{0,0}^n A_1(\boldsymbol{z}_{1:n})
    \right\}\\
    +&
    N^{-r_2}\mu_{0,0}^n B_1(\boldsymbol{z}_{1:n}, W_{1:n}) \\
    +&
    N^{-2r_1} \left\{
   \left[ n\mu_{0,0}^{n-1}\mu_{2,0} + \binom{n}{2}\mu_{0,0}^{n-2}\mu_{1,0}^2
   \right]\mathcal{I}(\boldsymbol{z}_{1:n}) 
    + n\mu_{0,0}^{n-1}\mu_{1,0}A_1(\boldsymbol{z}_{1:n}) + \mu_{0,0}^n A_2(\boldsymbol{z}_{1:n})
    \right\}\\
    +&
    N^{-2r_2}\left\{
    n\mu_{0,0}^{n-1}\mu_{0,2}\mathcal{I}(\boldsymbol{z}_{1:n})
    +
    \mu_{0,0}^n B_2(\boldsymbol{z}_{1:n},W_{1:n})
    \right\}\\
    +&
    N^{-r_1-r_2}\left\{
    n\mu_{0,0}^{n-1}\mu_{1,0}B_1(\boldsymbol{z}_{1:n},W_{1:n})+
    \mu_{0,0}^n \Theta(\boldsymbol{z}_{1:n}, W_{1:n})
    \right\}\\
    +&
    N^{-\frac{1}{2}}
    n\mu_{0,0}^{n-1}\sigma_{0,0}\mathcal{I}(\boldsymbol{z}_{1:n})U_{0,0}
    \\
    +&
    N^{-\frac{1}{2}-r_1}\left\{
    \left[ 
    n\mu_{0,0}^{n-1}\sigma_{1,0}U_{1,0} +
    2\binom{n}{2}\mu_{0,0}^{n-2}\mu_{1,0}\sigma_{0,0}U_{0,0}
    \right]\mathcal{I}(\boldsymbol{z}_{1:n})+
    n\mu_{0,0}^{n-1}\sigma_{0,0}A_1(\boldsymbol{z}_{1:n})U_{0,0}
    \right\}
    \\
    +&
    N^{-\frac{1}{2}-r_2}\left\{
    n\mu_{0,0}^{n-1}\sigma_{0,1}U_{0,1}\mathcal{I}(\boldsymbol{z}_{1:n})
    +n\mu_{0,0}^{n-1}\sigma_{0,0}U_{0,0}B_1(\boldsymbol{z}_{1:n}, W_{1:n})
    \right\}\\
    +&
    N^{-1}
    \Bigg\{
    -\frac{n(n+1)}{2}\mu_{0,0}^n + \binom{n}{2}\mu_{0,0}^{n-2}\sigma_{0,0}^2 U_{0,0}^2
    +\mu_{0,0}^{n-1} \left(
    \sum_{i=1}^{n} i\frac{g(\boldsymbol{z}_i)}{f(\boldsymbol{z}_i)}{\left( 
    1 + \tau(\boldsymbol{z}_i)
    \right)}^{-1}
    \right)
    \Bigg\}\mathcal{I}(\boldsymbol{z}_{1:n})\\
    +&
    Rmd_6(\boldsymbol{X}_{(n+1):N}, W_{1:N}),
    \end{split}
\end{equation}
where
\begin{equation}\label{rmd6}
\begin{split}
    &Rmd_6(\boldsymbol{X}_{(n+1):N}, W_{1:N})\\
    =&\mathcal{I}(\boldsymbol{z}_{1:n})Rmd_5(\boldsymbol{X}_{(n+1):N}, W_{(n+1):N};n)\\
    +&
    N^{-r_1}A_1(\boldsymbol{z}_{1:n})\left(
    T_N^{n}(\boldsymbol{X}_{(n+1):N},W_{(n+1):N}) - \mu_{0,0}^n - N^{-r_1}n\mu_{0,0}^{n-1}\mu_{1,0}
    -N^{-\frac{1}{2}}n\mu_{0,0}^{n-1}\sigma_{0,0}U_{0,0}
    \right)\\
    +&N^{-r_2}B_1(\boldsymbol{z}_{1:n},W_{1:n})\left(
    T_N^{n}(\boldsymbol{X}_{(n+1):N},W_{(n+1):N}) - \mu_{0,0}^n - N^{-r_1}n\mu_{0,0}^{n-1}\mu_{1,0}
    -N^{-\frac{1}{2}}n\mu_{0,0}^{n-1}\sigma_{0,0}U_{0,0}
    \right)\\
    +&
    N^{-2r_1}A_2(\boldsymbol{z}_{1:n})\left(
    T_N^{n}(\boldsymbol{X}_{(n+1):N},W_{(n+1):N}) - \mu_{0,0}^n
    \right)\\
    +&
    N^{-2r_2}B_2(\boldsymbol{z}_{1:n},W_{1:n})\left(
    T_N^{n}(\boldsymbol{X}_{(n+1):N},W_{(n+1):N}) - \mu_{0,0}^n
    \right)\\
    +&N^{-(r_1+r_2)}\Theta(\boldsymbol{z}_{1:n},W_{1:n})\left(
    T_N^{n}(\boldsymbol{X}_{(n+1):N},W_{(n+1):N}) - \mu_{0,0}^n
    \right)\\
    +&
    N^{-1}\Big[
    \frac{n(n-1)}{2}\left(
    T_N^{n}(\boldsymbol{X}_{(n+1):N},W_{(n+1):N}) - \mu_{0,0}^n
    \right)+\\
    &\qquad
    \left(
    \sum_{i=1}^{n} i\frac{g(\boldsymbol{z}_i)}{f(\boldsymbol{z}_i)}{\left(1 + \tau(\boldsymbol{z}_i)\right)}^{-1}
    \right)
    \left(
    T_N^{n-1}(\boldsymbol{X}_{(n+1):N},W_{(n+1):N}) - \mu_{0,0}^{n-1}
    \right)
    \Big]\mathcal{I}(\boldsymbol{z}_{1:n})\\
    +&Rmd_3(\boldsymbol{X}_{(n+1):N},W_{1:N}).
\end{split}
\end{equation}

Recall that for $p = 1, 2, \cdots$, $T_N^p(\boldsymbol{X}_{(n+1):N}, W_{(n+1):N}) \stackrel{w.p. 1}{=} O(1)$ (see \Cref{tnp-asym}), we have, according to \Cref{rmd2} and \Cref{T2} respectively, $Rmd_2(\boldsymbol{X}_{(n+1):N}, W_{1:N}) \stackrel{w.p. 1}{=} O(N^{-1-\min\{r_1, r_2\}})$ and $T_2(\boldsymbol{X}_{(n+1):N}, W_{(n+1):N}) \stackrel{w.p. 1}{=} O(1)$. Thus, by \Cref{rmd3}, $Rmd_3(\boldsymbol{X}_{(n+1):N}, W_{1:N}) \stackrel{w.p. 1}{=} O(N^{-2\min\{r_1,r_2,\frac{1}{2}\}-\min\{r_1,r_2\}})$. Hence, $Rmd_6(\boldsymbol{X}_{(n+1):N}, W_{1:N}) \stackrel{w.p. 1}{=} O(N^{-3\min\{r_1, r_2, \frac{1}{2}\}}) = o(N^{-1})$ by the assumptions that $r_1 > \frac{1}{3}$ and $r_2 > \frac{1}{3}$ (see \Cref{assump-fhat}). The asymptotic expansion of ${\left(\prod_{k=1}^{n}
    s_{N,k}(\boldsymbol{X}_{(n+1):N}, W_{k:N}) \right)}^{-1}$ can thus be obtained by its Taylor expansion at $N = \infty$, i.e.,
    
\begin{equation}\label{inv-fNk}
\begin{split}
    &{\left(\prod_{k=1}^{n}
    s_{N,k}(\boldsymbol{X}_{(n+1):N}, W_{k:N}) \right)}^{-1} \\
    =&
    \mu_{0,0}^{-n}\mathcal{I}^{-1}(\boldsymbol{z}_{1:n})\\
    -&
    N^{-r_1}\left\{
    n\mu_{0,0}^{-n-1}\mu_{1,0}\mathcal{I}^{-1}(\boldsymbol{z}_{1:n})
    +
    \mu^{-n}_{0,0}\mathcal{I}^{-2}(\boldsymbol{z}_{1:n})A_1(\boldsymbol{z}_{1:n})
    \right\}\\
    -&
    N^{-r_2}\mu_{0,0}^{-n}\mathcal{I}^{-2}(\boldsymbol{z}_{1:n})B_1(\boldsymbol{z}_{1:n}, W_{1:n})\\
    -&
    N^{-\frac{1}{2}}n\mu_{0,0}^{-n-1}\sigma_{0,0}\mathcal{I}^{-1}(\boldsymbol{z}_{1:n})U_{0,0}\\
    +&
    N^{-2r_1}\Big\{
    -n\mu_{0,0}^{-n-1}\mu_{2,0}\mathcal{I}^{-1}(\boldsymbol{z}_{1:n})
    + \frac{1}{2}n(n+1)\mu_{0,0}^{-n-2}\mu_{1,0}^2\mathcal{I}^{-1}(\boldsymbol{z}_{1:n}) \\
    &\qquad
    +n\mu_{0,0}^{-n-1}\mu_{1,0}\mathcal{I}^{-2}(\boldsymbol{z}_{1:n})A_1(\boldsymbol{z}_{1:n}) - \mu_{0,0}^{-n}\mathcal{I}^{-2}(\boldsymbol{z}_{1:n})A_2(\boldsymbol{z}_{1:n})\\
    &
    \qquad +\mu_{0,0}^{-n} \mathcal{I}^{-3}(\boldsymbol{z}_{1:n})A_1^2(\boldsymbol{z}_{1:n})
    \Big\}\\
    +&
    N^{-2r_2}\Big\{
    -n\mu_{0,0}^{-n-1}\mu_{0,2}\mathcal{I}^{-1}(\boldsymbol{z}_{1:n}) - \mu_{0,0}^{-n}\mathcal{I}^{-2}(\boldsymbol{z}_{1:n})B_2(\boldsymbol{z}_{1:n},W_{1:n})\\
    &\qquad +
    \mu_{0,0}^{-n}\mathcal{I}^{-3}(\boldsymbol{z}_{1:n})B_1^2(\boldsymbol{z}_{1:n}, W_{1:n}) \Big \}\\
    +&
    N^{-r_1-r_2}\Big\{
    n\mu_{0,0}^{-n-1}\mu_{1,0}\mathcal{I}^{-2}(\boldsymbol{z}_{1:n})B_1(\boldsymbol{z}_{1:n},W_{1:n}) \\
    &\qquad
    -\mu_{0,0}^{-n}\mathcal{I}^{-2}(\boldsymbol{z}_{1:n})\Theta(\boldsymbol{z}_{1:n}, W_{1:n})
    +2\mu_{0,0}^{-n}\mathcal{I}^{-3}(\boldsymbol{z}_{1:n})A_1(\boldsymbol{z}_{1:n})B_1(\boldsymbol{z}_{1:n},W_{1:n})
    \Big\}\\
    +&
    N^{-\frac{1}{2}-r_1}\Big\{
    \left[
    -n\mu_{0,0}^{-n-1}\sigma_{1,0}U_{1,0} + n(n+1)\mu_{0,0}^{-n-2}\mu_{1,0}\sigma_{0,0}U_{0,0}
    \right]\mathcal{I}^{-1}(\boldsymbol{z}_{1:n})\\
    &
    \qquad
    +n\mu_{0,0}^{-n-1}\sigma_{0,0}\mathcal{I}^{-2}(\boldsymbol{z}_{1:n})A_1(\boldsymbol{z}_{1:n})U_{0,0}
    \Big\}\\
    +&
    N^{-\frac{1}{2}-r_2}\left\{
    -n\mu_{0,0}^{-n-1}\sigma_{0,1}\mathcal{I}^{-1}(\boldsymbol{z}_{1:n})U_{0,1}
    +
    n\mu_{0,0}^{-n-1}\sigma_{0,0}\mathcal{I}^{-2}(\boldsymbol{z}_{1:n})U_{0,0}B_1(\boldsymbol{z}_{1:n}, W_{1:n})
    \right\}\\
    +&
    N^{-1}
    \Big[ 
    \frac{n(n+1)}{2}\mu_{0,0}^{-n} 
    +\frac{1}{2}n(n+1)\mu_{0,0}^{-n-2}\sigma_{0,0}^2U_{0,0}^{2} \\
    &\qquad 
    -\mu_{0,0}^{-n-1}\left(
    \sum_{i=1}^{n}i\frac{g(\boldsymbol{z}_i)}{f(\boldsymbol{z}_i)}{\left( 
    1 + \tau(\boldsymbol{z}_i)
    \right)}^{-1}
    \right)
    \Big]\mathcal{I}^{-1}(\boldsymbol{z}_{1:n})
     \\
    +&
    Rmd_7(\boldsymbol{X}_{(n+1):N},W_{1:N}).
\end{split}
\end{equation}
To find the close-form expression for the remainder term $Rmd_7(\boldsymbol{X}_{(n+1):N},W_{1:N})$, we introduce the following additional notations. Denote each term in \Cref{f_Nk-asym-new} as follows 

\begin{align}
C_0
    =&\mu_{0,0}^n\mathcal{I}(\boldsymbol{z}_{1:n}) \label{C0}\\
    C_1 = &
    n\mu_{0,0}^{n-1}\mu_{1,0}\mathcal{I}(\boldsymbol{z}_{1:n})
    + \mu_{0,0}^n A_1(\boldsymbol{z}_{1:n})
    \label{C1}
   \\
    C_2(W_{1:n}) = &
    \mu_{0,0}^n B_1(\boldsymbol{z}_{1:n}, W_{1:n}) 
    \label{C2}
    \\
    C_3 = & 
   \left[ n\mu_{0,0}^{n-1}\mu_{2,0} + \binom{n}{2}\mu_{0,0}^{n-2}\mu_{1,0}^2
   \right]\mathcal{I}(\boldsymbol{z}_{1:n}) 
    + n\mu_{0,0}^{n-1}\mu_{1,0}A_1(\boldsymbol{z}_{1:n}) + \mu_{0,0}^n A_2(\boldsymbol{z}_{1:n})
    \label{C3}
    \\
    C_4(W_{1:n}) = &
    n\mu_{0,0}^{n-1}\mu_{0,2}\mathcal{I}(\boldsymbol{z}_{1:n})
    +
    \mu_{0,0}^n B_2(\boldsymbol{z}_{1:n},W_{1:n})
    \label{C4}
    \\
    C_5(W_{1:n}) = &
    n\mu_{0,0}^{n-1}\mu_{1,0}B_1(\boldsymbol{z}_{1:n},W_{1:n})+
    \mu_{0,0}^n \Theta(\boldsymbol{z}_{1:n}, W_{1:n})
    \label{C5}
    \\
    C_6(\boldsymbol{X}_{(n+1):N},W_{(n+1):N}) = &
    n\mu_{0,0}^{n-1}\sigma_{0,0}\mathcal{I}(\boldsymbol{z}_{1:n})U_{0,0}
    \label{C6}
    \\
    C_7(\boldsymbol{X}_{(n+1):N},W_{(n+1):N}) =&
    \left[ 
    n\mu_{0,0}^{n-1}\sigma_{1,0}U_{1,0} +
    2\binom{n}{2}\mu_{0,0}^{n-2}\mu_{1,0}\sigma_{0,0}U_{0,0}
    \right]\mathcal{I}(\boldsymbol{z}_{1:n})+
    n\mu_{0,0}^{n-1}\sigma_{0,0}A_1(\boldsymbol{z}_{1:n})U_{0,0}
    \label{C7}
    \\
    C_8(\boldsymbol{X}_{(n+1):N},W_{1:N}) =&
    n\mu_{0,0}^{n-1}\sigma_{0,1}U_{0,1}\mathcal{I}(\boldsymbol{z}_{1:n})
    +n\mu_{0,0}^{n-1}\sigma_{0,0}U_{0,0}B_1(\boldsymbol{z}_{1:n}, W_{1:n})
    \label{C8}
    \\
    C_9(\boldsymbol{X}_{(n+1):N},W_{(n+1):N}) =&
    \Bigg\{
    -\frac{n(n+1)}{2}\mu_{0,0}^n + 
    \nonumber\\
    &\qquad
    \binom{n}{2}\mu_{0,0}^{n-2}\sigma_{0,0}^2 U_{0,0}^2
    +\mu_{0,0}^{n-1} \left(
    \sum_{i=1}^{n} i\frac{g(\boldsymbol{z}_i)}{f(\boldsymbol{z}_i)}{\left( 
    1 + \tau(\boldsymbol{z}_i)
    \right)}^{-1}
    \right)
    \Bigg\}\mathcal{I}(\boldsymbol{z}_{1:n}).
    \label{C9}
\end{align}
Also we denote
\begin{equation}\label{snk-inv-main}
\begin{split}
    v_N(\boldsymbol{X}_{(n+1):N},W_{1:N}) &= C_0^{-1}\Big(
    N^{-r_1}C_1 +N^{-r_2}C_2(W_{1:n}) +N^{-2r_1}C_3 + N^{-2r_2}C_4(W_{1:n})+\\
    &\qquad
    N^{-r_1-r_2}C_5(W_{1:n})+ N^{-\frac{1}{2}}C_6(\boldsymbol{X}_{(n+1):N},W_{(n+1):N})+\\
    &\qquad
    N^{-\frac{1}{2}-r_1}C_7(\boldsymbol{X}_{(n+1):N},W_{(n+1):N}) + N^{-\frac{1}{2}-r_2}C_8(\boldsymbol{X}_{(n+1):N},W_{1:N})+\\
    &\qquad
    N^{-1}C_9(\boldsymbol{X}_{(n+1):N},W_{(n+1):N})
    \Big).
\end{split}
\end{equation}
Then, we have
\begin{equation}\label{rmd7}
\begin{split}
    Rmd_7(\boldsymbol{X}_{(n+1):N},W_{1:N}) &= 
    -C_0^{-2}Rmd_6(\boldsymbol{X}_{(n+1):N},W_{1:N})+\\
    &\qquad
    C_0^{-1}\Big\{ {\left[v_N(\boldsymbol{X}_{(n+1):N},W_{1:N})+C_0^{-1}Rmd_6(\boldsymbol{X}_{(n+1):N},W_{1:N})\right]}^2 -\\
    &\qquad \qquad
    N^{-2r_1}C_0^{-2}C_1^2 - N^{-2r_2}C_0^{-2}C_2^2(W_{1:n})-\\
    &\qquad \qquad
    N^{-1}C_0^{-2}C_6^2(\boldsymbol{X}_{(n+1):N},W_{(n+1):N})
    - N^{-r_1-r_2}2C_0^{-2}C_1C_2(W_{1:n})-\\
    &\qquad \qquad
    N^{-r_1-\frac{1}{2}}2C_0^{-2}C_1C_6(\boldsymbol{X}_{(n+1):N},W_{(n+1):N})
    -\\
    &\qquad \qquad
    N^{-r_2-\frac{1}{2}}2C_0^{-2}C_2(W_{1:n})C_6(\boldsymbol{X}_{(n+1):N},W_{(n+1):N})
    \Big\}-\\
    &\qquad C_0^{-1}\frac{{\left[v_N(\boldsymbol{X}_{(n+1):N},W_{1:N})+C_0^{-1}Rmd_6(\boldsymbol{X}_{(n+1):N},W_{1:N})\right]}^3}{1+\left[v_N(\boldsymbol{X}_{(n+1):N},W_{1:N})+C_0^{-1}Rmd_6(\boldsymbol{X}_{(n+1):N},W_{1:N})\right]}\\
    &\stackrel{w.p. 1}{=}
    O(N^{-3\min\{r_1, r_2, \frac{1}{2}\}}).
\end{split}
\end{equation}

Now by \Cref{ptws-err-ds}, we have 
\begin{equation}\label{final-resul-ds}
    \begin{split}
        &\frac{p_N^n(\boldsymbol{z}_1, \cdots, \boldsymbol{z}_n)}{\prod_{k=1}^{n} g(\boldsymbol{z}_k)}-1 \\
        =& 
    E_{(\boldsymbol{X}_{(n+1):N},W_{1:N})} \left[ 
    {\left(
    \prod_{k=1}^{n}
    s_{N,k}(\boldsymbol{X}_{(n+1):N}, W_{k:N})
    \right)}^{-1}
    \right]
    -1\\
    =&
    \mu_{0,0}^{-n}\mathcal{I}^{-1}(\boldsymbol{z}_{1:n})-1
    \\
    -&
    N^{-r_1}\left\{
    n\mu_{0,0}^{-n-1}\mu_{1,0}\mathcal{I}^{-1}(\boldsymbol{z}_{1:n})
    +
    \mu^{-n}_{0,0}\mathcal{I}^{-2}(\boldsymbol{z}_{1:n})A_1(\boldsymbol{z}_{1:n})
    \right\}\\
    +&
    N^{-2r_1}\Big\{
    -n\mu_{0,0}^{-n-1}\mu_{2,0}\mathcal{I}^{-1}(\boldsymbol{z}_{1:n})
    + \frac{1}{2}n(n+1)\mu_{0,0}^{-n-2}\mu_{1,0}^2\mathcal{I}^{-1}(\boldsymbol{z}_{1:n}) \\
    &\qquad
    +n\mu_{0,0}^{-n-1}\mu_{1,0}\mathcal{I}^{-2}(\boldsymbol{z}_{1:n})A_1(\boldsymbol{z}_{1:n}) - \mu_{0,0}^{-n}\mathcal{I}^{-2}(\boldsymbol{z}_{1:n})A_2(\boldsymbol{z}_{1:n})\\
    &
    \qquad +\mu_{0,0}^{-n} \mathcal{I}^{-3}(\boldsymbol{z}_{1:n})A_1^2(\boldsymbol{z}_{1:n})
    \Big\}\\
    +&
    N^{-2r_2}\Big\{
    -n\mu_{0,0}^{-n-1}\mu_{0,2}\mathcal{I}^{-1}(\boldsymbol{z}_{1:n})  +
    \mu_{0,0}^{-n}\mathcal{I}^{-3}(\boldsymbol{z}_{1:n})\sum_{j=1}^n b_1^2(\boldsymbol{z}_j)\prod_{\stackrel{k=1}{k \neq j}}^{n}{\left(1 + \tau(\boldsymbol{z}_k)\right)}^2 \Big \}\\
    +&
    N^{-1}
    \Big[ 
    \frac{n(n+1)}{2}\mu_{0,0}^{-n} +
    \frac{1}{2}n(n+1)\mu_{0,0}^{-n-2}\sigma_{0,0}^2\\
    &\qquad 
    -\mu_{0,0}^{-n-1}\left(
    \sum_{i=1}^{n}i\frac{g(\boldsymbol{z}_i)}{f(\boldsymbol{z}_i)}{\left( 
    1 + \tau(\boldsymbol{z}_i)
    \right)}^{-1}
    \right)
    \Big]\mathcal{I}^{-1}(\boldsymbol{z}_{1:n})
    \\
    +&
    E_{(\boldsymbol{X}_{(n+1):N},W_{1:N})}[Rmd_7(\boldsymbol{X}_{(n+1):N},W_{1:N})],
    \end{split}
\end{equation}
where we used the fact that $W_{1:N}$ are i.i.d as $\mathcal{N}(0,1)$ and are also independent of $\boldsymbol{X}_{1:N}$,  
and the facts that $E_{(\boldsymbol{X}_{(n+1):N},W_{1:N})}[U_{0,0}] = E_{(\boldsymbol{X}_{(n+1):N},W_{1:N})}[U_{0,1}] = E_{(\boldsymbol{X}_{(n+1):N},W_{1:N})}[U_{1,0}] = 0$ and $E_{(\boldsymbol{X}_{(n+1):N},W_{1:N})}[U_{0,0}^2] = 1$. By \Cref{reminder-part-pf}, $E_{(\boldsymbol{X}_{(n+1):N},W_{1:N})}[Rmd_7(\boldsymbol{X}_{(n+1):N},W_{1:N})] = o(N^{-1})$, which completes the proof.
\end{proofpart}
\end{proof}

\begin{proposition}\label{reminder-part-pf}
    \color{black}
    Let $\boldsymbol{X}_{1:N}$, $W_{1:N}$ be as in \Cref{ds-convergence}. Considering \Cref{rmd7}, we have \newline $E_{(\boldsymbol{X}_{(n+1):N},W_{1:N})}[Rmd_7(\boldsymbol{X}_{(n+1):N},W_{1:N})] = o(N^{-1})$.
\end{proposition}

\begin{proof}[Proof of \Cref{reminder-part-pf}]
\color{black}
    By \Cref{rmd7},
\begin{equation}\label{e-rmd7}
    \begin{split}
       & E_{(\boldsymbol{X}_{(n+1):N},W_{1:N})}[Rmd_7(\boldsymbol{X}_{(n+1):N},W_{1:N})]\\
       &=
       -C_0^2 E_{(\boldsymbol{X}_{(n+1):N},W_{1:N})}[Rmd_6(\boldsymbol{X}_{(n+1):N},W_{1:N})] +\\
       &\qquad
       C_0^{-1}E_{(\boldsymbol{X}_{(n+1):N},W_{1:N})}\Big[
       v_N^2(\boldsymbol{X}_{(n+1):N},W_{1:N})
       -
    N^{-2r_1}C_0^{-2}C_1^2 - N^{-2r_2}C_0^{-2}C_2^2(W_{1:n})-\\
    &\qquad \qquad
    N^{-1}C_0^{-2}C_6^2(\boldsymbol{X}_{(n+1):N},W_{(n+1):N})
    - N^{-r_1-r_2}2C_0^{-2}C_1C_2(W_{1:n})-\\
    &\qquad \qquad
    N^{-r_1-\frac{1}{2}}2C_0^{-2}C_1C_6(\boldsymbol{X}_{(n+1):N},W_{(n+1):N})
    -\\
    &\qquad \qquad
    N^{-r_2-\frac{1}{2}}2C_0^{-2}C_2(W_{1:n})C_6(\boldsymbol{X}_{(n+1):N},W_{(n+1):N})
       \Big]+\\
      &\qquad
      C_0^{-2} 2E_{(\boldsymbol{X}_{(n+1):N},W_{1:N})}\left[
v_N(\boldsymbol{X_{(n+1):N}},W_{1:N})Rmd_6(\boldsymbol{X}_{(n+1):N},W_{1:N})
      \right]+\\
      &\qquad
      C_0^{-3}E_{(\boldsymbol{X}_{(n+1):N},W_{1:N})}\left[
      Rmd_6^2(\boldsymbol{X}_{(n+1):N},W_{1:N})
      \right]-\\
      &\qquad
      E_{(\boldsymbol{X}_{(n+1):N},W_{1:N})}\left[
\frac{{\left[v_N(\boldsymbol{X}_{(n+1):N},W_{1:N})+C_0^{-1}Rmd_6(\boldsymbol{X}_{(n+1):N},W_{1:N})\right]}^3}{1+\left[v_N(\boldsymbol{X}_{(n+1):N},W_{1:N})+C_0^{-1}Rmd_6(\boldsymbol{X}_{(n+1):N},W_{1:N})\right]}
      \right].
    \end{split}
\end{equation}
By \Cref{e-rmd7-part1,e-rmd7-part2,e-rmd7-part3,e-rmd7-part4,e-rmd7-part5}, the fives terms in \Cref{e-rmd7} are of $o(N^{-1})$, so we conclude \Cref{e-rmd7} is so too.
\end{proof}

\begin{proposition}\label{e-rmd7-part1}
\color{black}
    Considering \Cref{rmd6}, we have $E_{(\boldsymbol{X}_{(n+1):N},W_{1:N})}[Rmd_6(\boldsymbol{X}_{(n+1):N},W_{1:N})]  = o(N^{-1}).$
\end{proposition}

\begin{proof}[Proof of \Cref{e-rmd7-part1}]
\color{black}
By \Cref{rmd6,tnp-asym},
\begin{equation}\label{e-rmd6}
    \begin{split}
        &E_{(\boldsymbol{X}_{(n+1):N},W_{1:N})}[Rmd_6(\boldsymbol{X}_{(n+1):N},W_{1:N})] \\
        =&E_{(\boldsymbol{X}_{(n+1):N},W_{(n+1):N})}[Rmd_5(\boldsymbol{X}_{(n+1):N},W_{(n+1):N};n)]\Bigg\{ \mathcal{I}(\boldsymbol{z}_{1:n}) + N^{-r_1}A_1(\boldsymbol{z}_{1:n}) + \\
        &\qquad \qquad
        N^{-2r_1}A_2(\boldsymbol{z}_{1:n})+
        N^{-1}\mathcal{I}(\boldsymbol{z}_{1:n})\frac{n(n-1)}{2} 
        \Bigg\}\\
        &+
        E_{(\boldsymbol{X}_{(n+1):N},W_{(n+1):N})}[Rmd_5(\boldsymbol{X}_{(n+1):N},W_{(n+1):N};n-1)]N^{-1}\mathcal{I}(\boldsymbol{z}_{1:n})\left( 
        \sum_{i=1}^{n}i\frac{g(\boldsymbol{z}_i)}{f(\boldsymbol{z}_i)}{\left( 1 + \tau(\boldsymbol{z}_i) \right)}^{-1}
        \right)\\
        &+
        E_{(\boldsymbol{X}_{(n+1):N},W_{1:N})}[Rmd_3(\boldsymbol{X}_{(n+1):N},W_{1:N})]
        \\
        &+
        N^{-r_1}A_1(\boldsymbol{z}_{1:n})\Bigg\{
        N^{-2r_1}\left[
        n\mu_{0,0}^{n-1}\mu_{2,0} + \binom{n}{2}\mu_{0,0}^{n-2}\mu_{1,0}^2
        \right]+\\
        &\qquad
        N^{-2r_2}n\mu_{0,0}^{n-1}\mu_{0,2}+
        N^{-1}\left[
        -n^2\mu_{0,0}^n+\binom{n}{2}\mu_{0,0}^{n-2}\sigma_{0,0}^2\left( 1 + O(N^{-1}) \right)
        \right]
        \Bigg\}\\
        &+
        N^{-2r_1}A_2(\boldsymbol{z}_{1:n})\Bigg\{
        N^{-r_1}n\mu_{0,0}^{n-1}\mu_{1,0}+
        N^{-2r_1}\left[
        n\mu_{0,0}^{n-1}\mu_{2,0} + \binom{n}{2}\mu_{0,0}^{n-2}\mu_{1,0}^2
        \right]+\\
        &\qquad
        N^{-2r_2}n\mu_{0,0}^{n-1}\mu_{0,2}+
        N^{-1}\left[
        -n^2\mu_{0,0}^n+\binom{n}{2}\mu_{0,0}^{n-2}\sigma_{0,0}^2\left( 1 + O(N^{-1}) \right)
        \right]
        \Bigg\}\\
        &+
        N^{-1}\mathcal{I}(\boldsymbol{z}_{1:n})\frac{n(n-1)}{2}\Bigg\{
        N^{-r_1}n\mu_{0,0}^{n-1}\mu_{1,0}+
        N^{-2r_1}\left[
        n\mu_{0,0}^{n-1}\mu_{2,0} + \binom{n}{2}\mu_{0,0}^{n-2}\mu_{1,0}^2
        \right]+\\
        &\qquad
        N^{-2r_2}n\mu_{0,0}^{n-1}\mu_{0,2}+
        N^{-1}\left[
        -n^2\mu_{0,0}^n+\binom{n}{2}\mu_{0,0}^{n-2}\sigma_{0,0}^2\left( 1 + O(N^{-1}) \right)
        \right]
        \Bigg\}\\
        &+
        N^{-1}\mathcal{I}(\boldsymbol{z}_{1:n})\left( 
        \sum_{i=1}^{n}i\frac{g(\boldsymbol{z}_i)}{f(\boldsymbol{z}_i)}{(1+\tau(\boldsymbol{z}_i))}^{-1}
        \right)
        \Bigg\{
        N^{-r_1}(n-1)\mu_{0,0}^{n-2}\mu_{1,0}+\\
        &\qquad
        N^{-2r_1}\left[
        (n-1)\mu_{0,0}^{n-2}\mu_{2,0} + \binom{n-1}{2}\mu_{0,0}^{n-3}\mu_{1,0}^2
        \right]+
        N^{-2r_2}(n-1)\mu_{0,0}^{n-2}\mu_{0,2}+\\
        &\qquad
        N^{-1}\left[
        -n(n-1)\mu_{0,0}^{n-1}+\binom{n-1}{2}\mu_{0,0}^{n-3}\sigma_{0,0}^2\left( 1 + O(N^{-1}) \right)
        \right]
        \Bigg\}\\
        =&E_{(\boldsymbol{X}_{(n+1):N},W_{(n+1):N})}[Rmd_5(\boldsymbol{X}_{(n+1):N},W_{(n+1):N};n)]\Bigg\{ \mathcal{I}(\boldsymbol{z}_{1:n}) + N^{-r_1}A_1(\boldsymbol{z}_{1:n}) + \\
        &\qquad \qquad
        N^{-2r_1}A_2(\boldsymbol{z}_{1:n})+
        N^{-1}\mathcal{I}(\boldsymbol{z}_{1:n})\frac{n(n-1)}{2} 
        \Bigg\}\\
        &+
        E_{(\boldsymbol{X}_{(n+1):N},W_{(n+1):N})}[Rmd_5(\boldsymbol{X}_{(n+1):N},W_{(n+1):N};n-1)]N^{-1}\mathcal{I}(\boldsymbol{z}_{1:n})\left( 
        \sum_{i=1}^{n}i\frac{g(\boldsymbol{z}_i)}{f(\boldsymbol{z}_i)}{\left( 1 + \tau(\boldsymbol{z}_i) \right)}^{-1}
        \right)\\
        &+
        E_{(\boldsymbol{X}_{(n+1):N},W_{1:N})}[Rmd_3(\boldsymbol{X}_{(n+1):N},W_{1:N})]
        \\
        &+O(N^{-\min\{r_1+2\min\{r_1, r_2, \frac{1}{2}\}, 2r_2+1\}})
    \end{split}
\end{equation}
where we used the facts that $E_{(\boldsymbol{X}_{(n+1):N},W_{1:N})}[U_{j,k-j}] = 0$, for finite $k = 0, 1, \cdots$ and $j = 0, 1, \cdots, k$ and that $W_1, \cdots, W_N$ are i.i.d standard normal variables.

Now we study the term $E_{(\boldsymbol{X}_{(n+1):N},W_{1:N})}[Rmd_3(\boldsymbol{X}_{(n+1):N},W_{1:N})]$. By \Cref{tnp-asym}, we have 
\begin{equation}\label{e-tNp-with-rmd5}
\color{black}
    E_{(\boldsymbol{X}_{(n+1):N},W_{(n+1):N})}[T_N^p(\boldsymbol{X}_{(n+1):N},W_{(n+1):N})] = O(1)+E_{(\boldsymbol{X}_{(n+1):N},W_{(n+1):N})}[Rmd_5(\boldsymbol{X}_{(n+1):N},W_{(n+1):N};p)].
\end{equation}
Thus by \Cref{T0,T1,T2,rmd1} one can get
\begin{align}
    E_{(\boldsymbol{X}_{(n+1):N},W_{1:N})}[T_1(W_{1:n})] &= O(1)\label{e-T1}
    \\
    E_{(\boldsymbol{X}_{(n+1):N},W_{1:N})}[T_2(\boldsymbol{X
    }_{(n+1):N},W_{(n+1):N})] &= O(1) + E_{(\boldsymbol{X}_{(n+1):N},W_{(n+1):N})}[Rmd_5(\boldsymbol{X
    }_{(n+1):N},W_{(n+1):N};n)]+\nonumber \\
    &\qquad
    N^{-1}
    E_{(\boldsymbol{X}_{(n+1):N},W_{(n+1):N})}[Rmd_5(\boldsymbol{X
    }_{(n+1):N},W_{(n+1):N};n-1)]
     \label{e-T2}\\
    E_{(\boldsymbol{X}_{(n+1):N},W_{1:N})}[Rmd_1(W_{1:n})] &= O(N^{-3r_1}) \label{e-rmd1}.
\end{align}

Using \Cref{momentsSRNormal,Rziwi,}, one can see that 
\begin{equation}\label{e-rziwi}
\color{black}
E_{W_i}[R(\boldsymbol{z}_i, W_i)] = -\frac{N^{-r_1}a_1(\boldsymbol{z}_i)}{1 + \tau(\boldsymbol{z}_i)+N^{-r_1}a_1(\boldsymbol{z}_i)} + O(N^{-2r_2}) = O(N^{-\min\{r_1,2r_2\}})
\end{equation}
Also,
\begin{equation}\label{exp-prodgamma}
\begin{split}
    &E_{W_{1:n}}\left[\sum_{\stackrel{\{k_1, \cdots, k_i\}\subset \{1, \cdots, n\}}{k_1 < k_2 < \cdots < k_i}}
\prod_{j=1}^{i} \gamma_{k_j}(\boldsymbol{z}_{k_j:n},W_{k_j:n})\right]\\
=&
\sum_{\stackrel{\{k_1, \cdots, k_i\}\subset \{1, \cdots, n\}}{k_1 < k_2 < \cdots < k_i}}E_{W_{1:n}}\left[\prod_{j=1}^{i} \gamma_{k_j}(\boldsymbol{z}_{k_j:n},W_{k_j:n})\right]\\
=&
\color{black}{
\sum_{\stackrel{\{k_1, \cdots, k_i\}\subset \{1, \cdots, n\}}{k_1 < k_2 < \cdots < k_i}}E_{W_{1:n}}\left[\prod_{j=1}^{i}
\left(
\sum_{t=k_j}^{n}\frac{g(\boldsymbol{z}_t)}{f(\boldsymbol{z}_t)(1+\tau(\boldsymbol{z}_t))} \left(
1 - \frac{N^{-r_1}a_1(\boldsymbol{z}_t) + N^{-r_2}b_1(\boldsymbol{z}_t)W_t}{1+\tau(\boldsymbol{z}_t)+N^{-r_1}a_1(\boldsymbol{z}_t) + N^{-r_2}b_1(\boldsymbol{z}_t)W_t}
\right)
\right)
\right]
}
\\
=& \sum_{\stackrel{\{k_1, \cdots, k_i\}\subset \{1, \cdots, n\}}{k_1 < k_2 < \cdots < k_i}}E_{W_{1:n}}\left[\prod_{j=1}^{i}
\left(
\sum_{t=k_j}^n \frac{g(\boldsymbol{z}_t)}{f(\boldsymbol{z}_t)} \frac{1}{1 + \tau(\boldsymbol{z}_t) + N^{-r_1}a_1(\boldsymbol{z}_t) + N^{-r_2}b_1(\boldsymbol{z}_t) W_t}
\right)
\right] \\
=& \sum_{\stackrel{\{k_1, \cdots, k_i\}\subset \{1, \cdots, n\}}{k_1 < k_2 < \cdots < k_i}}
\sum_{\{r_1, \cdots, r_i\}\in \{k_1:n\}\times \cdots \times \{k_i:n \}}\\
&\qquad 
E_{W_{r_{1:i}}}\left[
\prod_{j=1}^i \frac{g(\boldsymbol{z}_{r_j})}{f(\boldsymbol{z}_{r_j})} \frac{1}{1 + \tau(\boldsymbol{z}_{r_j}) + N^{-r_1}a_1(\boldsymbol{z}_{r_j}) + N^{-r_2}b_1(\boldsymbol{z}_{r_j}) W_{r_j}}
\right]\\
=& \sum_{\stackrel{\{k_1, \cdots, k_i\}\subset \{1, \cdots, n\}}{k_1 < k_2 < \cdots < k_i}}
\sum_{\{r_1, \cdots, r_i\}\in \{k_1:n\}\times \cdots \times \{k_i:n \}}\\
&\qquad 
\left(\prod_{j=1}^i \frac{g(\boldsymbol{z}_{r_j})}{f(\boldsymbol{z}_{r_j})}  \right)
\prod_{j=1}^i E_{W_{r_j}}\left[{\left( \frac{1}{1 + \tau(\boldsymbol{z}_{r_j}) + N^{-r_1}a_1(\boldsymbol{z}_{r_j}) + N^{-r_2}b_1(\boldsymbol{z}_{r_j}) W_{r_j}}
\right)}^{\left(\sum_{t=j}^{i} \boldsymbol{I}_{\{r_t = r_j\}}\right) \left(\prod_{t=1}^{j-1} \boldsymbol{I}_{\{r_t \neq r_j\}} \right)}\right] \\
=&O(1),
\end{split}
\end{equation}
where we define $\prod_{t=1}^{0} \boldsymbol{I}_{\{r_t \neq r_j\}} \myeq 1$. \textcolor{black}{The second equality holds by \Cref{Rziwi,gammak}. In the fourth equality we exchanged the order of the production and summation operators.} In the last but one equality we used the fact that $W_{i}$, $i = 1, \cdots, n$ are i.i.d \textcolor{black}{and the term  $\left(\sum_{t=j}^{i} \boldsymbol{I}_{\{r_t = r_j\}}\right) \left(\prod_{t=1}^{j-1} \boldsymbol{I}_{\{r_t \neq r_j\}} \right)$ simply counts how many times the index $r_j$ appears in the set $\{r_1, \cdots, r_i\}$,} and in the last equality, we applied \Cref{momentsSRNormal}.

Hence,
\begin{equation}\label{e-rmd2}
\begin{split}
    &E_{(\boldsymbol{X}_{(n+1):N},W_{1:N})}[Rmd_2(\boldsymbol{X}_{(n+1):N},W_{1:N})] \\=& 
N^{-1} E_{(\boldsymbol{X}_{(n+1):N},W_{(n+1):N})}[T_N^{n-1}(\boldsymbol{X}_{(n+1):N},W_{(n+1):N})] \sum_{i=1}^{n}i
\frac{g(\boldsymbol{z}_i)}{f(\boldsymbol{z}_i)(1+\tau(\boldsymbol{z}_i))}E_{W_i}[R(\boldsymbol{z}_i,W_i)]+\\
&\qquad 
\sum_{i=2}^{n} N^{-i}E_{(\boldsymbol{X}_{(n+1):N},W_{(n+1):N})}[
T_N^{n-i}(\boldsymbol{X}_{(n+1):N},W_{(n+1):N})]
E_{W_{1:n}}\left[\sum_{\stackrel{\{k_1, \cdots, k_i\}\subset \{1, \cdots, n\}}{k_1 < k_2 < \cdots < k_i}}
\prod_{j=1}^{i} \gamma_{k_j}(\boldsymbol{z}_{k_j:n},W_{k_j:n})\right]\\
=&
O(N^{-\min\{r_1, 2r_2\}-1})\left\{
O(1) + E_{(\boldsymbol{X}_{(n+1):N},W_{(n+1):N})}[Rmd_5(\boldsymbol{X}_{(n+1):N},W_{(n+1):N};n-1)]
\right\}+\\
&\qquad
\sum_{i=2}^{n} O(N^{-i})\left\{
O(1) + E_{(\boldsymbol{X}_{(n+1):N},W_{(n+1):N})}[Rmd_5(\boldsymbol{X}_{(n+1):N},W_{(n+1):N};n-i)]
\right\}\\
=&
O(N^{-\min\{r_1, 2r_2,1\}-1})+\\
&\qquad
O(N^{-\min\{r_1, 2r_2\}-1})E_{(\boldsymbol{X}_{(n+1):N},W_{(n+1):N})}[Rmd_5(\boldsymbol{X}_{(n+1):N},W_{(n+1):N};n-1)]+\\
&\qquad
\sum_{i=2}^{n} O(N^{-i})E_{(\boldsymbol{X}_{(n+1):N},W_{(n+1):N})}[Rmd_5(\boldsymbol{X}_{(n+1):N},W_{(n+1):N};n-i)],
\end{split}
\end{equation}
\textcolor{black}{where the first equality holds by \Cref{rmd2}; the second equality holds by \Cref{e-tNp-with-rmd5,e-rziwi,exp-prodgamma}.}

To figure out the asymptotic order of $E_{(\boldsymbol{X}_{(n+1):N},W_{1:N})}[Rmd_3(\boldsymbol{X}_{(n+1):N},W_{1:N})]$, we first observe the following facts. For $j = 1, \cdots, n$,
\begin{equation}\label{e-xRziwi}
\begin{split}
    E_{W_j}[W_j R(\boldsymbol{z}_j, W_j)]& = -E_{W_j}\left[ 
   W_j \frac{N^{-r_1}a_1(\boldsymbol{z}_j)+N^{-r_2}b_1(\boldsymbol{z}_j)W_j}{1 + \tau(\boldsymbol{z}_j)+N^{-r_1}a_1(\boldsymbol{z}_j) + N^{-r_2}b_1(\boldsymbol{z}_j)W_j}\right] \\
    &=
    -E_{W_j}[W_j]  + (1 + \tau(\boldsymbol{z}_j))E_{W_j}\left[
    W_j
    \frac{1}{1 + \tau(\boldsymbol{z}_j)+N^{-r_1}a_1(\boldsymbol{z}_j) + N^{-r_2}b_1(\boldsymbol{z}_j)W_j}
    \right] 
     \\
     &=
     -(1 + \tau(\boldsymbol{z}_j))N^{-r_2}\abs{b_1(\boldsymbol{z}_j)}E_{W_j}\left[
     {\left( 
     \frac{1}{1 + \tau(\boldsymbol{z}_j)+N^{-r_1}a_1(\boldsymbol{z}_j) + N^{-r_2}b_1(\boldsymbol{z}_j)W_j}
     \right)}^2
     \right]  \\
     &= O(N^{-r_2}).
     \end{split}
\end{equation}
Here (see \Cref{Rziwi} for the definition of $R(\boldsymbol{z}_j, W_j)$) in the second but one equality, we used the fact that $W_j \sim \mathcal{N}(0,1)$ and applied \Cref{xSRNormal} and in the last equality, we applied \Cref{momentsSRNormal}. 
And for  $i = 1, \cdots, n$, $\{k_1, \cdots, k_i\}\subset \{1, \cdots, n\}$, $k_1 < k_2 < \cdots < k_i$, and $\{s_1, \cdots, s_l\} \subset \{1, \cdots, n\}$ (\textcolor{black}{note that $s_1, \cdots, s_l$ are mutually different})
\begin{equation}\label{e-xgammak}
\begin{split}
    &E_{W_{1:n}} \left[\left(\prod_{t=1}^{l} W_{s_t} \right)\prod_{t=1}^{i}\gamma_{k_t}(\boldsymbol{z}_{k_t:n}, W_{k_t:n}) \right]\\
    =& 
\sum_{\{r_1, \cdots, r_i\}\in \{k_1:n\}\times \cdots \times \{k_i:n \}}\left(\prod_{t=1}^i \frac{g(\boldsymbol{z}_{r_t})}{f(\boldsymbol{z}_{r_t})}  \right) \left(\prod_{t=1}^{l}{\left(E[W_{s_t}]\right)}^{\boldsymbol{I}_{\{s_t \notin \{r_1, \cdots, r_i\}\}}}\right)\\
&\qquad 
\prod_{m=1}^i E_{W_{r_m}}\left[{\left( \left(\prod_{t=1}^{l} W_{s_t}^{\boldsymbol{I}_{\{r_m=s_t\}}}\right) {\left( \frac{1}{1 + \tau(\boldsymbol{z}_{r_m}) + N^{-r_1}a_1(\boldsymbol{z}_{r_m}) + N^{-r_2}b_1(\boldsymbol{z}_{r_m}) W_{r_m}}
\right)}^{\left(\sum_{t=m}^{i} \boldsymbol{I}_{\{r_t = r_m\}}\right) }\right)}^{\prod_{t=1}^{m-1} \boldsymbol{I}_{\{r_t \neq r_m\}}}\right]\\
=& O(N^{-lr_2}),
\end{split}
\end{equation}
where in the first equality, we reused the derivation in \Cref{exp-prodgamma} and used the fact that $W_{1:n}$ are i.i.d $N(0,1)$; in the second equality, we applied \Cref{xSRNormal,momentsSRNormal}. The definition of $\gamma_{k_t}(\boldsymbol{z}_{k_t:n}, W_{k_t:n})$ (for $t = 1, \cdots, n$) can be found in \Cref{gammak}. 

Above yields,
\begin{equation}\label{e-b1rmd2}
\begin{split}
    &E_{(\boldsymbol{X}_{(n+1):N},W_{1:N})}\left[ 
    B_1(\boldsymbol{z}_{1:n}, W_{1:n}) Rmd_2(\boldsymbol{X}_{(n+1):N},W_{1:N})
    \right]\\
    =&E_{(\boldsymbol{X}_{(n+1):N},W_{1:N})}\Bigg[  \\
    &\qquad
    \sum_{j=1}^{n} W_j b_1(\boldsymbol{z}_j) 
    \left(\prod_{\stackrel{k=1}{k \neq j}}^{n} \left( 1 + \tau(\boldsymbol{z}_k) \right)\right)
    \Big\{
N^{-1} T_N^{n-1}(\boldsymbol{X}_{(n+1):N},W_{(n+1):N}) \sum_{i=1}^{n}i
\frac{g(\boldsymbol{z}_i)}{f(\boldsymbol{z}_i)(1+\tau(\boldsymbol{z}_i))}R(\boldsymbol{z}_i,W_i)+\\
&\qquad 
\sum_{i=2}^{n} N^{-i}
T_N^{n-i}(\boldsymbol{X}_{(n+1):N},W_{(n+1):N})
\sum_{\stackrel{\{k_1, \cdots, k_i\}\subset \{1, \cdots, n\}}{k_1 < k_2 < \cdots < k_i}}
\prod_{j=1}^{i} \gamma_{k_j}(\boldsymbol{z}_{k_j:n},W_{k_j:n})
    \Big\}
    \Bigg]\\
    =&
    \sum_{j=1}^{n} b_1(\boldsymbol{z}_j) \left(\prod_{\stackrel{k=1}{k \neq j}}^{n} \left( 1 + \tau(\boldsymbol{z}_k) \right)\right)
    \Bigg\{
N^{-1} E_{(\boldsymbol{X}_{(n+1):N},W_{(n+1):N})}\left[T_N^{n-1}(\boldsymbol{X}_{(n+1):N},W_{(n+1):N})\right] \times \\
&\qquad \left(
E_{W_j}[W_j]\sum_{\stackrel{i=1}{i \neq j}}^{n}i
\frac{g(\boldsymbol{z}_i)}{f(\boldsymbol{z}_i)(1+\tau(\boldsymbol{z}_i))}E_{W_i}[R(\boldsymbol{z}_i,W_i)] + \frac{g(\boldsymbol{z}_j)}{f(\boldsymbol{z}_j)(1+\tau(\boldsymbol{z}_j))}E_{W_j}[W_jR(\boldsymbol{z}_j,W_j)]\right)+\\
&\qquad 
\sum_{i=2}^{n} N^{-i} E_{(\boldsymbol{X}_{(n+1):N},W_{(n+1):N})}\left[
T_N^{n-i}(\boldsymbol{X}_{(n+1):N},W_{(n+1):N})\right] \times\\
&\qquad
\sum_{\stackrel{\{k_1, \cdots, k_i\}\subset \{1, \cdots, n\}}{k_1 < k_2 < \cdots < k_i}}
E_{W_{1:n}}\left[W_j\prod_{t=1}^{i} \gamma_{k_t}(\boldsymbol{z}_{k_t:n},W_{k_t:n})\right]
    \Bigg\}\\
    =&
    O(N^{-1-r_2})\left\{
O(1) + E_{(\boldsymbol{X}_{(n+1):N},W_{(n+1):N})}[Rmd_5(\boldsymbol{X}_{(n+1):N},W_{(n+1):N};n-1)]
\right\}+\\
&\qquad
\sum_{i=2}^{n} O(N^{-i-r_2})\left\{
O(1) + E_{(\boldsymbol{X}_{(n+1):N},W_{(n+1):N})}[Rmd_5(\boldsymbol{X}_{(n+1):N},W_{(n+1):N};n-i)]
\right\},
\end{split}
\end{equation}
\textcolor{black}{where the first equality holds by definitions in \Cref{B1-def,rmd2}; the second equality holds by using the assumption that $W_{1:N}$ are i.i.d (see \Cref{assump-fhat}); the third equality holds by \Cref{e-tNp-with-rmd5,e-xRziwi,e-xgammak}.}

And,
\begin{equation}\label{e-b2rmd2}
    \begin{split}
    &E_{(\boldsymbol{X}_{(n+1):N},W_{1:N})}\left[ 
    B_2(\boldsymbol{z}_{1:n}, W_{1:n}) Rmd_2(\boldsymbol{X}_{(n+1):N},W_{1:N})
    \right]\\
        =&
        E_{(\boldsymbol{X}_{(n+1):N},W_{1:N})}\Bigg[ 
    \sum_{\stackrel{\{s,t\}\subset \{1,\cdots, n\}}{s<t}} W_s W_t b_1(\boldsymbol{z}_s) b_1(\boldsymbol{z}_t) 
    \left(
    \prod_{\stackrel{k=1}{k \notin \{s,t\}}}^{n} \left( 1 + \tau(\boldsymbol{z}_k)
    \right)
    \right)
    \Big\{\\
    &\qquad
N^{-1} T_N^{n-1}(\boldsymbol{X}_{(n+1):N},W_{(n+1):N}) \sum_{i=1}^{n}i
\frac{g(\boldsymbol{z}_i)}{f(\boldsymbol{z}_i)(1+\tau(\boldsymbol{z}_i))}R(\boldsymbol{z}_i,W_i)+\\
&\qquad 
\sum_{i=2}^{n} N^{-i}
T_N^{n-i}(\boldsymbol{X}_{(n+1):N},W_{(n+1):N})
\sum_{\stackrel{\{k_1, \cdots, k_i\}\subset \{1, \cdots, n\}}{k_1 < k_2 < \cdots < k_i}}
\prod_{j=1}^{i} \gamma_{k_j}(\boldsymbol{z}_{k_j:n},W_{k_j:n})
    \Big\}
    \Bigg]\\
    =&
    \sum_{\stackrel{\{s,t\}\subset \{1,\cdots, n\}}{s<t}}  b_1(\boldsymbol{z}_s) b_1(\boldsymbol{z}_t) 
    \left(
    \prod_{\stackrel{k=1}{k \notin \{s,t\}}}^{n} \left( 1 + \tau(\boldsymbol{z}_k)
    \right)
    \right)
    \Big\{
    \sum_{i=2}^{n} N^{-i}
E_{(\boldsymbol{X}_{(n+1):N},W_{(n+1):N})}\left[T_N^{n-i}(\boldsymbol{X}_{(n+1):N},W_{(n+1):N})\right] \times \\
&\qquad
\sum_{\stackrel{\{k_1, \cdots, k_i\}\subset \{1, \cdots, n\}}{k_1 < k_2 < \cdots < k_i}}
 E_{W_{1:n}}\left[W_s W_t \prod_{j=1}^{i}\gamma_{k_j}(\boldsymbol{z}_{k_j:n},W_{k_j:n})\right]
    \Big\}\\
    =&
    \sum_{i=2}^{n} O(N^{-i-2r_2})\left\{
O(1) + E_{(\boldsymbol{X}_{(n+1):N},W_{(n+1):N})}[Rmd_5(\boldsymbol{X}_{(n+1):N},W_{(n+1):N};n-i)]
\right\},
    \end{split}
\end{equation}
\textcolor{black}{where the first equality holds by \Cref{B2-def,rmd2}; in the second equality we used the fact that $W_{1:N}$ are i.i.d as standard normal; the last equality holds by \Cref{e-xgammak,e-tNp-with-rmd5}.}

Also,
\begin{equation}\label{e-abrmd2}
    \begin{split}
        &E_{(\boldsymbol{X}_{(n+1):N},W_{1:N})}\left[ 
    \Theta(\boldsymbol{z}_{1:n}, W_{1:n}) Rmd_2(\boldsymbol{X}_{(n+1):N},W_{1:N})
    \right]\\
        =& E_{(\boldsymbol{X}_{(n+1):N},W_{1:N})}\Bigg[
\sum_{\stackrel{\{s,t\}\subset \{1,\cdots, n\}}{s<t}}
    \left[ 
    W_t a_1(\boldsymbol{z}_s) b_1(\boldsymbol{z}_t) 
    + W_s a_1 (\boldsymbol{z}_t)
    b_1(\boldsymbol{z}_s)
    \right]
    \left(
    \prod_{\stackrel{k=1}{k \notin \{s,t \}}}^{n} \left( 1 + \tau(\boldsymbol{z}_k) \right)
    \right)
     \Big\{\\
    &\qquad
N^{-1} T_N^{n-1}(\boldsymbol{X}_{(n+1):N},W_{(n+1):N}) \sum_{i=1}^{n}i
\frac{g(\boldsymbol{z}_i)}{f(\boldsymbol{z}_i)(1+\tau(\boldsymbol{z}_i))}R(\boldsymbol{z}_i,W_i)+\\
&\qquad 
\sum_{i=2}^{n} N^{-i}
T_N^{n-i}(\boldsymbol{X}_{(n+1):N},W_{(n+1):N})
\sum_{\stackrel{\{k_1, \cdots, k_i\}\subset \{1, \cdots, n\}}{k_1 < k_2 < \cdots < k_i}}
\prod_{j=1}^{i} \gamma_{k_j}(\boldsymbol{z}_{k_j:n},W_{k_j:n})
    \Big\}
        \Bigg]\\
        =&  O(N^{-1-r_2})\left\{
O(1) + E_{(\boldsymbol{X}_{(n+1):N},W_{(n+1):N})}[Rmd_5(\boldsymbol{X}_{(n+1):N},W_{(n+1):N};n-1)]
\right\}+\\
&\qquad
\sum_{i=2}^{n} O(N^{-i-r_2})\left\{
O(1) + E_{(\boldsymbol{X}_{(n+1):N},W_{(n+1):N})}[Rmd_5(\boldsymbol{X}_{(n+1):N},W_{(n+1):N};n-i)]
\right\},
    \end{split}
\end{equation}
where \textcolor{black}{the first equality holds by \Cref{AB-def,rmd2}; and the second equality holds by \Cref{e-tNp-with-rmd5,e-xRziwi,e-xgammak}}.

Hence,
\begin{equation}\label{e-t1rmd2}
\begin{split}
    &E_{(\boldsymbol{X}_{(n+1):N},W_{1:N})}\left[ 
    T_1(W_{1:n}) Rmd_2(\boldsymbol{X}_{(n+1):N},W_{1:N})
    \right]\\
    =& E_{(\boldsymbol{X}_{(n+1):N},W_{1:N})}\Bigg[
    \Big(
    \mathcal{I}(\boldsymbol{z}_{1:n}) +  N^{-r_1}A_1(\boldsymbol{z}_{1:n}) + N^{-r_2}B_1(\boldsymbol{z}_{1:n}, W_{1:n})+\\
    &\qquad
    N^{-2r_1}A_2(\boldsymbol{z}_{1:n}) + N^{-2r_2}B_2(\boldsymbol{z}_{1:n}, W_{1:n}) + \\
    &\qquad
    N^{-r_1-r_2}\Theta(\boldsymbol{z}_{1:n},W_{1:n}) \Big)
    Rmd_2(\boldsymbol{X}_{(n+1):N},W_{1:N})
    \Bigg]\\
    =& \left( \mathcal{I}(\boldsymbol{z}_{1:n}) +  N^{-r_1}A_1(\boldsymbol{z}_{1:n}) + N^{-2r_1}A_2(\boldsymbol{z}_{1:n})\right) E_{(\boldsymbol{X}_{(n+1):N},W_{1:N})}\left[ 
    Rmd_2(\boldsymbol{X}_{(n+1):N},W_{1:N})
    \right] +\\
    &\qquad 
    N^{-r_2}E_{(\boldsymbol{X}_{(n+1):N},W_{1:N})}\left[ 
B_1(\boldsymbol{z}_{1:n}, W_{1:n}) Rmd_2(\boldsymbol{X}_{(n+1):N},W_{1:N})
    \right] + \\
    &\qquad
    N^{-2r_2}E_{(\boldsymbol{X}_{(n+1):N},W_{1:N})}\left[ 
B_2(\boldsymbol{z}_{1:n}, W_{1:n})Rmd_2(\boldsymbol{X}_{(n+1):N},W_{1:N})
    \right]+\\
    &\qquad
    N^{-r_1-r_2}E_{(\boldsymbol{X}_{(n+1):N},W_{1:N})}\left[
    \Theta(\boldsymbol{z}_{1:n},W_{1:n})Rmd_2(\boldsymbol{X}_{(n+1):N},W_{1:N})
    \right]\\
    =& O(N^{-\min\{r_1, 2r_2, 1\}-1}) +\\
    &\qquad
    O(N^{-\min\{r_1, 2r_2\}-1})E_{(\boldsymbol{X}_{(n+1):N},W_{(n+1):N})}[Rmd_5(\boldsymbol{X}_{(n+1):N},W_{(n+1):N};n-1)]+\\
    &\qquad
    O(N^{-2r_2})\sum_{i=2}^n O(N^{-i})E_{(\boldsymbol{X}_{(n+1):N},W_{(n+1):N})}[Rmd_5(\boldsymbol{X}_{(n+1):N},W_{(n+1):N};n-i)],
\end{split}
\end{equation}
\textcolor{black}{where the first equality holds by \Cref{T1}; the second equality holds by direct computation from the first one; and the third equality holds by \Cref{e-rmd2,e-b1rmd2,e-b2rmd2,e-abrmd2}.}

And
\begin{equation}\label{e-rmd1rmd2-0}
\begin{split}
&E_{(\boldsymbol{X}_{(n+1):N},W_{1:N})}\left[ 
Rmd_1(W_{1:n})
    Rmd_2(\boldsymbol{X}_{(n+1):N},W_{1:N})\right]\\
   = &E_{(\boldsymbol{X}_{(n+1):N},W_{1:N})}\Bigg[ 
   \Bigg(
    \sum_{t=3}^n
\sum_{\stackrel{\{i_1,\cdots,i_t\}\subset \{1,\cdots,n\}}{i_1 < i_2 <\cdots<i_t}}  \\
&\qquad
\left\{
\prod_{j=1}^{t}\left[ N^{-r_1}a_1(\boldsymbol{z}_{i_j}) + N^{-r_2} b_1(\boldsymbol{z}_{i_j})W_{i_j} \right]
\right\} 
\prod_{\stackrel{j =1}{j \notin \{i_1,\cdots,i_t\}}}^n (1 + \tau(\boldsymbol{z}_j))
\Bigg) \times\\
&\qquad
\Big\{
N^{-1} T_N^{n-1}(\boldsymbol{X}_{(n+1):N},W_{(n+1):N}) \sum_{i=1}^{n}i
\frac{g(\boldsymbol{z}_i)}{f(\boldsymbol{z}_i)(1+\tau(\boldsymbol{z}_i))}R(\boldsymbol{z}_i,W_i)+\\
&\qquad 
\sum_{i=2}^{n} N^{-i}
T_N^{n-i}(\boldsymbol{X}_{(n+1):N},W_{(n+1):N})
\sum_{\stackrel{\{k_1, \cdots, k_i\}\subset \{1, \cdots, n\}}{k_1 < k_2 < \cdots < k_i}}
\prod_{l=1}^{i} \gamma_{k_l}(\boldsymbol{z}_{k_l:n},W_{k_l:n})
    \Big\}
    \Bigg]\\
    =&  \sum_{t=3}^n
\sum_{\stackrel{\{i_1,\cdots,i_t\}\subset \{1,\cdots,n\}}{i_1 < i_2 <\cdots<i_t}} \left( \prod_{\stackrel{j =1}{j \notin \{i_1,\cdots,i_t\}}}^n (1 + \tau(\boldsymbol{z}_j))\right) \times \\
&\qquad \Bigg(
N^{-1} E_{(\boldsymbol{X}_{(n+1):N},W_{(n+1):N})}\left[T_N^{n-1}(\boldsymbol{X}_{(n+1):N},W_{(n+1):N})\right] \times \\
&\qquad
\sum_{i=1}^{n}i
\frac{g(\boldsymbol{z}_i)}{f(\boldsymbol{z}_i)(1+\tau(\boldsymbol{z}_i))}E_{W_{1:n}}\left[R(\boldsymbol{z}_i,W_i)\left\{
\prod_{j=1}^{t}\left[ N^{-r_1}a_1(\boldsymbol{z}_{i_j}) + N^{-r_2} b_1(\boldsymbol{z}_{i_j})W_{i_j} \right]
\right\} \right] +\\
&\qquad
\sum_{i=2}^{n} N^{-i}
E_{(\boldsymbol{X}_{(n+1):N},W_{(n+1):N})}\left[
T_N^{n-i}(\boldsymbol{X}_{(n+1):N},W_{(n+1):N})\right] \times
\\
&\qquad
\sum_{\stackrel{\{k_1, \cdots, k_i\}\subset \{1, \cdots, n\}}{k_1 < k_2 < \cdots < k_i}} E_{W_{1:n}} \left[
\left\{\prod_{l=1}^{i} \gamma_{k_l}(\boldsymbol{z}_{k_l:n},W_{k_l:n})\right\}
\left\{
\prod_{j=1}^{t}\left[ N^{-r_1}a_1(\boldsymbol{z}_{i_j}) + N^{-r_2} b_1(\boldsymbol{z}_{i_j})W_{i_j} \right]
\right\} \right]
\Bigg),
\end{split}
\end{equation}
\textcolor{black}{where the first equality holds by \Cref{rmd1,rmd2}, and the second equality re-arranges terms in the first one.}

For $t = 3, \cdots, n$, $i = 1, \cdots, n$, $\{i_1, \cdots, i_t\} \subset \{1, \cdots, n\}$, $i_1 < i_2 < \cdots < i_t$, 
it holds that
\begin{equation}\label{e-rmd1rmd2-1}
\begin{split}
   &E_{W_{1:n}}\left[R(\boldsymbol{z}_i,W_i)\left\{
\prod_{j=1}^{t}\left[ N^{-r_1}a_1(\boldsymbol{z}_{i_j}) + N^{-r_2} b_1(\boldsymbol{z}_{i_j})W_{i_j} \right]
\right\} \right]\\
=& \left(N^{-r_1}a_1(\boldsymbol{z}_{i})E_{W_i}\left[
R(\boldsymbol{z}_i,W_i)\right] + N^{-r_2} b_1(\boldsymbol{z}_{i})E_{W_i}\left[
R(\boldsymbol{z}_i,W_i)W_{i}\right]
\right)\times \\
&\qquad
\prod_{\stackrel{j=1}{i_j \neq i}}^{t}E_{W_{i_j}}\left[
\left(
N^{-r_1}a_1(\boldsymbol{z}_{i_j}) + N^{-r_2} b_1(\boldsymbol{z}_{i_j})W_{i_j}
\right)
\right]\\
=& O(N^{-tr_1-\min\{r_1, 2r_2\}}+ N^{-(t-1)r_1-2r_2}),
\end{split}
\end{equation}
\textcolor{black}{where in the first equality we used the assumption that $W_{1:n}$ are i.i.d, and in the second equality we used \Cref{e-rziwi,e-xRziwi} and the assumption that $W_{1:n}$ are i.i.d $\mathcal{N}(0,1)$.}

For $t = 3, \cdots, n$,  $\{i_1, \cdots, i_t\} \subset \{1, \cdots, n\}$, $i_1 < i_2 < \cdots < i_t$, $i = 2, \cdots, n$, $\{k_1, \cdots, k_i\} \subset \{1, \cdots, n\}$, $k_1 < \cdots < k_i$, 
it holds that
\begin{equation}\label{e-rmd1rmd2-2}
\begin{split}
&E_{W_{1:n}} \left[
\left\{\prod_{l=1}^{i} \gamma_{k_l}(\boldsymbol{z}_{k_l:n},W_{k_l:n})\right\}
\left\{
\prod_{j=1}^{t}\left[ N^{-r_1}a_1(\boldsymbol{z}_{i_j}) + N^{-r_2} b_1(\boldsymbol{z}_{i_j})W_{i_j} \right]
\right\} \right]\\
=&E_{W_{1:n}} \Bigg[
\left\{\prod_{l=1}^{i} \gamma_{k_l}(\boldsymbol{z}_{k_l:n},W_{k_l:n})\right\} 
\Bigg(
N^{-tr_1}a_1^t(\boldsymbol{z}_{i_j})
+\\
&\qquad
\sum_{j=1}^{t}N^{-(t-j)r_1 - jr_2} 
\sum_{\stackrel{\{s_1, \cdots, s_j\} \subset \{1, \cdots, t \}}{s_1 < \cdots < s_j}}
\left( \prod_{l=1}^{j} b_1(\boldsymbol{z}_{i_{s_l}}) \right)
\left( \prod_{\stackrel{l=1}{l \notin \{s_1, \cdots, s_j \}}}^{t} a_1(\boldsymbol{z}_{i_l})\right)
\left( \prod_{l=1}^j W_{i_{s_l}}\right)
\Bigg)
\Bigg]\\
=&N^{-tr_1}a_1^t(\boldsymbol{z}_{i_j})E_{W_{1:n}}
\left[ 
\prod_{l=1}^{i} \gamma_{k_l}(\boldsymbol{z}_{k_l:n},W_{k_l:n})
\right] + \sum_{j=1}^{t}N^{-(t-j)r_1 - jr_2} 
\sum_{\stackrel{\{s_1, \cdots, s_j\} \subset \{1, \cdots, t \}}{s_1 < \cdots < s_j}}\\
&\qquad
\left( \prod_{l=1}^{j} b_1(\boldsymbol{z}_{i_{s_l}}) \right)
\left( \prod_{\stackrel{l=1}{l \notin \{s_1, \cdots, s_j \}}}^{t} a_1(\boldsymbol{z}_{i_l})\right)
E_{W_{1:n}}
\left[ 
\left( \prod_{l=1}^j W_{i_{s_l}}\right)
\left\{\prod_{l=1}^{i} \gamma_{k_l}(\boldsymbol{z}_{k_l:n},W_{k_l:n})\right\}
\right]\\
=& \color{black}{O(N^{-t\min\{r_1,2r_2\}})}
,
\end{split}
\end{equation}
where \textcolor{black}{the first equality rewrites $\prod_{j=1}^{t}\left[ N^{-r_1}a_1(\boldsymbol{z}_{i_j}) + N^{-r_2} b_1(\boldsymbol{z}_{i_j})W_{i_j} \right]$ by separating the terms involving $W_{1:n}$ with those do not; and the second equality uses \Cref{exp-prodgamma,e-xgammak}.}

Thus, 
\begin{equation}\label{e-rmd1rmd2}
    \begin{split}
        &E_{(\boldsymbol{X}_{(n+1):N},W_{1:N})}\left[ 
Rmd_1(W_{1:n})
    Rmd_2(\boldsymbol{X}_{(n+1):N},W_{1:N})\right]\\
    =&
    \sum_{t=3}^n
\sum_{\stackrel{\{i_1,\cdots,i_t\}\subset \{1,\cdots,n\}}{i_1 < i_2 <\cdots<i_t}} \left( \prod_{\stackrel{j =1}{j \notin \{i_1,\cdots,i_t\}}}^n (1 + \tau(\boldsymbol{z}_j))\right) \times \\
&\qquad \Bigg(
N^{-1} E_{(\boldsymbol{X}_{(n+1):N},W_{(n+1):N})}\left[T_N^{n-1}(\boldsymbol{X}_{(n+1):N},W_{(n+1):N})\right] \times \\
&\qquad
\sum_{i=1}^{n}i
\frac{g(\boldsymbol{z}_i)}{f(\boldsymbol{z}_i)(1+\tau(\boldsymbol{z}_i))}O(N^{-tr_1-\min\{r_1, 2r_2\}}+ N^{-(t-1)r_1-2r_2}) +\\
&\qquad
\sum_{i=2}^{n} N^{-i}
E_{(\boldsymbol{X}_{(n+1):N},W_{(n+1):N})}\left[
T_N^{n-i}(\boldsymbol{X}_{(n+1):N},W_{(n+1):N})\right] \times
\\
&\qquad
\sum_{\stackrel{\{k_1, \cdots, k_i\}\subset \{1, \cdots, n\}}{k_1 < k_2 < \cdots < k_i}}O(N^{-t\min\{r_1,2r_2\}})
\Bigg)\\
=&O(N^{-3r_1 - \min\{r_1, 2r_2\}-1}+N^{-2r_1-2r_2-1}) \times \\
&\qquad
\left(O(1)+E_{(\boldsymbol{X}_{(n+1):N},W_{(n+1):N})}[Rmd_5(\boldsymbol{X}_{(n+1):N},W_{(n+1):N};n-1)]\right)\\
&\qquad
+\sum_{i=2}^{n} O(N^{-i-3\min\{r_1,2r_2\}})
\left(O(1)+E_{(\boldsymbol{X}_{(n+1):N},W_{(n+1):N})}[Rmd_5(\boldsymbol{X}_{(n+1):N},W_{(n+1):N};n-i)]\right),
\end{split}
\end{equation}
\textcolor{black}{where the first equality holds by \Cref{e-rmd1rmd2-0,e-rmd1rmd2-1,e-rmd1rmd2-2}, and the second equality holds by \Cref{e-tNp-with-rmd5}.}

By \Cref{rmd3,T0,rmd0,e-T1,e-T2,e-rmd1,e-rmd2,e-t1rmd2,e-rmd1rmd2}, we have
\begin{equation}\label{e-rmd3}
    \begin{split}
        &E_{(\boldsymbol{X}_{(n+1):N},W_{1:N})}[Rmd_3(\boldsymbol{X}_{(n+1):N},W_{1:N})]\\
        =&
        T_0 E_{(\boldsymbol{X}_{(n+1):N},W_{1:N})}\left[T_1(W_{1:n}) Rmd_2(\boldsymbol{X}_{(n+1):N},W_{1:N})\right] + \\
        &\qquad
        T_0 E_{W_{1:n}}[Rmd_1(W_{1:n})] E_{(\boldsymbol{X}_{(n+1):N},W_{(n+1):N})}\left[T_2(\boldsymbol{X}_{(n+1):N},W_{(n+1):N})\right] +\\
    &\qquad
    T_0 E_{(\boldsymbol{X}_{(n+1):N},W_{1:N})}\left[Rmd_1(W_{1:n}) Rmd_2(\boldsymbol{X}_{(n+1):N},W_{1:N})\right]+\\
    &\qquad
    Rmd_0 E_{W_{1:n}}[T_1(W_{1:n})]
    E_{(\boldsymbol{X}_{(n+1):N},W_{(n+1):N})}\left[
    T_2(\boldsymbol{X}_{(n+1):N},W_{(n+1):N})\right]
    + \\
    &\qquad
    Rmd_0 E_{(\boldsymbol{X}_{(n+1):N},W_{1:N})}\left[T_1(W_{1:n}) Rmd_2(\boldsymbol{X}_{(n+1):N},W_{1:N})\right]+
    \\
    &\qquad
    Rmd_0 E_{W_{1:n}}[Rmd_1(W_{1:n})] E_{(\boldsymbol{X}_{(n+1):N},W_{(n+1):N})}\left[T_2(\boldsymbol{X}_{(n+1):N},W_{(n+1):N})\right]
    + \\
    &\qquad
    Rmd_0 E_{(\boldsymbol{X}_{(n+1):N},W_{1:N})}\left[Rmd_1(W_{1:n}) Rmd_2(\boldsymbol{X}_{(n+1):N},W_{1:N})\right]\\
    =& O(N^{-\min\{r_1, 2r_2, 1 \} -1}) + O(N^{-3r_1}) + \\
    &\qquad
    \left( 
    O(N^{-3r_1}) + O(N^{-2})
    \right)
    E_{(\boldsymbol{X}_{(n+1):N},W_{(n+1):N})}[Rmd_5(\boldsymbol{X}_{(n+1):N},W_{(n+1):N};n)]
    \\
    &\qquad \left( 
    O(N^{-\min\{r_1, 2r_2\}-1}) + O(N^{-3})
    \right)E_{(\boldsymbol{X}_{(n+1):N},W_{(n+1):N})}[Rmd_5(\boldsymbol{X}_{(n+1):N},W_{(n+1):N};n-1)] +\\
    &\qquad
    O(N^{-2r_2}+N^{-3\min\{r_1,2r_2\}})\sum_{i=2}^{n}O(N^{-i})
    E_{(\boldsymbol{X}_{(n+1):N},W_{(n+1):N})}[Rmd_5(\boldsymbol{X}_{(n+1):N},W_{(n+1):N};n-i)],
    \end{split}
\end{equation}
\textcolor{black}{where the first equality holds by \Cref{rmd3}, and the second equality holds by \Cref{e-t1rmd2,e-rmd1rmd2,T0,T1,T2,rmd0,rmd1,rmd2}.}

\Cref{e-rmd6,e-rmd3} together yield
\begin{equation}\label{e-rmd6-final}
    \begin{split}
        &E_{(\boldsymbol{X}_{(n+1):N},W_{1:N})}[Rmd_6(\boldsymbol{X}_{(n+1):N},W_{1:N})]\\
        =&O(N^{-\min\{r_1, 2r_2,1\}-1}) + O(N^{-r_1-2\min\{ r_1, r_2, \frac{1}{2}\}})+\\
        &\qquad
        O(1)E_{(\boldsymbol{X}_{(n+1):N},W_{(n+1):N})}[Rmd_5(\boldsymbol{X}_{(n+1):N},W_{(n+1):N};n)]+\\
        &\qquad
        O(N^{-1})E_{(\boldsymbol{X}_{(n+1):N},W_{(n+1):N})}[Rmd_5(\boldsymbol{X}_{(n+1):N},W_{(n+1):N};n-1)]+\\
        &\qquad
         O(N^{-2r_2}+N^{-3\min\{r_1,2r_2\}})\sum_{i=2}^{n}O(N^{-i})
    E_{(\boldsymbol{X}_{(n+1):N},W_{(n+1):N})}[Rmd_5(\boldsymbol{X}_{(n+1):N},W_{(n+1):N};n-i)].
    \end{split}
\end{equation}

Now we need the asymptotic expansion of $E_{(\boldsymbol{X}_{(n+1):N},W_{(n+1):N})}[Rmd_5(\boldsymbol{X}_{(n+1):N},W_{(n+1):N};p)]$, for $p = 1, 2, \cdots$ to conclude the order of $E_{(\boldsymbol{X}_{(n+1):N},W_{1:N})}[Rmd_6(\boldsymbol{X}_{(n+1):N},W_{1:N})]$.

First by definition of $T_N^p(\boldsymbol{X}_{(n+1):N}, W_{(n+1):N})$ (see first line of \Cref{tN-series}) and \Cref{assump-fhat}, one has
\begin{equation}\label{e-tN-0}
    \begin{split}
        &E_{(\boldsymbol{X}_{(n+1):N},W_{(n+1):N})}[T_N^p(\boldsymbol{X}_{(n+1):N},W_{(n+1):N})]\\
        =&E_{(\boldsymbol{X}_{(n+1):N},W_{(n+1):N})}
        \left[
        N^{-p}
        {\left(
        \sum_{i=n+1}^{N}\frac{g(\boldsymbol{X}_i)}{\hat{f}_N(\boldsymbol{X}_i)}
        \right)}^p
        \right]\\
        =&
        N^{-p}\sum_{\stackrel{\sum_{j={n+1}}^{N}p_j = p}{{\{p_j\}}_{j=n+1}^{N} \subset \mathbb{Z}_{\geq 0}}} \binom{p}{p_{n+1},\cdots,p_N} \prod_{\stackrel{i=n+1}{p_i > 0}}^{N}
        E_{(\boldsymbol{X}_{1},W_{1})}\left[
        {\left(
        \frac{g(\boldsymbol{X}_1)}{\hat{f}_N(\boldsymbol{X}_1)}
        \right)}^{p_i}
        \right]
        ,
    \end{split}
\end{equation}
where in the second equality, $\mathbb{Z}_{\geq 0}$ denotes the set of non-negative integers, and we used the assumption that ${(\boldsymbol{X}_i,W_i)}_{i=n+1}^N$ are i.i.d \textcolor{black}{and the multinomial theorem}.

By \Cref{assump-fhat}, Law of Total Expectations and \Cref{momentsSRNormal}, we have, for each $i = n+1, \cdots, N$ with $p_i > 0$,
\begin{equation}\label{e-g-fhat-p-0}
    \begin{split}
        &E_{(\boldsymbol{X}_{1},W_{1})}\left[
        {\left(
        \frac{g(\boldsymbol{X}_1)}{\hat{f}_N(\boldsymbol{X}_1)}
        \right)}^{p_i}
        \right]\\
        =&
         E_{\boldsymbol{X}_{1}}
        \left[
        \frac{g^{p_i}(\boldsymbol{X}_1)}{f^{p_i}(\boldsymbol{X}_1)}
        E_{W_1|\boldsymbol{X}_{1}}
        \left[
        {\left(
        \frac{1}{1 + \tau(\boldsymbol{X}_1) + N^{-r_1}a_1(\boldsymbol{X}_1) + N^{-r_2}b_1(\boldsymbol{X}_1)W_1}
        \right)}^{p_i}
        \right]
        \right]\\
        \sim &E_{\boldsymbol{X}_{1}}
        \Bigg[
        \frac{g^{p_i}(\boldsymbol{X}_1)}{f^{p_i}(\boldsymbol{X}_1)} \times\\
        &\qquad
        \sum_{k=0}^{\infty}
        N^{-k2r_2} \alpha_{p_i}(k+p_i-1) 
        \frac{b_1^{2k}(\boldsymbol{X}_1)}{{(1+\tau(\boldsymbol{X}_1) + N^{-r_1}a_1(\boldsymbol{X}_1))}^{2k+p_i}}
        \Bigg]\\
        =& \sum_{k=0}^{\infty}
        N^{-k2r_2} \alpha_{p_i}(k+p_i-1)\\
        &\qquad
        E_{\boldsymbol{X}_{1}}
        \Bigg[
        \frac{g^{p_i}(\boldsymbol{X}_1)}{f^{p_i}(\boldsymbol{X}_1)}
        \frac{b_1^{2k}(\boldsymbol{X}_1)}{{(1+\tau(\boldsymbol{X}_1) + N^{-r_1}a_1(\boldsymbol{X}_1))}^{2k+p_i}}
        \Bigg]
        \\
        =&
        \sum_{k=0}^{\infty}
        N^{-k2r_2} \alpha_{p_i}(k+p_i-1)E_{\boldsymbol{X}_{1}}
        \Bigg[
        \frac{g^{p_i}(\boldsymbol{X}_1)}{f^{p_i}(\boldsymbol{X}_1)}b_1^{2k}(\boldsymbol{X}_1)\times \\
        &\qquad
        \sum_{t=0}^{\infty}\binom{t+2k+p_i-1}{2k+p_i-1}{(-1)}^{t}N^{-tr_1}
        \frac{a_1^t(\boldsymbol{X}_1)}{{(1+\tau(\boldsymbol{X}_1))}^{t+2k+p_i}}
        \Bigg]
        ,
    \end{split}
\end{equation}
where the third equality holds due to Tonelli's Theorem and the fact that $\abs{1 + \tau(\boldsymbol{X}_1) + N^{-r_1}a_1(\boldsymbol{X}_1)} > 1 - \varepsilon - \alpha > 0$ (see \Cref{assump-fhat}); and in the last equality we assume that $N$ is close to $\infty$.

Considering that for  fixed $k$ and $N$, by \Cref{assump-fhat},
\begin{equation}
\begin{split}
    &\sum_{t=0}^{\infty}
    E_{\boldsymbol{X}_1}
        \left[\left|
        \frac{g^{p_i}(\boldsymbol{X}_1)}{f^{p_i}(\boldsymbol{X}_1)}
        b_1^{2k}(\boldsymbol{X}_1)
        \binom{t+2k+p_i-1}{2k+p_i-1} {(-1)}^t N^{-tr_1}
        \frac{a_1^t(\boldsymbol{X}_1)}{{\left(1+\tau(\boldsymbol{X}_1)\right)}^{t+2k+p_i}}
        \right|
        \right] \\
        <& \sum_{t=0}^{\infty}
    E_{\boldsymbol{X}_1}
    \left[
    \frac{g^{p_i}(\boldsymbol{X}_1)}{f^{p_i}(\boldsymbol{X}_1)}
    \right]
    {\beta}^{2k}
    \binom{t+2k+p_i-1}{2k+p_i-1} N^{-tr_1}
    \frac{\alpha^t}{{(1 - \varepsilon)}^{t+2k+p_i}}\\
    =&
    E_{\boldsymbol{X}_1}
    \left[
    \frac{g^{p_i}(\boldsymbol{X}_1)}{f^{p_i}(\boldsymbol{X}_1)}
    \right]
    {\beta}^{2k}
    \sum_{t=0}^{\infty}
    \binom{t+2k+p_i-1}{2k+p_i-1} N^{-tr_1}
    \frac{\alpha^t}{{(1 - \varepsilon)}^{t+2k+p_i}}\\
    =&
    E_{\boldsymbol{X}_1}
    \left[
    \frac{g^{p_i}(\boldsymbol{X}_1)}{f^{p_i}(\boldsymbol{X}_1)}
    \right]
    \frac{{\beta}^{2k}}{{(1 - \varepsilon)}^{2k+p_i}}
    {\left( \frac{1}{1-N^{-r_1}\frac{\alpha}{1-\varepsilon}}\right)}^{2k+p_i} < \infty.
\end{split}
\end{equation}
Thus applying Fubini's Theorem to \Cref{e-g-fhat-p-0},
we have

\begin{equation}\label{e-g-fhat-p-1}
    \begin{split}
        &E_{(\boldsymbol{X}_{1},W_{1})}\left[
        {\left(
        \frac{g(\boldsymbol{X}_1)}{\hat{f}_N(\boldsymbol{X}_1)}
        \right)}^{p_i}
        \right]\\
        \sim &
        \sum_{k=0}^{\infty}\sum_{t=0}^{\infty}
        N^{-tr_1-k2r_2} \alpha_{p_i}(k+p_i-1)
        \binom{t+2k+p_i-1}{2k+p_i-1}{(-1)}^{t} \times \\
        &\qquad
        E_{\boldsymbol{X}_{1}}
        \left[
        \frac{g^{p_i}(\boldsymbol{X}_1)}{f^{p_i}(\boldsymbol{X}_1)}b_1^{2k}(\boldsymbol{X}_1)
        \frac{a_1^t(\boldsymbol{X}_1)}{{(1+\tau(\boldsymbol{X}_1))}^{t+2k+p_i}}
        \right].
    \end{split}
\end{equation}

Plugging \Cref{e-g-fhat-p-1} back into \Cref{e-tN-0} yields
\begin{equation}\label{e-tN-p-1}
    \begin{split}
        &E_{(\boldsymbol{X}_{(n+1):N},W_{(n+1):N})}[T_N^p(\boldsymbol{X}_{(n+1):N},W_{(n+1):N})]\\
        \sim &N^{-p}\sum_{\stackrel{\sum_{j={n+1}}^{N}p_j = p}{{\{p_j\}}_{j=n+1}^{N} \subset \mathbb{Z}_{\geq 0}}} \binom{p}{p_{n+1},\cdots,p_N} \prod_{\stackrel{i=n+1}{p_i > 0}}^{N} \Bigg\{\\
        &\qquad
        \sum_{k=0}^{\infty}\sum_{t=0}^{\infty}
        N^{-tr_1-k2r_2} \alpha_{p_i}(k+p_i-1)
        \binom{t+2k+p_i-1}{2k+p_i-1} {(-1)}^{t} \times \\
        &\qquad
        E_{\boldsymbol{X}_{1}}
        \left[
        \frac{g^{p_i}(\boldsymbol{X}_1)}{f^{p_i}(\boldsymbol{X}_1)}b_1^{2k}(\boldsymbol{X}_1)
        \frac{a_1^t(\boldsymbol{X}_1)}{{(1+\tau(\boldsymbol{X}_1))}^{t+2k+p_i}}
        \right]
        \Bigg\}
        \\
        =& N^{-p}\sum_{\stackrel{\sum_{j={n+1}}^{N}p_j = p}{{\{p_j\}}_{j=n+1}^{N} \subset \mathbb{Z}_{\geq 0}}} \binom{p}{p_{n+1},\cdots,p_N} \sum_{k=0}^{\infty}\sum_{t=0}^{\infty} N^{-tr_1-k2r_2} {(-1)}^{t} \times \\
        &\qquad
        \sum_{\substack{\sum_{j={n+1}}^{N}k_j = k\\{\{k_j\}}_{j=n+1}^{N} \subset \mathbb{Z}_{\geq 0}\\k_i = 0\;\text{if}\;p_i=0, \forall i}}
        \sum_{\substack{\sum_{j={n+1}}^{N}t_j = t\\{\{t_j\}}_{j=n+1}^{N} \subset \mathbb{Z}_{\geq 0}\\t_i = 0\;\text{if}\;p_i=0, \forall i}} \prod_{\stackrel{i=n+1}{p_i > 0}}^{N}\\
        &\qquad
        \Bigg\{ \alpha_{p_i}(k_i+p_i-1)
        \binom{t_i+2k_i+p_i-1}{2k_i+p_i-1}
        E_{\boldsymbol{X}_{1}}
        \left[
        \frac{g^{p_i}(\boldsymbol{X}_1)}{f^{p_i}(\boldsymbol{X}_1)}
        \frac{a_1^{t_i}(\boldsymbol{X}_1)b_1^{2k_i}(\boldsymbol{X}_1)}{{(1+\tau(\boldsymbol{X}_1))}^{t_i+2k_i+p_i}}
        \right]
        \Bigg\}
        ,
    \end{split}
\end{equation}
\textcolor{black}{where the last equality was obtained by expanding the finite product of infinite summations (in the first step) as summations of finite products.}

Denote \newline $\mathcal{E}(p_i,k_i,t_i) \myeq \alpha_{p_i}(k_i+p_i-1)
        \binom{t_i+2k_i+p_i-1}{2k_i+p_i-1}
        E_{\boldsymbol{X}_{1}}
        \left[
        \frac{g^{p_i}(\boldsymbol{X}_1)}{f^{p_i}(\boldsymbol{X}_1)}
        \frac{a_1^{t_i}(\boldsymbol{X}_1)b_1^{2k_i}(\boldsymbol{X}_1)}{{(1+\tau(\boldsymbol{X}_1))}^{t_i+2k_i+p_i}}
        \right]$, for $i = n+1, \cdots, N$. \textcolor{black}{\Cref{e-tN-p-1} can be written as follows,}
\begin{equation}\label{e-tN-p}
    \begin{split}
        &E_{(\boldsymbol{X}_{(n+1):N},W_{(n+1):N})}[T_N^p(\boldsymbol{X}_{(n+1):N},W_{(n+1):N})]\\
        \sim&
        N^{-p}\binom{N-n}{p}p!\mu_{0,0}^p + N^{-p}\binom{N-n}{p}p!p N^{-r_1}\mu_{0,0}^{p-1}\mu_{1,0} +\\
        &
        N^{-p}\binom{N-n}{p}p!N^{-2r_1}\left[p\mu_{0,0}^{p-1}\mu_{2,0}
        +\binom{p}{2} \mu_{0,0}^{p-2}\mu_{1,1}^2
        \right]+\\
        &
        N^{-p}\binom{N-n}{p}p!pN^{-2r_2}\mu_{0,0}^{p-1}\mu_{0,2}
        +\\
        & N^{-p}\binom{N-n}{p-1}\frac{p!}{2}(p-1)\left(\sigma_{0,0}^2\mu_{0,0}^{p-2}
        +\mu_{0,0}^p
        \right) +\\
        &
        N^{-p}\binom{N-n}{p} p!\left(\sum_{\substack{k=0\\t=3,4,\cdots}
        } + \sum_{\substack{k=2,3,\cdots\\t=0}} + \sum_{k=1}^{\infty}\sum_{t=1}^{\infty}  \right)
         N^{-tr_1-k2r_2}  \times \\
        &\qquad 
        \sum_{\substack{\sum_{j={n+1}}^{n+p}k_j = k\\{\{k_j\}}_{j=n+1}^{n+p} \subset \mathbb{Z}_{\geq 0}\\k_{n+p+1} = \cdots = k_N = 0}}
        \sum_{\substack{\sum_{j={n+1}}^{n+p}t_j = t\\{\{t_j\}}_{j=n+1}^{n+p} \subset \mathbb{Z}_{\geq 0}\\t_{n+p+1} = \cdots = t_N = 0}} 
        \prod_{i=n+1}^{n+p} \mu_{t_i, 2k_i}
        +\\
        & N^{-p}\binom{N-n}{p-1}\frac{p!}{2}(p-1)
        \left(\sum_{\substack{k=0\\t=1,2,\cdots}
        } + \sum_{\substack{k=1,2,\cdots\\t=0}} + \sum_{k=1}^{\infty}\sum_{t=1}^{\infty}  \right)
         N^{-tr_1-k2r_2}  \times \\
        &\qquad 
        \sum_{\substack{\sum_{j={n+1}}^{n+p-1}k_j = k\\{\{k_j\}}_{j=n+1}^{n+p-1} \subset \mathbb{Z}_{\geq 0}\\k_{n+p} = \cdots = k_N = 0}}
        \sum_{\substack{\sum_{j={n+1}}^{n+p-1}t_j = t\\{\{t_j\}}_{j=n+1}^{n+p-1} \subset \mathbb{Z}_{\geq 0}\\t_{n+p} = \cdots = t_N = 0}} {(-1)}^{t_{n+1}}
         \mathcal{E}(2, k_{n+1}, t_{n+1})
        \prod_{i=n+2}^{n+p-1} \mu_{t_i, 2k_i}+\\
        &
        N^{-p}\sum_{j=2}^{\floor{\frac{p}{2}}}\binom{N-n}{p-j}\frac{p!}{2^j} C_2(p,j) \sum_{k=0}^{\infty} \sum_{t=0}^{\infty} N^{-tr_1-k2r_2}
        \sum_{\substack{\sum_{j={n+1}}^{n+p-j}k_j = k\\{\{k_j\}}_{j=n+1}^{n+p-j} \subset \mathbb{Z}_{\geq 0}\\k_{n+p-j+1} = \cdots = k_N = 0}}
        \sum_{\substack{\sum_{j={n+1}}^{n+p-j}t_j = t\\{\{t_j\}}_{j=n+1}^{n+p-j} \subset \mathbb{Z}_{\geq 0}\\t_{n+p-j+1} = \cdots = t_N = 0}}
        \\
        &\left(\prod_{i=n+1}^{n+j}{(-1)}^{t_{i}}\mathcal{E}
        (2,k_i,t_i)\right) 
        \left(
        \prod_{i=n+j+1}^{n+p-j}\mu_{t_i,2k_i}
        \right)+\\
        &
        N^{-p}\sum_{m=3}^p \sum_{j=\floor{\frac{p}{m}}+\text{mod}(p,m)}^{p-m+1}\binom{N-n}{j} \sum_{\substack{\sum_{i=n+1}^{n+j}p_i=p\\{\{p_i\}}_{i=n+1}^{n+j} \subset \mathbb{Z}_{\geq 1}\\ \max {\{p_i\}}_{i=n+1}^{n+j} = m\\p_{n+j+1} = \cdots = p_N = 0}} \binom{p!}{p_{n+1}!\cdots p_{n+j}!}
        \sum_{k=0}^{\infty}\sum_{t=0}^{\infty} N^{-tr_1-k2r_2} {(-1)}^{t} \times \\
        &\qquad
        \sum_{\substack{\sum_{j={n+1}}^{n+j}k_j = k\\{\{k_j\}}_{j=n+1}^{n+j} \subset \mathbb{Z}_{\geq 0}}}
        \sum_{\substack{\sum_{j={n+1}}^{n+j}t_j = t\\{\{t_j\}}_{j=n+1}^{n+j} \subset \mathbb{Z}_{\geq 0}}} \prod_{i=n+1}^{n+j}\mathcal{E}(p_i,k_i,t_i)
    \end{split}
\end{equation}
where we denote $\text{mod}(p,m) \myeq p - \floor{\frac{p}{m}}$, and \newline
$
    C_2(p,j) \myeq \left|\{p_1,\cdots,p_{p-j}\} \in \mathbb{Z}_{\geq 1} : p_1+\cdots+p_{p-j} = p, \left|{\{p_{m} = 2 \}}_{m=1}^{p-j}\right| = j, \max\{p_1,\cdots,p_{p-j} \} = 2 \}\right|
$.

Considering the fact that for $j = 1, \cdots, p$, $N^{-p}\binom{N-n}{j} = N^{-(p-j))}\frac{1}{j!}\left(1 - N^{-1}\sum_{i=0}^{j-1} (n+i) + O(N^{-2}) \right)$,
taking expectation to both sides of \Cref{tnp-asym}, and saving the algebra, we get 
\begin{equation}\label{e-rmd5}
\begin{split}
        &E_{(\boldsymbol{X}_{(n+1):N},W_{(n+1):N})}[Rmd_5(\boldsymbol{X}_{(n+1):N},W_{(n+1):N}; p)]\\
    \sim&
        O(N^{-2})\mu_{0,0}^p + N^{-r_1-1}\left(-\sum_{i=0}^{p-1}(n+i) +O(N^{-1})\right)p \mu_{0,0}^{p-1}\mu_{1,0} +\\
        &
        N^{-2r_1-1}\left(-\sum_{i=0}^{p-1}(n+i) +O(N^{-1})\right)\left[p\mu_{0,0}^{p-1}\mu_{2,0}
        +\binom{p}{2} \mu_{0,0}^{p-2}\mu_{1,1}^2
        \right]+\\
        &
        N^{-2r_2-1}\left(-\sum_{i=0}^{p-1}(n+i) +O(N^{-1})\right)p\mu_{0,0}^{p-1}\mu_{0,2}
        +\\
        & N^{-2}\left(-\sum_{i=0}^{p-2}(n+i) +O(N^{-1})\right)\binom{p}{2}\left(\sigma_{0,0}^2\mu_{0,0}^{p-2}
        +\mu_{0,0}^p
        \right) -O(N^{-2})\binom{p}{2}\sigma_{0,0}^2\mu_{0,0}^{p-2}+\\
        &
        \left(\sum_{\substack{k=0\\t=3,4,\cdots}
        } + \sum_{\substack{k=2,3,\cdots\\t=0}} + \sum_{k=1}^{\infty}\sum_{t=1}^{\infty}  \right)
         N^{-tr_1-k2r_2}\left(1 - N^{-1}\sum_{i=0}^{p-1}(n+i)+O(N^{-2})\right)  \times \\
        &\qquad 
        \sum_{\substack{\sum_{j={n+1}}^{n+p}k_j = k\\{\{k_j\}}_{j=n+1}^{n+p} \subset \mathbb{Z}_{\geq 0}\\k_{n+p+1} = \cdots = k_N = 0}}
        \sum_{\substack{\sum_{j={n+1}}^{n+p}t_j = t\\{\{t_j\}}_{j=n+1}^{n+p} \subset \mathbb{Z}_{\geq 0}\\t_{n+p+1} = \cdots = t_N = 0}} 
        \prod_{i=n+1}^{n+p} \mu_{t_i, 2k_i}
        +\\
        & 
        \left(\sum_{\substack{k=0\\t=1,2,\cdots}
        } + \sum_{\substack{k=1,2,\cdots\\t=0}} + \sum_{k=1}^{\infty}\sum_{t=1}^{\infty}  \right)
         N^{-tr_1-k2r_2-1}\left(1 - N^{-1}\sum_{i=0}^{p-2}(n+i)+O(N^{-2})\right)\binom{p}{2}  \times \\
        &\qquad 
        \sum_{\substack{\sum_{j={n+1}}^{n+p-1}k_j = k\\{\{k_j\}}_{j=n+1}^{n+p-1} \subset \mathbb{Z}_{\geq 0}\\k_{n+p} = \cdots = k_N = 0}}
        \sum_{\substack{\sum_{j={n+1}}^{n+p-1}t_j = t\\{\{t_j\}}_{j=n+1}^{n+p-1} \subset \mathbb{Z}_{\geq 0}\\t_{n+p} = \cdots = t_N = 0}} {(-1)}^{t_{n+1}}
         \mathcal{E}(2, k_{n+1}, t_{n+1})
        \prod_{i=n+2}^{n+p-1} \mu_{t_i, 2k_i}+\\
        &
        \sum_{j=2}^{\floor{\frac{p}{2}}} \sum_{k=0}^{\infty} \sum_{t=0}^{\infty} N^{-tr_1-k2r_2-j}\frac{1}{(p-j)!} 
        \left(
        1-N^{-1}\sum_{i=0}^{p-j-1}(n+i) +O(N^{-2})
        \right)
        \frac{p!}{2^j} C_2(p,j)\\
        &\qquad
        \sum_{\substack{\sum_{j={n+1}}^{n+p-j}k_j = k\\{\{k_j\}}_{j=n+1}^{n+p-j} \subset \mathbb{Z}_{\geq 0}\\k_{n+p-j+1} = \cdots = k_N = 0}}
        \sum_{\substack{\sum_{j={n+1}}^{n+p-j}t_j = t\\{\{t_j\}}_{j=n+1}^{n+p-j} \subset \mathbb{Z}_{\geq 0}\\t_{n+p-j+1} = \cdots = t_N = 0}}
        \left(\prod_{i=n+1}^{n+j}{(-1)}^{t_{i}}\mathcal{E}
        (2,k_i,t_i)\right) 
        \left(
        \prod_{i=n+j+1}^{n+p-j}\mu_{t_i,2k_i}
        \right)+\\
        &
        \sum_{m=3}^p \sum_{j=\floor{\frac{p}{m}}+\text{mod}(p,m)}^{p-m+1}
        \sum_{\substack{\sum_{i=n+1}^{n+j}p_i=p\\{\{p_i\}}_{i=n+1}^{n+j} \subset \mathbb{Z}_{\geq 1}\\ \max {\{p_i\}}_{i=n+1}^{n+j} = m\\p_{n+j+1} = \cdots = p_N = 0}} 
        \sum_{k=0}^{\infty}\sum_{t=0}^{\infty} N^{-tr_1-k2r_2-(p-j)}
        \frac{1}{j!}\left(
        1-N^{-1}\sum_{i=0}^{j-1}(n+i)+O(N^{-2})
        \right) \times\\
        &\qquad
        \binom{p!}{p_{n+1}!\cdots p_{n+j}!}
        {(-1)}^{t} 
        \sum_{\substack{\sum_{j={n+1}}^{n+j}k_j = k\\{\{k_j\}}_{j=n+1}^{n+j} \subset \mathbb{Z}_{\geq 0}}}
        \sum_{\substack{\sum_{j={n+1}}^{n+j}t_j = t\\{\{t_j\}}_{j=n+1}^{n+j} \subset \mathbb{Z}_{\geq 0}}} \prod_{i=n+1}^{n+j}\mathcal{E}(p_i,k_i,t_i),
    \end{split}
\end{equation}
which indicates $E_{(\boldsymbol{X}_{(n+1):N},W_{(n+1):N})}[Rmd_5(\boldsymbol{X}_{(n+1):N},W_{(n+1):N}; p)] = O(N^{-\min\{r_1+1,2,2r_2+1,3r_1,4r_2,r_1+2r_2\}}) = o(N^{-1})$. By \Cref{e-rmd6-final}, we have
\begin{equation}\label{e-rmd6-final-real}
    \begin{split}
        &E_{(\boldsymbol{X}_{(n+1):N},W_{1:N})}[Rmd_6(\boldsymbol{X}_{(n+1):N},W_{1:N})]\\
        =&
        O(N^{-\min\{r_1+1,2,2r_2+1,3r_1,4r_2,r_1+2r_2\}})\\
        =&o(N^{-1}).
    \end{split}
\end{equation}
\end{proof}

\begin{proposition}\label{e-rmd7-part2}
\color{black}
    The second term of \Cref{e-rmd7} is of $o(N^{-1})$, i.e., 
\begin{equation}\label{2nd-term-e-rmd7}
    \begin{split}
        &C_0^{-1}E_{(\boldsymbol{X}_{(n+1):N},W_{1:N})}\Big[
       v_N^2(\boldsymbol{X}_{(n+1):N},W_{1:N})
       -
    N^{-2r_1}C_0^{-2}C_1^2 - N^{-2r_2}C_0^{-2}C_2^2(W_{1:n})-\\
    &\qquad 
    N^{-1}C_0^{-2}C_6^2(\boldsymbol{X}_{(n+1):N},W_{(n+1):N})
    - N^{-r_1-r_2}2C_0^{-2}C_1C_2(W_{1:n})-\\
    &\qquad 
    N^{-r_1-\frac{1}{2}}2C_0^{-2}C_1C_6(\boldsymbol{X}_{(n+1):N},W_{(n+1):N})
    -\\
    &\qquad 
    N^{-r_2-\frac{1}{2}}2C_0^{-2}C_2(W_{1:n})C_6(\boldsymbol{X}_{(n+1):N},W_{(n+1):N})
       \Big] = o(N^{-1}).
    \end{split}
\end{equation}
\end{proposition}

\begin{proof}[Proof of \Cref{e-rmd7-part2}]
\color{black}
Since the second term of \Cref{e-rmd7} only involves finite number of elements, using \Cref{lemma-prod-U} and saving the algebra, one gets that its expectation is of $O(N^{-\min\{4r_2, 1+2r_2, 2, 3r_1, r_1+1,r_1+2r_2 \}}) = o(N^{-1})$, under \Cref{assump-fhat}.
\end{proof}

\begin{proposition}\label{e-rmd7-part3}
\color{black}
    The third term of \Cref{e-rmd7} is of $o(N^{-1})$, i.e., \newline $C_0^{-2} 2E_{(\boldsymbol{X}_{(n+1):N},W_{1:N})}\left[
v_N(\boldsymbol{X_{(n+1):N}},W_{1:N})Rmd_6(\boldsymbol{X}_{(n+1):N},W_{1:N})
      \right] = o(N^{-1})$.
\end{proposition}

\begin{proof}[Proof of \Cref{e-rmd7-part3}]
    \color{black}
Now we analyze the third term of \Cref{e-rmd7}. By \Cref{lemma-prod-U}, we have $\forall p \in \mathbb{Z}_{\geq 2}$, for $k_p = 0, 1, \cdots$ and $j_p = 0, \cdots, k_p$, 
\begin{equation}\label{e-prod-U}
\begin{split}
    E_{(\boldsymbol{X}_{(n+1):N}, W_{(n+1):N})}\left[ 
    \prod_{t=1}^{p} U_{j_t, k_t-j_t}(\boldsymbol{X}_{(n+1):N}, W_{(n+1):N}) \right]=
    &O(N^{-\frac{p}{2}+\floor{\frac{p}{2}}})\\
    =&
    \begin{cases}
    O(1), \text{if}\; p \; \text{even}\\
    O(N^{-\frac{1}{2}}), \text{if}\; p \; \text{odd}.
    \end{cases} 
\end{split}
\end{equation}

 By \Cref{tN-series} and \Cref{expan-tN-term1}, $\forall q \in \mathbb{Z}_{\geq 1}$, for $r_q = 0, 1, \cdots$ and $s_q = 0, \cdots, r_q$ (without ambiguity, we use notations $U_{s_t, r_t-s_t}(\boldsymbol{X}_{(n+1):N}, W_{(n+1):N})$ and $U_{s_t, r_t-s_t}$ interchangeably), 
\begin{equation}\label{e-tn-p}
\begin{split}
    &E_{(\boldsymbol{X}_{(n+1):N},W_{(n+1):N})}[\left(
    \prod_{t=1}^{q} U_{s_t, r_t-s_t}(\boldsymbol{X}_{(n+1):N}, W_{(n+1):N})
    \right) T_N^p(\boldsymbol{X}_{(n+1):N},W_{(n+1):N})]\\
    =&E_{(\boldsymbol{X}_{(n+1):N},W_{(n+1):N})}\Bigg[
    \left(
    \prod_{t=1}^{q} U_{s_t, r_t-s_t}
    \right) \times\\
    &\qquad
    {\left(
    \sum_{k=0}^{\infty} \sum_{j=0}^{k} N^{-jr_1-(k-j)r_2} \left(\sigma_{j, k-j}\frac{\sqrt{N-n}}{N}U_{j,k-j}+\frac{N-n}{N}\mu_{j,k-j}\right)\right)}^p
    \Bigg]\\
    =&
    E_{(\boldsymbol{X}_{(n+1):N},W_{(n+1):N})}\Bigg[
    \sum_{k=0}^{\infty} \sum_{\substack{i_1 + \cdots +i_p = k\\\{i_1, \cdots, i_p\} \subset \mathbb{Z}_{\geq 0}}}
    \left(
    \prod_{t=1}^{q} U_{s_t, r_t-s_t}
    \right)
    \\
    &\qquad
    \prod_{t=1}^{p} \left(\sum_{j=0}^{i_t} N^{-jr_1-(i_t-j)r_2} \left(\sigma_{j, i_t-j}\frac{\sqrt{N-n}}{N}U_{j,i_t-j} +\frac{N-n}{N}\mu_{j,i_t-j}\right)\right)\Bigg] \\
    =&
   E_{(\boldsymbol{X}_{(n+1):N},W_{(n+1):N})}\Bigg[
    \sum_{k=0}^{\infty} \sum_{\substack{i_1 + \cdots +i_p = k\\\{i_1, \cdots, i_p\} \subset \mathbb{Z}_{\geq 0}}}
    \sum_{j_1=0}^{i_1} \cdots \sum_{j_p=0}^{i_p}
    \left(
    \prod_{t=1}^{p} N^{-j_tr_1 - (i_t-j_t)r_2}
    \right)\\
    &\qquad
    \sum_{m_1=1}^{2} \cdots \sum_{m_p=1}^{2} 
    \left(
    \prod_{t=1}^{p} V_{m_t, i_t, j_t},
    \right)
    \left(
    \prod_{t=1}^{q} U_{s_t, r_t-s_t}
    \right)
     \Bigg]
\end{split}
\end{equation}
where for $t = 1, \cdots, p$,
\begin{equation}\label{V-4idx}
V_{m_t, i_t, j_t} =
    \begin{cases}
    \sigma_{j_t, i_t-j_t}\frac{\sqrt{N-n}}{N}U_{j_t,i_t-j_t} , \; m_t = 1\\
    \frac{N-n}{N}\mu_{j_t,i_t-j_t}, \; m_t = 2.
\end{cases}
\end{equation}
\textcolor{black}{Here, the first equality holds by the last equality of \Cref{tN-series} and first equality of \Cref{expan-tN-term1}; the second equality expands the power of infinite sum in the first step; the last equality expands the finite products of finite summations in the second step.}

Since
\begin{equation}\label{e-tnp-abs}
    \begin{split}
         &
    \sum_{k=0}^{\infty} E_{(\boldsymbol{X}_{(n+1):N},W_{(n+1):N})}\Bigg[
    \Bigg| \sum_{\substack{i_1 + \cdots +i_p = k\\\{i_1, \cdots, i_p\} \subset \mathbb{Z}_{\geq 0}}}
    \sum_{j_1=0}^{i_1} \cdots \sum_{j_p=0}^{i_p}
    \left(
    \prod_{t=1}^{p} N^{-j_tr_1 - (i_t-j_t)r_2}
    \right)\\
    &\qquad
    \sum_{m_1=1}^{2} \cdots \sum_{m_p=1}^{2} 
    \left(
    \prod_{t=1}^{p} V_{m_t, i_t, j_t},
    \right) \left(
    \prod_{t=1}^{q} U_{s_t, r_t-s_t}
    \right) \Bigg|
     \Bigg]\\
     \leq &
     \sum_{k=0}^{\infty} 
    \sum_{\substack{i_1 + \cdots +i_p = k\\\{i_1, \cdots, i_p\} \subset \mathbb{Z}_{\geq 0}}}
    \sum_{j_1=0}^{i_1} \cdots \sum_{j_p=0}^{i_p}
    \left(
    \prod_{t=1}^{p} N^{-j_tr_1 - (i_t-j_t)r_2}
    \right)\\
    &\qquad
    \sum_{m_1=1}^{2} \cdots \sum_{m_p=1}^{2} 
    E_{(\boldsymbol{X}_{(n+1):N},W_{(n+1):N})}\left[
    \left|
    \left(
    \prod_{t=1}^{p} V_{m_t, i_t, j_t},
    \right) \left(
    \prod_{t=1}^{q} U_{s_t, r_t-s_t}
    \right) \right|
     \right]\\
     \leq &
     \sum_{k=0}^{\infty} 
    \sum_{\substack{i_1 + \cdots +i_p = k\\\{i_1, \cdots, i_p\} \subset \mathbb{Z}_{\geq 0}}}
    \sum_{j_1=0}^{i_1} \cdots \sum_{j_p=0}^{i_p}
    \left(
    \prod_{t=1}^{p} N^{-j_tr_1 - (i_t-j_t)r_2}
    \right)\\
    &\qquad
    \sum_{m_1=1}^{2} \cdots \sum_{m_p=1}^{2} \left(
    E_{(\boldsymbol{X}_{(n+1):N},W_{(n+1):N})}\left[
    \left(
    \prod_{t=1}^{p} V_{m_t, i_t, j_t}^2
    \right)
    \left(
    \prod_{t=1}^{q} U^2_{s_t, r_t-s_t}
    \right)
     \right] + 1\right)
    \end{split}
\end{equation}

Now let 
\begin{equation}
\begin{split}
    A_k &\myeq \sum_{\substack{i_1 + \cdots +i_p = k\\\{i_1, \cdots, i_p\} \subset \mathbb{Z}_{\geq 0}}}
    \sum_{j_1=0}^{i_1} \cdots \sum_{j_p=0}^{i_p}
    \left(
    \prod_{t=1}^{p} N^{-j_tr_1 - (i_t-j_t)r_2}
    \right)
    \sum_{m_1=1}^{2} \cdots \sum_{m_p=1}^{2}  \\
    &\qquad
    \left(
    E_{(\boldsymbol{X}_{(n+1):N},W_{(n+1):N})}\left[
    \left(
    \prod_{t=1}^{p} V_{m_t, i_t, j_t}^2
    \right)
    \left(
    \prod_{t=1}^{q} U^2_{s_t, r_t-s_t}
    \right)
     \right] + 1\right)
\end{split}
\end{equation}

By \Cref{V-4idx} and \Cref{e-prod-U}, with $N$ going to infinity, $E_{(\boldsymbol{X}_{(n+1):N},W_{(n+1):N})}\left[
    \left(
    \prod_{t=1}^{p} V_{m_t, i_t, j_t}^2
    \right)
    \left(
    \prod_{t=1}^{q} U^2_{s_t, r_t-s_t}
    \right)
     \right] + 1 = O(1)$. Thus $A_k = O(N^{-k\min\{r_1,r_2\}})$, which indicates the quantity evaluated in \Cref{e-tnp-abs} is of $o(1)$ as $N$ goes to infinity. 

Now applying Fubini's theorem and using \Cref{tnp-asym}, we have
\begin{equation}\label{e-prodU-prod-Tnp}
    \begin{split}
        &E_{(\boldsymbol{X}_{(n+1):N},W_{(n+1):N})}[\left(
    \prod_{t=1}^{q} U_{s_t, r_t-s_t}(\boldsymbol{X}_{(n+1):N}, W_{(n+1):N})
    \right) T_N^p(\boldsymbol{X}_{(n+1):N},W_{(n+1):N})]\\
    =&
    \sum_{k=0}^{\infty} \sum_{\substack{i_1 + \cdots +i_p = k\\\{i_1, \cdots, i_p\} \subset \mathbb{Z}_{\geq 0}}}
    \sum_{j_1=0}^{i_1} \cdots \sum_{j_p=0}^{i_p}
    \left(
    \prod_{t=1}^{p} N^{-j_tr_1 - (i_t-j_t)r_2}
    \right)\\
    &\qquad
    \sum_{m_1=1}^{2} \cdots \sum_{m_p=1}^{2} 
     E_{(\boldsymbol{X}_{(n+1):N},W_{(n+1):N})}\left[
    \left(
    \prod_{t=1}^{p} V_{m_t, i_t, j_t},
    \right)
    \left(
    \prod_{t=1}^{q} U_{s_t, r_t-s_t}
    \right)
     \right]\\
     =&
      \left(\mu_{0,0}^p+ N^{-r_1} p\mu_{0,0}^{p-1}\mu_{1,0}
      + N^{-2r_1}\left\{p\mu_{0,0}^{p-1}\mu_{2,0} + \binom{p}{2}\mu_{0,0}^{p-2} \mu_{1,0}^2\right\}+
        N^{-2r_2} 
        p\mu_{0,0}^{p-1}\mu_{0,2}
      \right)\times\\
     &\qquad E_{(\boldsymbol{X}_{(n+1):N},W_{(n+1):N})}\left[\left(
    \prod_{t=1}^{q} U_{s_t, r_t-s_t}
    \right)
     \right]  
        \\
        &+
        N^{-\frac{1}{2}}p\mu_{0,0}^{p-1}\sigma_{0,0}E_{(\boldsymbol{X}_{(n+1):N},W_{(n+1):N})}\left[U_{0,0}\left(
    \prod_{t=1}^{q} U_{s_t, r_t-s_t}\right)\right]\\
        &+
        N^{-(r_1+\frac{1}{2})}\Bigg\{
        p\mu_{0,0}^{p-1}\sigma_{1,0}E_{(\boldsymbol{X}_{(n+1):N},W_{(n+1):N})}\left[U_{1,0}\left(
    \prod_{t=1}^{q} U_{s_t, r_t-s_t}\right)\right]
        +\\
        &\qquad
        \binom{p}{2} \mu_{0,0}^{p-2}2\mu_{1,0}\sigma_{0,0}E_{(\boldsymbol{X}_{(n+1):N},W_{(n+1):N})}\left[U_{0,0}\left(
    \prod_{t=1}^{q} U_{s_t, r_t-s_t}\right)\right]
        \Bigg\}\\
        &+
        N^{-(r_2+\frac{1}{2})}
        p\mu_{0,0}^{p-1}\sigma_{0,1}E_{(\boldsymbol{X}_{(n+1):N},W_{(n+1):N})}\left[U_{0,1}\left(
    \prod_{t=1}^{q} U_{s_t, r_t-s_t}
    \right)
     \right] 
        \\
        &+
        N^{-1}\Bigg\{
        -p\mu_{0,0}^pn E_{(\boldsymbol{X}_{(n+1):N},W_{(n+1):N})}\left[\left(
    \prod_{t=1}^{q} U_{s_t, r_t-s_t}
    \right)
     \right] \\
     &\qquad + \binom{p}{2}\mu_{0,0}^{p-2}\sigma_{0,0}^2
E_{(\boldsymbol{X}_{(n+1):N},W_{(n+1):N})}\left[U_{0,0}^2\left(
    \prod_{t=1}^{q} U_{s_t, r_t-s_t}
    \right)
     \right] 
        \Bigg\}\\
      &+E_{(\boldsymbol{X}_{(n+1):N},W_{(n+1):N})}\left[\left(
    \prod_{t=1}^{q} U_{s_t, r_t-s_t}
    \right) Rmd_5(\boldsymbol{X}_{(n+1):N},W_{(n+1):N};p)
     \right],
    \end{split}
\end{equation}
where (saving some algebra and using \Cref{e-prod-U}), 
\begin{equation}\label{e-uprod-rmd5}
    \begin{split}
       &E_{(\boldsymbol{X}_{(n+1):N},W_{(n+1):N})}\left[\left(
    \prod_{t=1}^{q} U_{s_t, r_t-s_t}
    \right) Rmd_5(\boldsymbol{X}_{(n+1):N},W_{(n+1):N};p)
     \right]\\
     =&O(N^{-2\min\{\frac{1}{2},r_1,r_2\}-\frac{1}{2}})+\\
     &\qquad
      \sum_{k=3}^{\infty} \sum_{\substack{i_1 + \cdots +i_p = k\\\{i_1, \cdots, i_p\} \subset \mathbb{Z}_{\geq 0}}}
    \sum_{j_1=0}^{i_1} \cdots \sum_{j_p=0}^{i_p}
    \left(
    \prod_{t=1}^{p} N^{-j_tr_1 - (i_t-j_t)r_2}
    \right) \times\\
    &\qquad \qquad
    \sum_{m_1=1}^{2} \cdots \sum_{m_p=1}^{2} 
     E_{(\boldsymbol{X}_{(n+1):N},W_{(n+1):N})}\left[
    \left(
    \prod_{t=1}^{p} V_{m_t, i_t, j_t},
    \right)
    \left(
    \prod_{t=1}^{q} U_{s_t, r_t-s_t}
    \right)
     \right]\\
     =&O(N^{-3\min\{r_1, r_2, \frac{1}{2}\}}).
    \end{split}
\end{equation}

Also, we see from \Cref{e-prodU-prod-Tnp,e-uprod-rmd5} that 
\begin{equation}\label{asym-e-prodU-prod-Tnp}
E_{(\boldsymbol{X}_{(n+1):N},W_{(n+1):N})}[\left(
    \prod_{t=1}^{q} U_{s_t, r_t-s_t}(\boldsymbol{X}_{(n+1):N}, W_{(n+1):N})
    \right) T_N^p(\boldsymbol{X}_{(n+1):N},W_{(n+1):N})] = 
    \begin{cases}
        O(N^{-\frac{1}{2}}), \; q = 1\\
        O(1), \; q \geq 2.
    \end{cases}
\end{equation}

Using above and saving the algebra, we have 
\begin{equation}\label{e-vn-rmd6}
\begin{split}
   & E_{(\boldsymbol{X}_{(n+1):N},W_{1:N})}\left[
v_N(\boldsymbol{X_{(n+1):N}},W_{1:N})Rmd_6(\boldsymbol{X}_{(n+1):N},W_{1:N})\right] \\
=& O(N^{-\min\{1+2r_1, 4r_1, 4r_2, 2r_2+1, \frac{1}{2}+3r_1, \frac{1}{2}+3r_2, 2\}})\\
=&
o(N^{-1}).
\end{split}
\end{equation}
\end{proof}

\begin{proposition}\label{e-rmd7-part4}
\color{black}
    The fourth term of \Cref{e-rmd7} is of $o(N^{-1})$, i.e., $C_0^{-3}E_{(\boldsymbol{X}_{(n+1):N},W_{1:N})}\left[
      Rmd_6^2(\boldsymbol{X}_{(n+1):N},W_{1:N})
      \right] = o(N^{-1})$.
\end{proposition}

\begin{proof}[Proof of \Cref{e-rmd7-part4}]\label{pf-part4}
 \color{black}

Using \Cref{rmd4,rmd5}, we can similarly compute for $m \in \mathbb{Z}_{\geq 1}$, \newline $E_{(\boldsymbol{X}_{(n+1):N},W_{1:N})}[Rmd_4^m(\boldsymbol{X}_{(n+1):N},W_{(n+1):N})] = O(N^{-3m\min\{r_1, r_2, \frac{1}{2}\}})$; and \newline $E_{(\boldsymbol{X}_{(n+1):N},W_{1:N})}[Rmd_5^m(\boldsymbol{X}_{(n+1):N},W_{(n+1):N};p)] = O(N^{-3m\min\{r_1, r_2, \frac{1}{2}\}})$ (note that this is consistent with, albeit less accurate than, the results we provided above \Cref{e-rmd6-final-real}).

Besides, \Cref{SRnormal,Rziwi,gammak} yields, for $p = 1, 2, \cdots$ and $m = 1, 2, \cdots$
\begin{equation}
    E_{W_i}[W_i^p R^m(z_i, W_i)] 
    =
    \begin{cases}
        O(N^{-m\min\{r_1, r_2\}}), \; p \; \text{even}\\
        O(N^{-mr_2}),  \; p \; \text{odd}.
    \end{cases}
\end{equation}
And \begin{equation}
    E_{W_{1:n}}\left[W_i^p {\left(\sum_{\stackrel{\{k_1, \cdots, k_i\} \subset \{1, \cdots, n \}}{k_1 < k_2 < \cdots < k_i}} \prod_{j=1}^{i} \gamma_{k_j}(\boldsymbol{z}_{k_j:n}, W_{k_j:n}) \right)}^m\right] = O(N^{-pr_2}).
\end{equation}

Thus we have 
\begin{equation}
    E_{(\boldsymbol{X}_{(n+1):N},W_{1:N})}[Rmd_3^2(\boldsymbol{X}_{(n+1):N},W_{1:N})] = O(N^{-\min\{2+2r_1, 2+2r_2, 6r_1, 2r_1+4r_2, 8r_2 \}}),
\end{equation}
which further gives (again, saving all the algebra for brevity)
\begin{equation}
    E_{(\boldsymbol{X}_{(n+1):N},W_{1:N})}[Rmd_6^2(\boldsymbol{X}_{(n+1):N},W_{1:N})] = O(N^{-\min\{6r_1, 6r_2, 3, 2+2r_1, 2+2r_2\}}) = o(N^{-1}).
\end{equation}

\end{proof}

\begin{proposition}\label{e-rmd7-part5}
    \color{black}
    The last term of \Cref{e-rmd7} is of $o(N^{-1})$, i.e., \newline $
      E_{(\boldsymbol{X}_{(n+1):N},W_{1:N})}\left[
\frac{{\left[v_N(\boldsymbol{X}_{(n+1):N},W_{1:N})+C_0^{-1}Rmd_6(\boldsymbol{X}_{(n+1):N},W_{1:N})\right]}^3}{1+\left[v_N(\boldsymbol{X}_{(n+1):N},W_{1:N})+C_0^{-1}Rmd_6(\boldsymbol{X}_{(n+1):N},W_{1:N})\right]}
      \right] = o(N^{-1})$. 
\end{proposition}

\begin{proof}[Proof of \Cref{e-rmd7-part5}]
    \color{black}
For $p = 1, 2, \cdots$, $E_{(\boldsymbol{X}_{(n+1):N},W_{1:N})}[Rmd_3^p(\boldsymbol{X}_{(n+1):N},W_{1:N})] = O(N^{-p\min\{r_1, r_2\}-2p\min\{r_1, r_2, \frac{1}{2}\}})$, and $E_{(\boldsymbol{X}_{(n+1):N},W_{1:N})}[Rmd_6^p(\boldsymbol{X}_{(n+1):N},W_{1:N})] = O(N^{-3p\min\{r_1, r_2, \frac{1}{2}\}}) < \infty$. By \Cref{snk-inv-main} and results in the proofs of \Cref{e-rmd7-part3,e-rmd7-part4}, $E_{(\boldsymbol{X}_{(n+1):N},W_{1:N})}[v_N^p((\boldsymbol{X}_{(n+1):N},W_{1:N}))] = O(N^{-p\min\{r_1, r_2, \frac{1}{2}\}})<\infty$. Since w.p. 1 \newline $Rmd_6(\boldsymbol{X}_{(n+1):N},W_{1:N}) = O(N^{-3\min\{r_1, r_2, \frac{1}{2}\}})$ and  $v_N = O(N^{-\min\{r_1, r_2, \frac{1}{2}\}})$, we have \newline $v_N(\boldsymbol{X}_{(n+1):N},W_{1:N})+C_0^{-1}Rmd_6(\boldsymbol{X}_{(n+1):N},W_{1:N}) = O(N^{-\min\{r_1, r_2, \frac{1}{2}\}})$. And we can expand the last term of \Cref{e-rmd7} at $0$ and obtain
\begin{equation}\label{e-last-reminder}
    \begin{split}
       & E_{(\boldsymbol{X}_{(n+1):N},W_{1:N})}\left[
\frac{{\left[v_N(\boldsymbol{X}_{(n+1):N},W_{1:N})+C_0^{-1}Rmd_6(\boldsymbol{X}_{(n+1):N},W_{1:N})\right]}^3}{1+\left[v_N(\boldsymbol{X}_{(n+1):N},W_{1:N})+C_0^{-1}Rmd_6(\boldsymbol{X}_{(n+1):N},W_{1:N})\right]}
      \right]\\
      =&
      E_{(\boldsymbol{X}_{(n+1):N},W_{1:N})}\left[
       \sum_{k=0}^{\infty} {(-1)}^k {(v_N(\boldsymbol{X}_{(n+1):N},W_{1:N})+C_0^{-1}Rmd_6(\boldsymbol{X}_{(n+1):N},W_{1:N}))}^{k+3}
       \right].
    \end{split}
\end{equation}

Since $\forall p \in \mathbb{Z}_{\geq 1}$ and $q \in \mathbb{Z}_{\geq 1}$, by Jesen's and H\"older's inequalities,
\begin{equation}
    \begin{split}
    &E_{(\boldsymbol{X}_{(n+1):N},W_{1:N})}\left[v_N^p(\boldsymbol{X}_{(n+1):N},W_{1:N})Rmd_6^q(\boldsymbol{X}_{(n+1):N},W_{1:N}))
        \right]\\
        \leq &
    \left|E_{(\boldsymbol{X}_{(n+1):N},W_{1:N})}\left[v_N^p(\boldsymbol{X}_{(n+1):N},W_{1:N})Rmd_6^q(\boldsymbol{X}_{(n+1):N},W_{1:N}))
        \right]\right|\\
        \leq &
        E_{(\boldsymbol{X}_{(n+1):N},W_{1:N})}\left[\left|v_N^p(\boldsymbol{X}_{(n+1):N},W_{1:N})Rmd_6^q(\boldsymbol{X}_{(n+1):N},W_{1:N}))\right|
        \right]\\
        \leq &
        \sqrt{
E_{(\boldsymbol{X}_{(n+1):N},W_{1:N})}[v_N^{2p}(\boldsymbol{X}_{(n+1):N},W_{1:N})]
        E_{(\boldsymbol{X}_{(n+1):N},W_{1:N})}[
        Rmd_6^{2q}(\boldsymbol{X}_{(n+1):N},W_{1:N}))
        ]
        }\\
        &=
        O(N^{-(p+3q)\min\{r_1, r_2, \frac{1}{2} \}}).
    \end{split}
\end{equation}
Thus, $E_{(\boldsymbol{X}_{(n+1):N},W_{1:N})}\left[v_N^p(\boldsymbol{X}_{(n+1):N},W_{1:N})Rmd_6^q(\boldsymbol{X}_{(n+1):N},W_{1:N}))
        \right] = O(N^{-(p+3q)\min\{r_1, r_2, \frac{1}{2} \}})$ and $\forall k = 2, 3, \cdots,$
\begin{equation}\label{sum-vn-rmd6-k}
\begin{split}
    &E_{(\boldsymbol{X}_{(n+1):N},W_{1:N})}\left[{(v_N(\boldsymbol{X}_{(n+1):N},W_{1:N})+C_0^{-1}Rmd_6(\boldsymbol{X}_{(n+1):N},W_{1:N}))}^{k}\right]\\
    =&\sum_{j=0}^{k}\binom{k}{j} C_0^{-(k-j)}E_{(\boldsymbol{X}_{(n+1):N},W_{1:N})}\left[
    v_N^j(\boldsymbol{X}_{(n+1):N},W_{1:N})Rmd_6^{k-j}(\boldsymbol{X}_{(n+1):N},W_{1:N}))
    \right]\\=&
    \sum_{j=0}^{k}\binom{k}{j} C_0^{-(k-j)}
    O(N^{-(j+3(k-j))\min\{r_1, r_2, \frac{1}{2} \}})\\
    =&O(N^{-k\min\{r_1, r_2, \frac{1}{2} \}}).
\end{split}
\end{equation}

Now by Jesen's inequality,
\begin{equation}
    \begin{split}
    &
       \sum_{k=0}^{\infty} E_{(\boldsymbol{X}_{(n+1):N},W_{1:N})}\left[ \left|{(v_N(\boldsymbol{X}_{(n+1):N},W_{1:N})+C_0^{-1}Rmd_6(\boldsymbol{X}_{(n+1):N},W_{1:N}))}^{k+3}\right|
       \right]\\
       \leq 
       &
       \sum_{k=0}^{\infty}
       \sqrt{
           E_{(\boldsymbol{X}_{(n+1):N},W_{1:N})}\left[ {(v_N(\boldsymbol{X}_{(n+1):N},W_{1:N})+C_0^{-1}Rmd_6(\boldsymbol{X}_{(n+1):N},W_{1:N}))}^{2k+6}
       \right]
       }\\
       \sim
       &
       O(N^{-3\min\{r_1, r_2, \frac{1}{2} \}})\sum_{k=0}^{\infty}
       O(N^{-k\min\{r_1, r_2, \frac{1}{2} \}})\\
       <&
       \infty,
    \end{split}
\end{equation}
for sufficiently large $N$.

By Fubini's Theorem and \Cref{e-last-reminder,sum-vn-rmd6-k}, we have
\begin{equation}
    \begin{split}
       & E_{(\boldsymbol{X}_{(n+1):N},W_{1:N})}\left[
\frac{{\left[v_N(\boldsymbol{X}_{(n+1):N},W_{1:N})+C_0^{-1}Rmd_6(\boldsymbol{X}_{(n+1):N},W_{1:N})\right]}^3}{1+\left[v_N(\boldsymbol{X}_{(n+1):N},W_{1:N})+C_0^{-1}Rmd_6(\boldsymbol{X}_{(n+1):N},W_{1:N})\right]}
      \right]\\
      =&
       \sum_{k=0}^{\infty} {(-1)}^k      E_{(\boldsymbol{X}_{(n+1):N},W_{1:N})}\left[{(v_N(\boldsymbol{X}_{(n+1):N},W_{1:N})+C_0^{-1}Rmd_6(\boldsymbol{X}_{(n+1):N},W_{1:N}))}^{k+3}
       \right] \\
       = &
       O(N^{-3\min\{r_1, r_2, \frac{1}{2} \}})\\
       =&o(N^{-1}).
    \end{split}
\end{equation}

Therefore, we conclude that by \Cref{e-rmd7}, $E_{(\boldsymbol{X}_{(n+1):N},W_{1:N})}\left[Rmd_7(\boldsymbol{X}_{(n+1):N},W_{1:N})\right] = o(N^{-1})$. 
\end{proof}

\printbibliography

\end{document}